\documentclass[aps,pre,twocolumn]{revtex4}
\usepackage{amsmath,amssymb,mathrsfs}
\usepackage{psfig}

\begin{document}

\title{Homogenization of Maxwell's Equations in Periodic Composites: Boundary Effects and Dispersion Relations} 

\author{Vadim A. Markel~\footnote{E-mail: vmarkel@mail.med.upenn.edu}}
\affiliation{Department of Radiology and Graduate Group in Applied Mathematics and Computational Science, University of Pennsylvania, Philadelphia, PA 19104}

\author{John C. Schotland~\footnote{E-mail: schotland@umich.edu}}
\affiliation{Department of Mathematics, University of Michigan, Ann Arbor, MI 48109}

\begin{abstract}
  We consider the problem of homogenizing the Maxwell equations for
  periodic composites. The analysis is based on Bloch-Floquet theory.
  We calculate explicitly the reflection coefficient for a half-space,
  and derive and implement a computationally-efficient
  continued-fraction expansion for the effective permittivity. Our
  results are illustrated by numerical computations for the case of
  two-dimensional systems. The homogenization theory of this paper is
  designed to predict various physically-measurable quantities rather
  than to simply approximate certain coefficients in a PDE.
\end{abstract}

\date{\today}
\maketitle

\section{Introduction}
\label{sec:intro}

Theories of electromagnetic homogenization of composite
materials---also known as effective medium theories (EMTs)---have a
history which dates to the time of J.C. Maxwell. Nevertheless, these
theories continue to attract attention and even controversy, as
evidenced by recent reviews~\cite{simovski_09_1,simovski_11_1} and
many references therein. In applied mathematics, the theory of
homogenization based on multiscale analysis of partial differential
equations is also
well-established~\cite{bensoussan_78_1,oleinik_book_92,milton_book_02,tartar_book_09}.
However, interest in EMTs has been steadily on the rise for the past
ten years with conceptually new approaches continuing to
appear~\cite{silveirinha_07_1,tsukerman_11_1,pors_11_1,silveirinha_11_1}.
This can be explained, perhaps, by noting that the tasks of relating
the existing mathematical theories to physical observables and of
determining the range of applicability of a given theory have not been
fully addressed, particularly, for the case of Maxwell's equations.
Indeed, in the past ten years or so, homogenization theories have been
applied to obtain ``extreme'' properties of electromagnetic
composites, including the phenomenon of strong ``artificial''
magnetism. At the same time, a significant experimental progress has
been recently made in manufacturing deeply-subwavelength (in the
visible spectral range) periodic metallic
nanostructures~\cite{feng_10_2,feng_11_1,feng_11_2}. The question is
whether the existing theories are directly applicable or accurate
enough to guide the experimental design of periodic nanostructures of
desirable properties. Another reason for the renewed interest in
homogenization theories is that, in addition to abstract mathematical
results, there is a need for efficient, stable computational methods.
Thus the question of how to construct physically-relevant and
computationally-effective EMTs and determine their limits of
applicability have not been completely settled.

This paper is an attempt to address the above issues for the case of
periodic composites; random media are not considered. The framework we
develop is based on the Bloch-Floquet expansion, which is a well-known
tool in homogenization
theory~\cite{milton_book_02,krokhin_02_1,krokhin_04_1,cherednichenko_07_1,guenneau_07_1,guenneau_11_1,craster_11_1}.
However, in several aspects, we go beyond the standard theory. In
particular, (i) we explicitly account for boundary effects and derive
a general expression for the reflection coefficient (many existing
homogenization theories consider infinite composites) (ii) we make use
of the integral equation formulation of scattering theory for the
Maxwell equations.  The resulting formulas for the effective medium
parameters (EMPs) have a different mathematical structure than those
derived from partial differential equations (iii) we develop a
computationally-efficient algorithm for calculating the EMPs. The
algorithm is based on a continued-fraction expansion of the
self-energy and is obtained from a new result on the resolvent of a
linear operator and (iv) a numerical study of stability and
convergence is performed for some test cases. Stability is
investigated by comparing the results for inclusions of the same
volume fraction but different shape and of the same shape but
different volume fractions.

It is useful to recognize that all EMTs can be classified as either
{\em standard} or {\em extended}. A standard EMT is obtained by taking
the limit $h\rightarrow 0$, where $h$ is the scale of the medium's
heterogeneity; in this paper, $h$ is the lattice spacing. In standard
theories, $h$ is viewed as a mathematically- and
physically-independent variable and the resulting EMPs are independent
of $h$, as long as the latter is small enough for the theory to be
applicable. Another feature of all standard theories is the so-called
law of unaltered ratios~\cite{bohren_09_1}, which states that, if a
composite medium is made of several constituents with permittivities
$\epsilon_j$ ($j=1,2,\ldots$) and if $\epsilon_j \rightarrow
\lambda\epsilon_j$ ($\lambda>0$), then the effective permittivity
$\bar{\epsilon}$ also scales as $\bar{\epsilon} \rightarrow
\lambda\bar{\epsilon}$.

Extended EMTs came to the fore (at least in the physics literature)
in~\cite{niklasson_81_1,doyle_89_1}. The basic idea of these papers is
to note that one can compute the exact electric and magnetic
polarizabilities, $\alpha_e$ and $\alpha_m$, of a spherical particle
through the use of the first Lorenz-Mie coefficients, $a_1$ and $b_1$,
even when the sphere in question is not small compared to the external
wavelength.  These polarizabilities can be used to construct an
``extended'' Maxwell-Garnett approximation. Since $a_1$ and $b_1$ are
not proportional to the sphere volume, except in the quasistatic
limit, the resultant EMTs contain the sphere radius explicitly. In
Refs.~\cite{nicorovici_95_1,nicorovici_95_2}, a counter-intuitive
effect of non-commuting limits was demonstrated. Specifically, it was
shown that insofar as the effective refractive index of a photonic
crystal is computed from the slope of the dispersion curve near the
$\Gamma$-point, different results are generally obtained depending on
which of the two limits, $h\rightarrow 0$ and $\epsilon_1 \rightarrow
\infty$ is taken first, where $\epsilon_1$ is the permittivity of one
of the components of the photonic crystal.  A related point is that a
complete theory of homogenization requires error estimates. That is,
it is essential to determine how the error in the homogenization limit
depends upon contrast.  Moreover, the reflection and transmission
properties of the composite medium have not been
considered~\cite{niklasson_81_1,doyle_89_1,nicorovici_95_1,nicorovici_95_2}.

In this paper, we develop a standard EMT. However, when considering
reflection and refraction at a planar interface, we derive formulas
for the reflection and transmission coefficients, which are valid for
finite values of $h$. Then we show that taking the limit $h\rightarrow
0$ results in the standard Fresnel coefficients. In this case, the
electric and magnetic properties of the medium constituents do not
mix, in agreement with~\cite{wellander_03_1}. That is, if we begin
with nonmagnetic inclusions, the resultant composite is also
nonmagnetic.  An extended EMT can be obtained by taking a different
limit, in which the permittivity of one of the constituents scales as
$1/h^2$~\cite{felbacq_05_1}. Here we note again the existence of the
effect of non-commuting
limits~\cite{nicorovici_95_1,nicorovici_95_2,poulton_04_1}, which
calls for additional scrutiny of the homogenization results thus
obtained.  In particular, one would expect that, in the limit
considered in~\cite{felbacq_05_1}, Fresnel formulas would also be
reproduced, but with a nontrivial magnetic permeability. We have not
been able to show that this is the case. In other words, it is not
clear whether the EMPs obtained from an extended EMT are independent
of the incidence angle or, more generally, of the type of incident
wave. This is in accord
with~\cite{bohren_86_1,menzel_10_1,menzel_10_2,simovski_10_1,andryieuski_10_1,paul_11_1},
which find that the conditions under which metamaterials exhibiting
strong magnetic resonances can be assigned purely local
(incidence-angle-independent) EMPs are rather restrictive. The same
point has been made in the recent review article~\cite{simovski_11_1}.

An additional feature by which EMTs can be classified is the physical model
of the medium. In the model of dipole lattices, the medium is thought
of as being composed of point particles which are completely
characterized by their polarizabilities (electric and, possibly,
magnetic) and whose shape and size do not enter into the problem
directly~\cite{sipe_74_1,draine_93_1,belov_05_1}.  Alternatively, one
can consider the space as a two-component continuous
medium~\cite{bergman_78_1,bergman_79_1,bergman_79_2}. The point-dipole
model is appealing because of its simplicity but leads to serious
mathematical problems. The so-called dipole sum (also known as the
lattice sum or the dipole self-energy), which plays a key role in this
model, diverges in the case of three-dimensional lattices.  While it
is true that even divergent series can be summed by means of applying
various mathematical tricks, the results obtained depend on the
particular trick used, a state of affairs that is not very satisfying.
Therefore, we will adopt from the start a model of a two-component
continuous medium. As the development in this paper progresses, it
will become apparent why the point-dipole model is inadequate.

The mathematical development in this paper begins by considering the
integral equation obeyed by the polarization field, which is
introduced in Sec.~\ref{sec:basic}. In Sec.~\ref{sec:inf_latt}, we
derive a homogenization theory of the standard type for infinite
periodic media. Reflection and refraction at a planar boundary is
considered in Sec.~\ref{sec:half-space}. In Sec.~\ref{sec:point-bulk},
we discuss the correspondence between the point-dipole model and the
continuous-medium model of this paper. One mathematically-novel
element of the theory developed herein is a continued-fraction
expansion of the effective permittivity, which is derived in
Sec.~\ref{sec:CF} and used in the numerical simulations of
Sec.~\ref{sec:num}. The expansion has its origins in a theorem on
resolvents of general linear operators (with no special symmetry
properties), which is stated in Sec.~\ref{sec:CF} and proved in the
appendices. A discussion and a brief summary of results are contained
in Secs.~\ref{sec:disc} and \ref{sec:summ}.

\section{Basic equations}
\label{sec:basic}

\begin{figure}
\centerline{\psfig{file=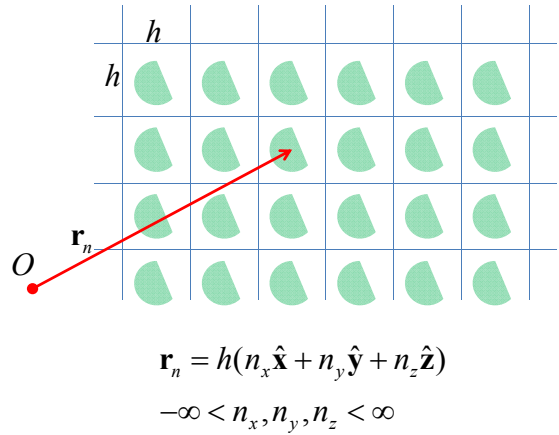,width=9cm,clip=}}
\caption{\label{fig:sketch_3D} (color online) Sketch of the geometry
  considered: an infinite 3D lattice.}
\end{figure}

The geometry of the problem we consider is sketched in
Fig.~\ref{fig:sketch_3D}. The medium consists of two intrinsically
non-magnetic constituents: a host medium of permittivity $\epsilon_b$
and periodically-arranged inclusions of permittivity $\epsilon_a$. In
practice, the host is often a transparent dielectric with ${\rm
  Re}\epsilon_b > 0$, $0<{\rm Im}\epsilon_b \ll {\rm Re}\epsilon_b$,
and the inclusions are metallic.  However, the theory of this paper
places no such restriction on the permittivities and only requires
that ${\rm Im}\epsilon_b>0$, ${\rm Im}\epsilon_a>0$. In the case when
the host medium is vacuum, we will take $\epsilon_b = 1 + i0$.  The
inclusions are arranged on a cubic lattice of period $h$.  The
position vector of the center of each unit cell is denoted by ${\bf
  r}_n$, where $n$ can be viewed as a composite index:
$n=(n_x,n_y,n_z)$ and ${\bf r}_n = h(\hat{\bf x}n_x + \hat{\bf y}n_y +
\hat{\bf z}n_z)$. Whenever a summation over $n$ (or a similar
composite index $m$) appears in the text, it is implied that the sum
runs over all three integer indexes. Inside the $n$th cell, the
spatial region $\Omega_n$ has the permittivity $\epsilon_a$, and the
rest of the cell has the background permittivity $\epsilon_b$.  All
regions $\Omega_n$ are identical and only differ by translation. It is
assumed that $\Omega_n$ can touch but not cross the cell boundaries.
No assumption on the connectivity of $\Omega_n$ is made.  The union of
all regions $\Omega_n$ is denoted by $\Omega_{\rm tot}$ and the volume
of each region by $V$:
\begin{equation}
\label{Omega_tot}
 \Omega_{\rm tot} = \bigcup_n \Omega_n  \ , \ \ \int_{\Omega_n} d^3r = V \ .
\end{equation}
We work in the frequency domain and the common factor $\exp(-i\omega
t)$ is suppressed. All frequency-dependent quantities, such as the
permittivities $\epsilon_b$ and $\epsilon_a$, are evaluated at the
frequency $\omega$.

The mathematical development in this paper begins with the integral
equation
\begin{equation}
\label{main_eq}
{\bf P}({\bf r}) = \frac{3\chi}{4\pi}\left[{\bf E}_i({\bf r}) +
  \int_{\Omega_{\rm tot}} G({\bf r},{\bf r}^\prime) {\bf P}({\bf r}^\prime)
  d^3r^\prime \right] \ , \ \ {\bf r} \in \Omega_{\rm tot} \ .
\end{equation}
\noindent 
Here ${\bf P}({\bf r})$ is the vector of ``polarization'', which is
related to the electric field ${\bf E}({\bf r})$ by
\begin{equation}
\label{P_def}
{\bf P}({\bf r}) = \frac{\epsilon({\bf r}) - \epsilon_b}{4\pi\epsilon_b}{\bf
  E}({\bf r}) \ ,
\end{equation}
\noindent
${\bf E}_i({\bf r})$ is the incident electric field, $G({\bf r},
{\bf r}^\prime)$ is the {\em regular} part of the free-space, retarded
Green's tensor, and
\begin{equation}
\label{chi_def}
\chi = \frac{\epsilon_a - \epsilon_b}{\epsilon_a + 2\epsilon_b} \ .
\end{equation}
\noindent
Note that ${\bf P}({\bf r})$ defined in (\ref{P_def}) is not the true
physical polarization, which is given by $[\epsilon({\bf r}) - 1] {\bf
  E}({\bf r})/4\pi$, but rather an auxiliary field; ${\bf P}({\bf r})$
vanishes in the host medium while the true polarization does not.

In what follows, we will make use of the spatial Fourier transform of the
Green's tensor, namely,
\begin{equation}
\label{GR_F}
G({\bf r},{\bf r}^\prime) = \frac{4\pi}{3}\int \frac{d^3
p}{(2\pi)^3} K({\bf p})\exp\left[ i {\bf p} \cdot ({\bf r} - {\bf
  r}^\prime) \right] \ ,
\end{equation}
\noindent
where
\begin{equation}
\label{K_def}
K({\bf p}) = \frac{2 k_b^2 + p^2 - 3{\bf p} \otimes {\bf p}}{p^2 -
  k_b^2} \ ,
\end{equation}
\noindent 
and
\begin{equation}
k_b^2 = \epsilon_b k \ , \ \ k = \frac{\omega}{c} \ . 
\end{equation}
\noindent
Here the wave number in the background medium is denoted by $k_b$ and
the wave number in vacuum by $k$. We note that the integral equation
(\ref{main_eq}) is equivalent to the pair of curl Maxwell equations
written in the frequency domain.

\section{Waves in infinite lattices}
\label{sec:inf_latt}

\subsection{Three-dimensional lattices}
\label{subsec:inf_3D}

Consider the propagation of a wave in a three-dimensional infinite
lattice. In this case, the incident field is absent and
Eq.~(\ref{main_eq}) must be satisfied for ${\bf E}_i=0$. We seek the
solution to Eq.~(\ref{main_eq}) in the form of a Bloch wave:
\begin{equation}
\label{BW_1}
{\bf P}({\bf r}) = \exp\left ( i{\bf q} \cdot {\bf r}_n\right) {\bf F}({\bf r} - {\bf
  r}_n) \ , \ \ \  {\bf r} \in \Omega_n \ .
\end{equation}
\noindent
Here ${\bf q}$ is the Bloch wave number and ${\bf F}({\bf r})$ is a
vector function. Equivalently, if we write ${\bf r} = {\bf r}_n + {\bf
  R}$, then
\begin{equation}
\label{BW_2}
  {\bf P}({\bf r}_n + {\bf R}) = \exp\left( {i{\bf q} \cdot {\bf
        r}_n}\right) {\bf F}({\bf R}) \ , \ \ \ {\bf R} \in \Omega \ .
\end{equation}
\noindent
In this formula, $\Omega \equiv \Omega_0$ is the region centered at
the origin of a rectangular reference frame. From the above relation, we find the
equation obeyed by ${\bf F}({\bf R})$:
\begin{equation}
\label{F_W}
{\bf F}({\bf R}) = \frac{3\chi}{4\pi}\int_\Omega W ({\bf
  R}, {\bf R}^\prime) {\bf F}({\bf R}^\prime) d^3 R^\prime \ ,
\end{equation}
\noindent
where
\begin{align}
\label{W_def}
W({\bf R}, {\bf R}^\prime) =
\sum_m G({\bf r}_n + {\bf R}, {\bf r}_m + {\bf R}^\prime)  \nonumber \\ 
\times \exp\left[ i {\bf q} \cdot ({\bf r}_m - {\bf r}_n) \right] \ .
\end{align}
\noindent
It can be seen that $W$ is independent of $n$. It should also be noted
that the summation in Eq.~(\ref{W_def}) runs over the entire lattice,
including the term $m=n$. In theories that consider point-like
particles, the dipole sum is defined as an incomplete lattice sum,
which excludes the term $m=n$. This makes application of the Poisson
summation formula problematic and unnecessarily complicates the
mathematics~\cite{belov_05_1}.

Returning to our derivation, we evaluate $W$ as
\begin{align}
\label{W_eval}
W({\bf R}, {\bf R}^\prime) = \frac{4\pi}{3} \int \frac{d^3
  p}{(2\pi)^3} K({\bf p}) \exp\left[i{\bf p} \cdot ({\bf R} - {\bf
    R}^\prime)\right] \nonumber \\
 \times \sum_m\exp\left[ i({\bf p} - {\bf q})\cdot({\bf
    r}_n - {\bf
    r}_m) \right] \nonumber \\
= \frac{4\pi}{3h^3} \sum_{\bf g} K({\bf q+g}) \exp\left[ i({\bf q} +
  {\bf g}) \cdot ({\bf R}-{\bf R}^\prime) \right] \ ,
\end{align}
\noindent
where 
\begin{equation}
\label{g_def}
{\bf g} = \frac{2\pi}{h}\left( \hat{\bf x}n_x + \hat{\bf y}n_y +
  \hat{\bf z}n_z \right) 
\end{equation}
\noindent
are the reciprocal lattice vectors and we have used the Poisson
summation formula 
\begin{equation}
\label{Poisson}
\sum_m\exp\left[i ({\bf p}-{\bf q}) \cdot ({\bf r}_m - {\bf r}_n) \right] =
  \left( \frac{2\pi}{h} \right)^3 \sum_{\bf g} \delta({\bf p} - {\bf q}
  - {\bf g}) \ .
\end{equation}
\noindent
The summation in Eqs.~(\ref{W_eval}),(\ref{Poisson}) is over the
complete set of reciprocal lattice vectors; equivalently, it can be
viewed as summation over the triplet of indexes $(n_x,n_y,n_z)$ which
appear in (\ref{g_def}).

The series in the right-hand side of (\ref{W_eval}) diverges when
${\bf R} = {\bf R}^\prime$. This is the well-known divergence of the
dipole sum~\cite{abajo_07_1} which hinders the analysis of waves in
lattices made of point-like polarizable particles. The model of
point-like dipoles is discussed in more detail in
Sec.~\ref{sec:point-bulk}. In the equations derived above, the
divergence is of no concern because $W({\bf R}, {\bf R}^\prime)$
appears only inside an integral and the singularity in question is
integrable.

Upon substitution of (\ref{W_eval}) into (\ref{F_W}), we obtain
\begin{align}
\label{F_R}
{\bf F}({\bf R}) = \frac{\chi}{h^3}\sum_{\bf g} K({\bf q}+{\bf g})
\exp\left[i ({\bf q}+{\bf g})\cdot {\bf R}\right] \nonumber \\
\times \int_\Omega {\bf F}({\bf
  R}^\prime) \exp\left[-i ({\bf q} + {\bf g}) \cdot {\bf R}^\prime \right] d^3R^\prime
\ .
\end{align}
\noindent
It follows from (\ref{F_R}) that ${\bf F}({\bf R})$ can be expanded as
\begin{equation}
\label{F_g_def}
{\bf F}({\bf R}) = \sum_{\bf g} {\bf F}_{\bf g} \exp\left[i ({\bf q} +
  {\bf g}) \cdot {\bf R} \right]
\end{equation}
\noindent
and that the expansion coefficients satisfy the system of equations
\begin{equation}
\label{eq_F}
{\bf F}_{\bf g} = \rho \chi K({\bf q}+{\bf g}) \sum_{{\bf g}^\prime}
M({\bf g} - {\bf g}^\prime) {\bf F}_{{\bf g}^\prime} \ ,
\end{equation}
\noindent
where $\rho = V /h^3$ is the volume fraction of inclusions and $M({\bf
  g})$ is defined by the expression
\begin{equation}
\label{M_def}
M({\bf g}) = \frac{1}{V} \int_\Omega \exp\left(- i {\bf g} \cdot
  {\bf R}\right) d^3 R \ .
\end{equation}
\noindent
Note that $M({\bf g})$ is defined only by the shape of the inclusions
and is invariant with respect to the coordinate rescaling ${\bf r}
\rightarrow \lambda {\bf r}$.  Some mathematical properties and
calculations of $M({\bf g})$ for special geometries are given in
Appendix~\ref{app:M}.

So far, we have simply restated the well known theorem of Floquet. The
eigenproblem~(\ref{eq_F}) defines the band structure of a photonic
crystal. It is well known that EMTs are not always applicable to
photonic crystals.  However, there exists a regime in which EMPs can
be reasonably introduced, and this regime will be explored below.
Namely, if $qh, k_b h \ll 1$, we can consider the cases ${\bf g}=0$
and ${\bf g} \neq 0$ in (\ref{eq_F}) separately.  This yields the
following equations:
\begin{subequations}
\label{sys_3D}
\begin{align}
\label{g=0}
{\bf F}_0 &= \rho\chi K({\bf q})\left[{\bf F}_0 + \sum_{\bf g\neq 0} M(-{\bf
    g}) {\bf F}_{\bf g} \right] \ , \\
{\bf F}_{\bf g} &= \rho\chi Q({\bf g}) \left[M({\bf g}) {\bf F}_0 +
  \sum_{{\bf g}^\prime \neq 0}M({\bf g} - {\bf g}^\prime) {\bf F}_{{\bf
      g}^\prime} \right] \ , \quad {\bf g}\neq 0 \nonumber  \ , \\
 \label{g=/=0}
\end{align}
\end{subequations}
\noindent
where
\begin{equation}
\label{Q_def}
Q({\bf g}) \equiv \lim_{h\rightarrow 0} K({\bf q}+{\bf
  g}) = 1 - 3 \hat{\bf g} \otimes \hat{\bf g}\ , \quad {\bf g}\neq 0 \ .
\end{equation}
\noindent
Here $\hat{\bf g}= {\bf g}/\vert{\bf g}\vert$ is a unit vector.

The derivation of Eqs.~(\ref{sys_3D}) is one of the key developments
of this paper. It can be seen that the equations in (\ref{g=/=0}) do
not contain the variables $k$ or ${\bf q}$, but are completely defined
by the geometry of inclusions and by the variable $\chi$. Moreover,
these equations are invariant with respect to the rescaling ${\bf r}
\rightarrow \lambda {\bf r}$. For any given value of ${\bf F}_0$,
(\ref{g=/=0}) can be solved uniquely as ${\bf F}_{\bf g} = A_{\bf
  g}{\bf F}_0$, where the tensors $A_{\bf g}$ depend on ${\bf g}$, the
shape of inclusions, and on $\chi$. Given this result, we can write
\begin{equation}
\label{Sigma_def}
\sum_{\bf g\neq 0} M(-{\bf g}) {\bf F}_{\bf g} = \sum_{\bf g\neq 0}
M(-{\bf g}) A_{\bf g}{\bf F}_0 = \Sigma {\bf F}_0 \ ,
\end{equation}
\noindent
where the tensor $\Sigma$ has all the properties of $A_{\bf g}$ and,
in addition, is independent of ${\bf g}$. It will be shown in
Sec.~\ref{sec:CF} that $\Sigma$ plays the role of the self-energy and
originates due to the electromagnetic interaction within and between
the inclusions. It will also be shown that $\Sigma$ can be computed as
a resolvent of a linear operator, which depends only on the shape of
inclusions.

Using the notation introduced in (\ref{Sigma_def}), we can rewrite
(\ref{g=0}) as
\begin{equation}
\label{F_0}
\left[1 - \rho\chi K({\bf q})(1 + \Sigma) \right] {\bf F}_0 = 0 \ .
\end{equation}
\noindent
This equation has nontrivial solutions if
\begin{equation}
\label{det=0}
\det\left[ 1 - \rho\chi K({\bf q}) (1 + \Sigma) \right] = 0 \ .
\end{equation}
\noindent
Here the quantity in the square brackets is a $3\times 3$ matrix. For
a fixed value of $k$ (that is, at a fixed frequency), the condition
(\ref{det=0}) is an algebraic equation with respect to the Cartesian
components of the Bloch vector ${\bf q}$. Roots of this equation,
computed at different values of $k$, determine the dispersion relation
${\bf q}(k)$. There can be more than one branch of the dispersion
relation corresponding to different polarization states. By
polarization of the mode, we mean here the direction of the vector
${\bf F}_0$. 

EMPs can be inferred by comparing these results to the polarization
states and dispersion relation in a homogeneous medium characterized
by tensor permittivity and permeability $\bar{\epsilon}$ and
$\bar{\mu}$. However, it is not possible to determine $\bar{\epsilon}$
and $\bar{\mu}$ {\em simultaneously and uniquely} from consideration
of the dispersion relation alone. For example, in an isotropic medium,
only the product of these two quantities (the squared refractive
index) can be unambiguously obtained. Indeed, the dispersion relation
in such a medium is invariant with respect to the transformation
$\bar{\epsilon} \rightarrow \xi \bar{\epsilon}$, $\bar{\mu}
\rightarrow \xi^{-1} \bar{\mu}$, where $\xi \neq 0$ is a complex
number. To determine $\bar{\epsilon}$ and $\bar{\mu}$ uniquely, one
must consider reflection and refraction at the medium boundary. This
will be done in Sec.~\ref{sec:half-space}. In particular, it will be
shown that, in order to obtain the correct Fresnel reflection
coefficients, one must set $\bar{\mu} = 1$.

To summarize the results of this section, the electromagnetic modes of a
medium can be found if the tensor $\Sigma$ is known. Computation of
the modes involves diagonalization of a $3\times 3$ matrix, while the
tensor $\Sigma$ is uniquely determined by the solution to
Eqs.~(\ref{g=/=0}). The latter is an infinite set of equations which
must be appropriately truncated in numerical computations. Thus, we
have reduced the homogenization problem to solving a set of algebraic
equations in which the shape of the inclusions appears only in the
functions $M({\bf g})$.

\subsection{Main homogenization result for three-dimensional composites
  with well-defined optical axes}
\label{subsec:iso}

The standard description of electromagnetic waves in anisotropic
crystals is based on the assumption that the tensors $\bar{\epsilon}$
and $\bar{\mu}$ commute and are simultaneously diagonalizable by a
rotation of the reference frame, with purely real Euler angles. The
axes of the reference frame in which $\bar{\epsilon}$ and $\bar{\mu}$
are diagonal are known as the optical axes. Moreover, standard
textbooks often specialize to the case $\bar{\mu}=1$, which is a very
good approximation in crystal optics~\cite{landau_ess_84}. In the most
general case, however, the tensors $\bar{\epsilon}$ and $\bar{\mu}$ do
not commute, which gives rise to two distinct sets of electric and
magnetic axes. Furthermore, $\bar{\epsilon}$ and $\bar{\mu}$ are
complex-valued, symmetric and, hence, non-Hermitian matrices. A purely
real rotation that diagonalizes any one of these two tensors may not
exist. A mathematically tractable dispersion relation for the most
general case has been derived only recently~\cite{itin_10_1}, and we
will use below one particular case of this result.

For the composite medium consisting of non-magnetic components, which
is considered in this paper, the situation is somewhat simpler. It can
be seen from Eq.~(\ref{det=0}) that a unique set of optical axes
exists if the tensor $\Sigma$ is diagonalizable by a real-angle
rotation of the reference frame. Thus, the issue of commutability of
two different tensors does not arise in this case. 

In this subsection, we assume that the optical axes of the composite
medium (that is, the principal axes of the tensor $\Sigma$) exist and,
moreover, coincide with the crystallographic axes of the medium. The
latter assumption is not really necessary but any composite can be cut
is such a way that it holds.  In particular, $\Sigma$ is diagonal in
the reference frame defined by the crystallographic axes (which is the
laboratory frame in this paper) if the inclusions are symmetric with
respect to reflections in each of the $xy$-, $xz$- and $yz$-planes.
The principal values of $\Sigma$, denoted by $\Sigma_{\alpha\alpha}$
($\alpha=x,y,z$), are not necessarily equal in this case.  The two
familiar examples of reflection-symmetric inclusions which result in
all three principal values being different are a general
parallelepiped and an ellipsoid with unequal semi-axes.  However, if
the inclusions also have cubic symmetry (which, in addition to
reflections, includes rotations about each axis by the angle $\pi/4$),
then $\Sigma$ is reduced to a scalar and the effective medium is
isotropic.

\subsubsection{General direction of propagation}
\label{subsubsec:q_general}

Let the tensor $\Sigma$ be diagonal in the rectangular frame $xyz$.
We then use the expression (\ref{K_def}) for $K({\bf q})$, evaluate
the determinant in Eq.~(\ref{det=0}), and obtain the following
equation:
\begin{equation}
\label{det=0_expand_comp}
\frac{\epsilon_b^2 \prod_{\alpha} \left[1 + 2\rho\chi (1 +
    \Sigma_{\alpha\alpha}) \right] }{\left( q^2 - k_b^2 \right)^2}
{\mathscr D}_c(k,{\bf q}) = 0 \ ,  
\end{equation}
\noindent
where
\begin{equation}
\label{Pc_def}
{\mathscr D}_c(k,{\bf q}) = k^4 - {\mathscr A}_c({\bf q}) k^2 +
{\mathscr B}_c({\bf q}) \ , \
\end{equation}
\noindent
and
\begin{subequations}
\label{ac_bc_def}
\begin{align}
{\mathscr A}_c({\bf q}) = &
  q_x^2 \left(\frac{1}{\eta_y} + \frac{1}{\eta_z}\right) 
 + q_y^2 \left(\frac{1}{\eta_x} +
  \frac{1}{\eta_z}\right) \nonumber \\
+ & q_z^2 \left(\frac{1}{\eta_x} + \frac{1}{\eta_y}\right) \ , 
\label{ac_def} \\
{\mathscr B}_c({\bf q}) = & \frac{q_x^4}{\eta_y\eta_z} +
\frac{q_y^4}{\eta_x\eta_z} +
\frac{q_z^4}{\eta_x\eta_y} + \frac{q_x^2 q_y^2}{\eta_z} \left(\frac{1}{\eta_x} +
  \frac{1}{\eta_y}\right) \nonumber \\
+ & \frac{q_x^2 q_z^2}{\eta_y}
\left(\frac{1}{\eta_x} + \frac{1}{\eta_z}\right) + \frac{q_y^2
  q_z^2}{\eta_x} \left(\frac{1}{\eta_y} + \frac{1}{\eta_z}\right) \ . 
\label{bc_def}
\end{align}
\end{subequations}
\noindent
The quantities $\eta_\alpha$ are given by
\begin{equation}
\label{eta_def}
\eta_\alpha = \epsilon_b \frac{1 + 2\rho\chi (1 +
  \Sigma_{\alpha\alpha})}{1 - \rho\chi (1 + \Sigma_{\alpha\alpha})} \
, \ \ \alpha=x,y,z
\end{equation}
\noindent
and the subscript in ${\mathscr D}_c$, ${\mathscr A}_c$ and ${\mathscr
  B}_c$ has been used to emphasize that these expressions are
applicable to composite media and have been obtained by evaluating the
left-hand side of (\ref{det=0}).

The set of dispersion relations
(\ref{det=0_expand_comp})-(\ref{ac_bc_def}) should be compared to the
analogous set of equations in a homogeneous medium characterized by
the effective tensors $\bar{\epsilon}$ and $\bar{\mu}$. Generally, the
dispersion relation in such media reads
\begin{subequations}
\begin{equation}
\label{det=0_hom_mu-inv}
\det\left[\left( {\bf q} \times \bar{\mu}^{-1} {\bf q}
  \times \right) + k^2 \bar{\epsilon} \right] = 0 \ ,
\end{equation}
\noindent
if $\bar{\mu}^{-1}$ exists, or
\begin{equation}
\label{det=0_hom_eps-inv}
\det\left[\left({\bf q} \times \bar{\epsilon}^{-1} {\bf q}
  \times \right) + k^2 \bar{\mu} \right] = 0 \ ,
\end{equation}
\end{subequations}
\noindent
if $\bar{\epsilon}^{-1}$ exists. If both $\bar{\mu}$ and
$\bar{\epsilon}$ are invertible, the two equations
(\ref{det=0_hom_mu-inv}) and (\ref{det=0_hom_eps-inv}) are identical.

For homogenization theory to be applicable, the effective medium
must have the same symmetry as the composite. It is evident,
therefore, that the principal axes of $\Sigma$ should coincide with
the optical axes of the effective medium. Denote the principal values
of $\bar{\epsilon}$ and $\bar{\mu}$ by $\bar{\epsilon}_{\alpha\alpha}$
and $\bar{\mu}_{\alpha\alpha}$. Let us further assume that $\bar{\mu}$
is invertible. In this case, Eq.~(\ref{det=0_hom_mu-inv}) takes the
following form:
\begin{equation}
\label{det=0_expand_homo}
k^2 \bar{\epsilon}_{xx} \bar{\epsilon}_{yy} \bar{\epsilon}_{zz}
{\mathscr D}_h(k,{\bf q}) = 0 \ , 
\end{equation}
\noindent
where
\begin{equation}
\label{Ph_def}
{\mathscr D}_h(k,{\bf q}) = k^4 - {\mathscr A}_h({\bf q}) k^2 +
{\mathscr B}_h({\bf q})
\end{equation}
\noindent
and
\begin{subequations}
\label{ah_bh_def}
\begin{align}
  {\mathscr A}_h({\bf q}) = & q_x^2 \left(\frac{1}{\bar{\epsilon}_{yy}
      \bar{\mu}_{zz}} + \frac{1}{\bar{\epsilon}_{zz} \bar{\mu}_{yy}}\right) +
  q_y^2 \left(\frac{1}{\bar{\epsilon}_{xx} \bar{\mu}_{zz}} +
    \frac{1}{\bar{\epsilon}_{zz} \bar{\mu}_{xx}}\right) \nonumber \\
  + & q_z^2 \left(\frac{1}{\bar{\epsilon}_{xx} \bar{\mu}_{yy}} +
    \frac{1}{\bar{\epsilon}_{yy} \bar{\mu}_{xx}}\right) \ ,
\label{ah_def} \\
{\mathscr B}_h({\bf q}) = &
\frac{q_x^4}{\bar{\epsilon}_{yy}\bar{\epsilon}_{zz}\bar{\mu}_{yy}\bar{\mu}_{zz}} +
\frac{q_y^4}{\bar{\epsilon}_{xx}\bar{\epsilon}_{zz}\bar{\mu}_{xx}\bar{\mu}_{zz}} +
\frac{q_z^4}{\bar{\epsilon}_{xx}\bar{\epsilon}_{yy}\bar{\mu}_{xx}\bar{\mu}_{yy}} \nonumber \\
+ & \frac{q_x^2 q_y^2}{\bar{\epsilon}_{zz}\bar{\mu}_{zz}}
\left(\frac{1}{\bar{\epsilon}_{xx} \bar{\mu}_{yy}} +
  \frac{1}{\bar{\epsilon}_{yy}\bar{\mu}_{xx}}\right) \nonumber \\
+ & \frac{q_x^2
  q_z^2}{\bar{\epsilon}_{yy}\bar{\mu}_{yy}}
\left(\frac{1}{\bar{\epsilon}_{xx}\bar{\mu}_{zz}} +
  \frac{1}{\bar{\epsilon}_{zz}\bar{\mu}_{xx} }\right) \nonumber \\
+ & \frac{q_y^2 q_z^2}{\bar{\epsilon}_{xx} \bar{\mu}_{xx}}
\left(\frac{1}{\bar{\epsilon}_{yy} \bar{\mu}_{zz}} +
  \frac{1}{\bar{\epsilon}_{zz} \bar{\mu}_{yy}}\right) \ .
\label{bh_def}
\end{align}
\end{subequations}
\noindent
Here the subscript in ${\mathscr D}_h$, ${\mathscr A}_h$ and
${\mathscr B}_h$ has been used to emphasize that these expressions are
applicable to homogeneous media. In the case $\bar{\mu}_{xx} =
\bar{\mu}_{yy} = \bar{\mu}_{zz} = 1$, (\ref{det=0_expand_homo})
reduces to the well-known Fresnel equation.

The prefactors in Eqs. (\ref{det=0_expand_comp}) and
(\ref{det=0_expand_homo}) are ``almost always'' nonzero, except in the
case of non-dissipative plasmas, which can support longitudinal waves.
This case will be considered by us separately. Assuming that the
prefactors are nonzero, the dispersion relations are ${\mathscr
  D}_c(k, {\bf q})=0$ for the composite medium and ${\mathscr D}_h(k,
{\bf q})=0$ for the homogeneous medium. We can introduce EMPs for the
composite by observing that these two dispersion relations become
identical if we set
\begin{equation}
\label{eff_pars}
\bar{\epsilon}_{\alpha\alpha} = \xi \eta_\alpha \ , \ \ \bar{\mu}_{\alpha\alpha} =
\frac{1}{\xi} \ ,
\end{equation}
\noindent
where $\xi \neq 0$ is an arbitrary complex number. As was already
mentioned, the non-uniqueness in the above definition of the EMPs can
not be removed by considering the dispersion relations alone.

Several remarks regarding the dispersion relations obtained above
should be made. First, in the general case, the functions ${\mathscr
  D}_c(k, {\bf q})$ and ${\mathscr D}_h(k, {\bf q})$ can not be
factorized into products of two quadratic forms in the variables $k$,
$q_x$, $q_y$ and $q_z$. However, such a factorization becomes possible
for special directions of propagation, when one or more of the
Cartesian components of ${\bf q}$ are zero. Examples will be given
below.

Second, the condition (\ref{eff_pars}), which guarantees that
${\mathscr D}_c(k, {\bf q}) = {\mathscr D}_h(k, {\bf q})$, requires
that the effective permeability $\bar{\mu}$ be a scalar. Any deviation
of $\bar{\mu}$ from a scalar will result in different laws of
dispersion in the composite and in the effective medium with no hope
of obtaining the same measurables from these two models. This
requirement that $\bar{\mu}$ be a scalar even in a strongly
anisotropic composite is difficult to justify on physical grounds,
unless, of course, $\bar{\mu}=1$.

Third, the dispersion relations (\ref{det=0}) (for a composite medium)
and (\ref{det=0_hom_mu-inv}),(\ref{det=0_hom_eps-inv}) (for a
homogeneous medium) appear to have very different mathematical
structure. The fact that they reduce to the same equation under the
simple condition (\ref{eff_pars}) is quite remarkable.

Thus, we have shown that, if orthogonal optical axes of the composite
medium can be defined, its dispersion relation ${\bf q}(k)$ and its
isofrequency surfaces [defined as the sets containing all ${\bf q}$
such that ${\mathscr D}_c(k,{\bf q})=0$ for each $k=\omega/c$] are
equivalent to those obtained in a homogeneous medium with EMPs
$\bar{\epsilon}$ and $\bar{\mu}$ given by (\ref{eff_pars}), where the
quantities $\eta_\alpha$ are defined in (\ref{eta_def}).

Since it will be proved below that the correct choice of the parameter
$\xi$ in (\ref{eff_pars}) is $\xi=1$, we now state the main
homogenization result of this paper pertaining to the principal values
of the EMPs:
\begin{equation}
\label{eps_eff}
\bar{\epsilon}_{\alpha\alpha} = \epsilon_b \frac{1 +
2 \rho \chi (1 + \Sigma_{\alpha\alpha}) } {1 - \rho \chi (1 +
\Sigma_{\alpha\alpha})} \ , \ \  \bar{\mu}_{\alpha\alpha} = 1 \ .
\end{equation}
\noindent
It can be seen that the Maxwell-Garnett mixing formula is obtained
from (\ref{eps_eff}) by setting $\Sigma=0$. Electromagnetic
interactions of inclusions in the composite result in a nonzero value
of $\Sigma$ and, correspondingly, in the deviation of the EMPs from
the predications of Maxwell-Garnett theory.

\subsubsection{Propagation along crystallographic axes}
\label{subsubsec:q_axes}

Consider a plane wave propagating along the $z$-axis, so that
$q_x=q_y=0$. In this case,
\begin{align}
  {\mathscr D}_c(k,{\bf q}) = & k^4 - k^2 q_z^2\left(\frac{1}{\eta_x}
    + \frac{1}{\eta_y}\right) + \frac{q_z^4}{\eta_x\eta_y} \nonumber
  \\
= & \left(k^2 - \frac{q_z^2}{\eta_x} \right) \left(k^2 -
    \frac{q_z^2}{\eta_y} \right) \ .
\label{Dc_1D}
\end{align}
Thus, ${\mathscr D}_c(k,{\bf q})$ is factorized into a product of two
quadratic forms, giving rise to two branches of the dispersion
relation: $q_z^2 = \eta_x k^2$ and $q_z^2 = \eta_y k^2$. Obviously,
these two branches correspond to $x$- and $y$-polarized modes. It can
be seen that, in agreement with (\ref{eff_pars}), the quantities
$\eta_\alpha$ give the effective squared refractive index for the
transverse modes of the composite.

In addition to the two transverse modes, a
lon\-gi\-tu\-dinally-polarized mode can also exist under certain
conditions. A mode with an arbitrary wave number $q_\alpha$, which
propagates and is polarized along the same axis $\alpha$, exists if
and only if
\begin{equation}
\label{eps_m_long}
1 + 2\rho\chi (1 + \Sigma_{\alpha\alpha}) = 0 \ .
\end{equation}
\noindent
Under this condition, the equality (\ref{det=0_expand_comp}) holds,
even if ${\mathscr D}_c(k,{\bf q}) \neq 0$.

Let us consider briefly the physical conditions for existence of the
longitudinal waves. From the property (\ref{B3}) (given in
Appendix~\ref{app:M}), it follows that $\lim_{\rho\rightarrow
  1}\Sigma_{\alpha\alpha} = 0$.  Consequently, the longitudinal waves
exist in the high-density limit if $1+2\chi=0$, which is only possible
if $\epsilon_a = 0$. This is the well-known condition for longitudinal
waves in non-dissipative plasma. The low-density limit can not be
considered so easily because $\Sigma_{\alpha\alpha}$ does not approach
zero when $\rho \rightarrow 0$ (see the Sec.~\ref{subsec:low-dens})
and can, in fact, diverge for certain values of $\chi$. However, we
can use the reciprocity substitution $\rho \leftrightarrow 1 - \rho$,
$\epsilon_a \leftrightarrow \epsilon_b$ to see that, in the
low-density limit, the condition for existence of the longitudinal
waves is $\epsilon_b = 0$. Quite analogously, longitudinal waves can
be obtained by considering the dispersion relation
(\ref{det=0_expand_homo}) and setting one of the principal values
$\bar{\epsilon}_{\alpha\alpha}$ to zero.

\subsubsection{Propagation in a crystallographic plane}
\label{subsubsec:q_plane}

We now discuss the case when ${\bf q}$ lies in the $xz$-plane.
Problems of this type can arise when one considers reflection and
refraction at the interface $z=0$, where the $xz$-plane is the plane
of incidence, as is shown in Fig.~\ref{fig:sketch_2D}. Under the
condition $q_y=0$, we have
\begin{align}
{\mathscr D}_c(k,{\bf q}) = k^4 - k^2 & \left[ q_x^2\left(\frac{1}{\eta_y}
    + \frac{1}{\eta_z}\right) + q_z^2\left(\frac{1}{\eta_x}
    + \frac{1}{\eta_y}\right) \right] \nonumber \\
+ \frac{q_x^4}{\eta_y\eta_z} + & 
\frac{q_z^4}{\eta_x\eta_y} + \frac{q_x^2 q_z^2}{\eta_y}\left(\frac{1}{\eta_x}
    + \frac{1}{\eta_z}\right) \nonumber \\
= \left[k^2 - \left(\frac{q_x^2}{\eta_y} +
    \frac{q_z^2}{\eta_y} \right) \right] & \left[k^2 - \left(\frac{q_x^2}{\eta_z} + \frac{q_z^2}{\eta_x}
  \right) \right]   \ . 
\label{Dc_3D}
\end{align}
Thus, ${\mathscr D}_c(k,{\bf q})$ is factorized into a product of two
quadratic forms, which correspond to the s- and p-polarized modes.

By equating the first factor in (\ref{Dc_3D}) to zero, we obtain the
dispersion relation for the s-polarized wave:
\begin{equation}
\label{disp_2D_s}
\frac{q_z^2}{\eta_y} + \frac{q_x^2}{\eta_y} = k^2 \ .
\end{equation}
The vector ${\bf F}_0$ of the s-polarized wave is aligned with the
$y$-axis and is, therefore, perpendicular to the plane of incidence.

By equating the second factor in (\ref{Dc_3D}) to zero, we obtain the
dispersion relation for the p-polarized wave:
\begin{equation}
\label{disp_2D_p}
\frac{q_z^2}{\eta_x} + \frac{q_x^2}{\eta_z} = k^2  \ .
\end{equation}
We can now find the vector ${\bf F}_0$ for the p-polarized wave by
considering the nontrivial solutions to (\ref{F_0}). It can be easily
seen that ${\bf F}_0$ lies in this case in the plane of incidence (its
projection onto the $y$-axis is zero), and the $x$ and $z$ components
of ${\bf F}_0$, $F_{0x}$ and $F_{0z}$, satisfy the following relation
(details of derivation are given in Appendix~\ref{app:p-pol}):
\begin{equation}
\label{Fx_over_Fz}
\frac{F_{0x}}{F_{0z}} = - \frac{1 + 2 \rho\chi(1 + \Sigma_{zz})}{1 + 2
  \rho\chi(1 + \Sigma_{xx})} \frac{q_z}{q_x} \ .
\end{equation}
\noindent
Eq.~(\ref{Fx_over_Fz}) will be used below in Sec.~\ref{sec:half-space}
to compute the half-space reflection coefficient for the p-polarized
incident wave.

\subsection{Low-density and low-contrast limits}
\label{subsec:low-dens}

Iteration of Eq.(\ref{g=/=0}) results in the following expansion for
the self-energy:
\begin{align}
\label{Sigma_Born}
\Sigma = \rho\chi \sum_{{\bf g}\neq 0} M(-{\bf g}) Q({\bf g}) M({\bf
  g})+ (\rho\chi)^2 \nonumber \\
\times \sum_{{\bf g},{\bf g}^\prime\neq 0} 
M(-{\bf g}) Q({\bf g}) M({\bf g} - {\bf g}^\prime) Q({\bf g}^\prime)
M({\bf g}^\prime) + \cdots
\end{align}
\noindent
It is important to note that this expansion should be used with
caution. Indeed, if $\chi$ is of the order of unity or larger, the
series in (\ref{Sigma_Born}) does not converge, even for arbitrarily
small values of the density $\rho$. This result may seem unexpected,
but it is easily understood by observing that the functions $M({\bf
  g})$ depend on $\rho$ and obey the sum rules (\ref{B1}).

In Sec.~\ref{sec:CF}, a more useful (and always convergent) expansion
of $\Sigma$ will be derived. Here we note that the functions $M({\bf
  g})$ are independent of $\chi$. Therefore, (\ref{Sigma_Born}) is the
formal expansion of $\Sigma$ into the powers of $\chi$. Thus, in the
low-contrast limit ($\chi \rightarrow 0$), we have $\Sigma \rightarrow
\rho \chi \sigma_1$, where $\sigma_1 = \sum_{{\bf g}\neq 0} M(-{\bf
  g}) Q({\bf g}) M({\bf g})$. In the case of three-dimensional
inclusions with cubic symmetry, $\sigma_1$ is identically zero. Then
the first non-vanishing term in the low-contrast expansion of $\Sigma$
is given by $(\rho\chi)^2 \sigma_2$, where $\sigma_2$ grows naturally
out of the second term in the right-hand side of (\ref{Sigma_Born}).

\subsection{Two-dimensional lattices}
\label{subsec:2D}

Consider a medium in which $\epsilon = \epsilon(x,y)$ is independent
of $z$. As above, we assume that $\epsilon(x,y)$ is periodic on a
square lattice with lattice step $h$. The homogenization theory for
this medium can be obtained either by considering a three-dimensional
lattice with unequal steps $h_x$, $h_y$, $h_z$ and taking the limit
$h_z \rightarrow 0$, or by following the derivations of
Sec.~\ref{subsec:inf_3D}, taking account of the modified geometry.
The results obtained are very similar to those in the 3D case, with
some obvious modifications. Specifically, we arrive at
Eqs.~(\ref{g=0}),(\ref{g=/=0}) in which, however, we must take ${\bf
  g} = (2\pi/h)(\hat{\bf x}n_x + \hat{\bf y}n_y)$. Additionally, in
the integrals (\ref{M_def}), $\Omega$ must be understood as a
two-dimensional region (the intersection of an inclusion with the
$xy$-plane), $V$ as the area of $\Omega$, and $d^3R$ is replaced by
$d^2R$.  The definition of $Q({\bf g})$ (\ref{Q_def}) remains
unchanged, but $Q({\bf g})$ is now a $2\times 2$ tensor.

Consider a wave propagating in the $xy$-plane and polarized along the
$z$-axis. In this case, ${\bf F}_{\bf g} = \hat{\bf z} F_{\bf g}$,
where $F_{\bf g}$ is a scalar and $\Sigma$ can be found analytically
in general. Indeed, we have in this case $Q({\bf g}){\bf F}_{{\bf
    g}^\prime} = {\bf F}_{{\bf g}^\prime}$, $Q({\bf g}){\bf F}_0 =
{\bf F}_0$, and Eq.~(\ref{g=/=0}) becomes
\begin{equation}
\label{g=/=0_2D}
F_{\bf g} = \rho\chi \left[M({\bf g}) F_0 + \sum_{{\bf g}^\prime \neq
0}M({\bf g} - {\bf g}^\prime) F_{{\bf g}^\prime} \right] \ , \quad
{\bf g}\neq 0 \ .
\end{equation}
\noindent
The solution to this equation is
\begin{equation}
F_{\bf g} = \frac{\rho\chi}{1 - (1-\rho)\chi} M({\bf g}) F_0 \ ,
\end{equation}
\noindent
where some of the properties (\ref{B1}) have been used (keeping in
mind that the term ${\bf g}=0$ must be excluded from the summation). We
then have
\begin{equation}
\label{S_2D}
\Sigma_{zz} = \frac{\rho\chi}{1 - (1-\rho)\chi} \sum_{{\bf g} \neq 0} M(-{\bf
  g})M({\bf g}) = \frac{(1-\rho)\chi}{1 - (1-\rho)\chi} \ .
\end{equation}
\noindent
It can be seen from the above equation that $\Sigma_{zz}$ does not
approach zero when $\rho\rightarrow 0$, as was discussed in
Sec.~\ref{subsec:low-dens}. Upon substitution of (\ref{S_2D}) into
(\ref{eps_eff}), we find that
\begin{equation}
\label{eps_eff_Z}
\bar{\epsilon}_{zz} = (1-\rho)\epsilon_b + \rho\epsilon_a = \langle
\epsilon \rangle \ .
\end{equation}
\noindent
Thus, the effective permittivity for $z$ polarization is given by the
arithmetic average of $\epsilon(x,y)$. This is in agreement with
Krokhin {\em al.}~\cite{krokhin_02_1,krokhin_04_1}.

\subsection{Concept of the smooth field}
\label{subsec:smooth}

The result (\ref{eps_eff_Z}) for a $z$-polarized wave could have been
anticipated. To understand better why the effective permittivity in
this case is given by an arithmetic average, it is instructive to
consider the concept of the {\em smooth field}. The smooth field ${\bf
  S}({\bf r})$ changes slowly on the characteristic scale defined by
the heterogeneities in the medium. As a result, one can factorize
spatial averages of ${\bf S}({\bf r})$ multiplied by any
rapidly-varying function. For example, we can write $\langle {\bf S}
\epsilon \rangle = \langle {\bf S} \rangle \langle \epsilon \rangle$,
etc.

Let us recall some well-known results for 1D periodically-layered
media~\cite{markel_10_1}. The effective permittivity of such media is
$\bar{\epsilon}_\parallel = \langle \epsilon \rangle$ for waves
polarized parallel to the layers and $\bar{\epsilon}_\perp =
\langle \epsilon^{-1} \rangle^{-1}$ for waves polarized
perpendicularly to the layers. These two results can be obtained quite
expeditiously by applying the concept of the smooth field. In the case
of tangential polarization, the electric field ${\bf E}$ is smooth.
This follows from the boundary condition which requires that the
tangential components of the electric field be continuous at all
interfaces.  Consequently, we can write
\begin{equation}
\langle {\bf D} \rangle = \langle \epsilon {\bf E} \rangle = \langle
\epsilon \rangle \langle {\bf E} \rangle \ ,
\end{equation}
\noindent
from which it follows that $\bar{\epsilon}_\parallel = \langle
\epsilon \rangle$. For perpendicular polarization, the field ${\bf D}$
is smooth. We then write
\begin{equation}
\langle {\bf E} \rangle = \langle \epsilon^{-1} {\bf D} \rangle =
\langle \epsilon^{-1} \rangle \langle {\bf D} \rangle
\end{equation}
\noindent
and $\bar{\epsilon}_\perp = \langle \epsilon^{-1} \rangle^{-1}$.

Similar considerations can be applied to the 2D problem of
Sec.~\ref{subsec:2D}. For waves polarized along the $z$-axis, the
field ${\bf E}$ is smooth, which results in $\bar{\epsilon}_{zz} =
\langle \epsilon \rangle$, in agreement with (\ref{eps_eff_Z}).

One can also consider a more general smooth field of the form ${\bf S}
= p{\bf E} + (1-p){\bf D} = [p + (1-p)\epsilon]{\bf E}$, where $p$ is
a mixing parameter. Here we consider the 3D case and assume that ${\bf
  S}$ is smooth for any polarization state. Application of the smooth
field principle results in the following equalities:
\begin{subequations}
\label{E-D_av}
\begin{align}
\langle {\bf E} \rangle = \langle {\bf S} \rangle \left\langle 
1/[p + (1-p)\epsilon] \right\rangle \ , \\
\langle {\bf D} \rangle = \langle {\bf S} \rangle \left\langle 
\epsilon/[p + (1-p)\epsilon] \right\rangle \ ,
\end{align}
\end{subequations}
\noindent
from which we find the effective permittivity to be
\begin{equation}
\label{D_av_over_E_av}
\bar{\epsilon}_{\alpha\beta} = 
\delta_{\alpha\beta} \frac{\left\langle \epsilon/[\epsilon + p/(1-p)]
\right\rangle}{\left\langle 1 /[\epsilon + p/(1-p)] \right\rangle} \ .
\end{equation}
\noindent
Eq.~(\ref{D_av_over_E_av}) is, in fact, the Maxwell-Garnett formula.
Although this form is rarely used, the Maxwell-Garnett effective
permittivity can be written as
\begin{equation}
\label{MG_av}
\bar{\epsilon}_{\rm MG} = \frac{\left\langle \epsilon/(\epsilon + 2\epsilon_b)
\right\rangle}{\left\langle 1 /(\epsilon + 2\epsilon_b) \right\rangle} \ .
\end{equation}
\noindent
We see that (\ref{D_av_over_E_av}) and (\ref{MG_av}) coincide if
$p=2\epsilon_b/(1 + 2\epsilon_b)$.

Thus, the Maxwell-Garnett EMT assumes that the field ${\bf S} =
[(\epsilon + 2\epsilon_b)/(1 + 2\epsilon_b)]{\bf E}$ is smooth. Since
the mixing parameter $p$ depends on the permittivity of the host
medium, Eq.~(\ref{MG_av}) is not invariant with respect to the
substitution $\epsilon_a \leftrightarrow \epsilon_b$ and $\rho
\leftrightarrow 1-\rho$. The homogenization formula (\ref{eps_eff})
derived in this paper, however, is fully symmetric. Note that
Bruggeman's EMT is also symmetric but can not be easily written in
terms of averages.  Therefore, it is not clear which form of the
smooth field Bruggeman's approximation assumes. In general, the smooth
field does not need to be a linear functional of ${\bf E}$ and ${\bf
  D}$.

\section{Reflection and refraction at a half-space boundary}
\label{sec:half-space}

An infinite lattice is a mathematical abstraction. All experimental
media are bounded, and the physical effects which occur at the
boundary are often important. For instance, as mentioned above, it is
not possible to determine simultaneously and uniquely the effective
permittivity and permeability of a medium from the bulk dispersion
relation alone.

The problem of reflection and refraction of a wave at a flat interface
is considered in this section. The goals are three-fold. First, we
will derive the limit in which the correct expression for the Fresnel
reflection coefficient is obtained. This will turn out to be the same
limit as was used in Sec.~\ref{subsec:inf_3D}. Second, we will show
that the correct expression for the reflection coefficients results
only if we take $\xi=1$ in (\ref{eff_pars}), from which it follows
that $\bar{\mu}=1$. Third, we will provide additional mathematical
justification for the results of Sec.~\ref{subsec:inf_3D}. Indeed, the
derivations of that section contain one dubious step.  Namely, the
applicability of the Poisson summation formula (\ref{Poisson}) can be
questioned because the variable ${\bf q}$ is complex. Strictly
speaking, the series in the left-hand side of (\ref{Poisson}) diverges
for an infinite lattice.  The problem can be fixed, in principle, by
considering real-valued ${\bf q}$'s and then analytically-continuing
the summation result to the whole complex plane. In this section, no
such complication will arise since all series in question are
convergent.

\subsection{General setup}
\label{subsec:gen_halfspace}

\begin{figure}
\centerline{\psfig{file=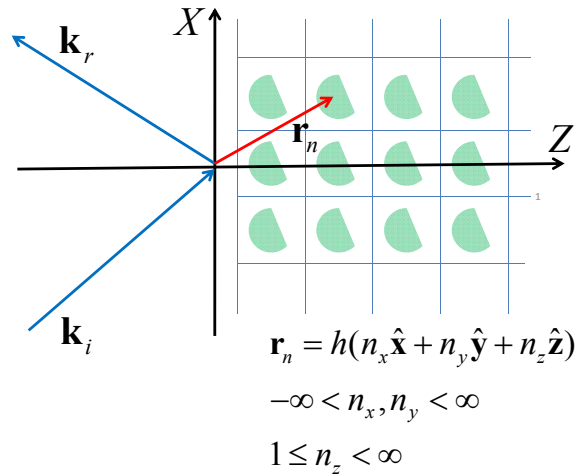,width=9cm,clip=}}
\caption{\label{fig:sketch_2D} (color online) Sketch of the geometry
  considered: reflection and refraction at a half-space boundary.}
\end{figure}

The geometry considered in this section is sketched in
Fig.~\ref{fig:sketch_2D}. The medium occupies the right half-space and
the left half-space has the background permittivity $\epsilon_b$. It
would be more appropriate to consider the case when the left
half-space is vacuum and the right half-space is a two-component
mixture, so that there are three different components in the problem.
This, however, requires the use of the half-space Green's
tensor~\cite{maradudin_75_1} -- a step that is not conceptually
difficult, yet mathematically involved.  Here we restrict
consideration to only two components. This includes the cases when the host
medium is vacuum and also when the incident beam is first refracted
from vacuum into a homogeneous medium of permittivity $\epsilon_b \neq
1$ (at a planar interface that is located at $z=z_1\ll -h$ and is not
considered explicitly) and then into a heterogeneous medium which is a
mixture of $a$- and $b$-type components.

Physically, the $z$ coordinate of the effective medium boundary can be
stated only approximately, within an interval of width $\sim h$. It
will prove mathematically convenient to place the boundary on the
plane $z=0$, and the centers of the left-most cells on the plane
$z=h$, as shown in Fig.~\ref{fig:sketch_2D}. In the EMT developed
below, the half-space $z>0$ is assumed to be filled with an effective
medium.

A wave can not propagate in a semi-infinite medium without an external
source. Therefore, we must solve the integral equation (\ref{main_eq})
with a nonzero incident field ${\bf E}_i$ which we will take to be a
plane wave. We will also find that, under appropriate conditions, a
uniquely-defined reflected plane wave ${\bf E}_r$ exists in the region
$z<0$. The incident and the reflected waves are given by
\begin{subequations}
\begin{align}
\label{E_i_def}
{\bf E}_i({\bf r}) &= {\bf A}_i \exp({\bf k}_i \cdot {\bf r}) \ , \ \
 -\infty < z < \infty \ , \\
\label{E_r_def}
{\bf E}_r({\bf r}) &= {\bf A}_r \exp({\bf k}_r \cdot {\bf r}) \ , \ \
 -\infty < z < 0 \ .
\end{align}
\end{subequations}
\noindent
Note that the incident wave is defined in the whole space but
Eq.~(\ref{main_eq}) is only defined for ${\bf r}\in\Omega_{\rm tot}$.
The wave numbers of the incident and the reflected waves can be
written as
\begin{equation}
\label{k_i_r}
{\bf k}_i = {\bf k}_\perp + \hat{\bf z} k_{iz} \ , \ \ {\bf k}_r = {\bf
  k}_\perp - \hat{\bf z} k_{iz}  \ .
\end{equation}
\noindent
Henceforth, the subscript ``$\perp$'' will be used to denote
projections of vectors onto the $xy$-plane. Note that ${\bf k}_\perp
\cdot \hat{\bf z} = 0$ and
\begin{equation}
\label{k_ir_2}
k_i^2 = k_r^2 = k_\perp^2 + k_{iz}^2 = k_b^2 = k^2 \epsilon_b =
\left( \frac{\omega}{c} \right)^2 \epsilon_b \ .
\end{equation}
\noindent
It is important to note that the vector ${\bf k}_\perp$ is purely
real. A complex-valued ${\bf k}_\perp$ would imply a wave that is
evanescent in a direction parallel to the interface. This would
necessitate the presence of additional interfaces; such a possibility
is not considered here. The vector ${\bf k}_\perp$ is real-valued even
if the host medium is absorbing. Indeed, we should keep in mind that
the incident wave enters the host medium from vacuum and that the
tangential component of the wave vector is conserved at any planar
interface, even if one of the media is absorbing. However, the
$z$-projection of ${\bf k}_i$ does not need to be real. In a
transparent host ($\epsilon_b > 0$), the incident wave is evanescent
and $k_{iz}$ is purely imaginary if $k_\perp > k_b$; in an absorbing
host, $k_{iz}$ is, generally, complex.

Note that the reflected wave (\ref{E_r_def}) does not enter
Eq.~(\ref{main_eq}) because it is identically zero in $\Omega_{\rm
  tot}$. The reflected wave is computed {\em a posteriori} once the
polarization field ${\bf P}$ is found. Then the amplitudes
${\bf A}_r$ and ${\bf A}_i$ can be used to determine the reflection
coefficient.

To solve Eq.~(\ref{main_eq}) in the presence of the incident field, we
decompose ${\bf P}$ as
\begin{equation}
\label{P_BS}
{\bf P} = {\bf P}_B + {\bf P}_S \ , 
\end{equation}
\noindent
where ${\bf P}_B$ is the Bloch wave of the form (\ref{BW_1})
and ${\bf P}_S$ is an additional wave that originates due to
the presence of the surface. We seek the condition under which
\begin{align}
\label{E_Ewald}
{\bf E}_{\rm EO}({\bf r}) & \equiv \int_{\Omega_{\rm tot}} G({\bf
  r},{\bf r}^\prime){\bf P}_B({\bf r}^\prime) d^3 r^\prime \nonumber \\
&= {\bf E}_B({\bf r}) + {\bf E}_{\rm ext}({\bf r}) + {\bf E}_S({\bf  r}) \ , 
\end{align}
\noindent
where in $\Omega_{\rm tot}$
\begin{subequations}
\label{EB_PB}
\begin{align}
\label{EB_PB_a}
{\bf E}_B({\bf r}) &= \frac{4\pi}{3\chi } {\bf P}_B({\bf r}) \ , \\
\label{EB_PB_b}
{\bf E}_{\rm ext}({\bf r}) &= - {\bf E}_i({\bf r})  \ ,
\end{align}
\end{subequations}
\noindent
If (\ref{P_BS})-(\ref{EB_PB}) hold, then Eq.~(\ref{main_eq}) becomes
\begin{align}
\label{surf_eq}
 {\bf P}_S({\bf r}) = \frac{3\chi}{4\pi} 
\left[{\bf E}_S({\bf r}) + 
  \int_{\Omega_{\rm tot}} G({\bf r},{\bf r}^\prime) {\bf P}_S({\bf r}^\prime)
  d^3r^\prime \right] \ , \nonumber \\ 
{\bf r} \in \Omega_{\rm tot} \ .
\end{align}
\noindent
Note that Eq.~(\ref{surf_eq}) contains only quantities which are
associated with the surface wave.

Eq.~(\ref{E_Ewald}) is the mathematical formulation of the Ewald-Oseen
extinction theorem and we will refer to ${\bf E}_{\rm EO}$ as
to the Ewald-Oseen field. We will see that one can determine the
reflection coefficient from the conditions (\ref{EB_PB}). We will also
see that the surface wave is exponentially localized near the
interface and does not contribute to either reflection or transmission
coefficients if
\begin{equation}
\label{no_diffr}
\left({\bf k}_\perp + {\bf g}_\perp\right)^2 >  k_b^2  
\ \ \ \forall  {\bf g}_\perp\neq 0 \ .
\end{equation}
\noindent
Inequality (\ref{no_diffr}) is weaker than what is required for
homogenization.  It is merely the condition that there is no Bragg
diffraction in the medium; if (\ref{no_diffr}) is violated, the
conventional reflection and transmission coefficients can not be
defined. If, however, (\ref{no_diffr}) holds, we do not need to solve
Eq.~(\ref{surf_eq}) explicitly; it suffices to know that the surface
wave does not contribute to any measurement performed sufficiently far
from the interface.

\subsection{Evaluation of the Ewald-Oseen field}
\label{subsec:Ewald}

To compute the Ewald-Oseen field, we proceed along the lines of
Sec.~\ref{subsec:inf_3D} to arrive at the following expression:
\begin{align}
\label{E_Ewald_rn}
& {\bf E}_{\rm EO}({\bf r}) =
\frac{4\pi}{3}\int\frac{d^3p}{(2\pi)^3} K({\bf p}) \int_{\Omega} d^3 R
{\bf F}({\bf R})  \nonumber \\
& \times \exp\left[ i{\bf p} \cdot ({\bf r} - {\bf R}) \right] \sum_m
\exp\left[i({\bf q} - {\bf p}) \cdot {\bf r}_m \right] \ . 
\end{align}
\noindent
So far, no restrictions on ${\bf r}$ have been placed. In particular,
${\bf r}$ can be either in the right or left half-space. However, when
we later substitute the result of integration into Eqs.~(\ref{EB_PB}),
${\bf r}$ will be restricted to $\Omega_{\rm tot}$.

The sum over $m$ in (\ref{E_Ewald_rn}) can be evaluated as follows.
First, we expand the summation as
\begin{align}
\label{S_sum_m_1}
& \sum_m \exp\left[i({\bf q} - {\bf p}) \cdot {\bf r}_m \right] =
\nonumber \\
& \sum_{m_x,m_y=-\infty}^\infty \exp\left[i(q_x - p_x)h m_x +
 i(q_y - p_y)h m_y\right] \nonumber \\
& \times \sum_{m_z=1}^\infty \exp\left[i(q_z - p_z)h m_z\right] \ .
\end{align}
\noindent
From symmetry considerations, we know that ${\bf q}_\perp = {\bf
  k}_\perp$. This property is a manifestation of momentum conservation
and will be confirmed below by considering the conditions
(\ref{EB_PB}). Since, as discussed above, ${\bf k}_\perp$ is purely
real, $q_x$ and $q_y$ are also real. Therefore, we can compute the
sums over $m_x$ and $m_y$ using the Poisson sum formula. Further, the
half-range sum over $m_z$ converges absolutely because the transmitted
wave decays into the medium and, correspondingly, ${\rm Im}q_z >0$.
We, therefore, have
\begin{align}
\label{S_sum_m_2}
\sum_m \exp\left[i({\bf q} - {\bf p}) \cdot {\bf r}_m \right]
\nonumber \\
 = \left(\frac{2\pi}{h}\right)^2 f(p_z) & \sum_{{\bf g}_\perp}
 \delta({\bf p}_\perp - {\bf q}_\perp - {\bf g}_\perp) \ , 
\end{align}
\noindent
where
\begin{subequations}
\label{f_pz}
\begin{align}
f(p_z) \equiv
\sum_{m_z=1}^\infty & \exp\left[i\left(q_z-p_z\right)h m_z\right]
\nonumber \\
\label{f_pz_a}
& = \frac{1}{\exp[i(p_z-q_z)h] - 1} \\
\label{f_pz_b}
& = \frac{2\pi}{h} \sum_{g_z} \frac{(2\pi i)^{-1}}{p_z - q_z - g_z} \ .
\end{align}
\end{subequations}
\noindent
Here the well-known Laurant expansion of the function $1/[\exp(iz)-1]$
has been used. The equality (\ref{f_pz_b}) is an important
observation. It will allow us to evaluate the Ewald-Oseen field.

\begin{figure}
\centerline{\psfig{file=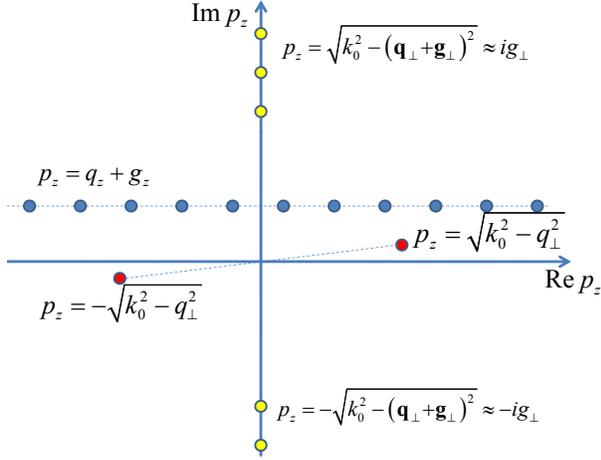,width=9cm,clip=t}}
\caption{\label{fig:poles} (color online) Poles of the integrand of
  Eq.~(\ref{E_Ewald_rn_1}) in the complex $p_z$-plane.}
\end{figure}

We now proceed by substituting (\ref{S_sum_m_2}) into
(\ref{E_Ewald_rn}), which yields
\begin{align}
\label{E_Ewald_rn_1}
{\bf E}_{\rm EO}({\bf r}) = \frac{4\pi}{3h^2} \sum_{{\bf g}_\perp}
\int_{-\infty}^\infty\frac{dp_z}{2\pi} f(p_z) K({\bf q}_\perp + {\bf
  g}_\perp + \hat{\bf z}p_z) \nonumber \\
\times \int_{\Omega} d^3 R {\bf F}({\bf R}) \exp\left[ i({\bf
    q}_\perp + {\bf g}_\perp +\hat{\bf z}p_z) \cdot ({\bf r} - {\bf
    R}) \right] \ .
\end{align}
\noindent
The integral over $p_z$ can be computed by contour integration since
all the poles and residues of the integrand are known. The positions
of the poles in the complex $p_z$-plane are shown in
Fig.~\ref{fig:poles}. The poles at $p_z = q_z + g_z$ are the
singularities of the function $f(p_z)$. Since $q_z$ has a positive
imaginary part and all $g_z$'s are real-valued, these poles lie in the
upper half-plane. The remaining poles are the singularities of $K({\bf
  q}_\perp + {\bf g}_\perp + \hat{\bf z}p_z)$, which is viewed here as
a function of $p_z$. From the definition (\ref{K_def}), we find that
these singularities are located at $p_z = \pm {\mathcal P}_{{\bf
    g}_\perp}$, where
\begin{equation}
\label{K_poles}
{\mathcal P}_{{\bf g}_\perp} =  \sqrt{ k_b^2 - ({\bf q}_\perp + {\bf g}_\perp)^2} \ .
\end{equation}
\noindent
These poles can be considered separately for ${\bf g}_\perp = 0$ and
${\bf g}_\perp \neq 0$. The two poles corresponding to ${\bf g}_\perp
= 0$ are $p_z = \pm {\mathcal P}_0 = \pm \sqrt{ k_b^2 - q_\perp^2}$.
The poles with ${\bf g}_\perp \neq 0$ have large (either positive or
negative) imaginary parts if $h k_b, hq_\perp \ll 1$,
in which case they can be written, approximately, as ${\mathcal
  P}_{{\bf g}_\perp} \approx i g_\perp$.

Note that in the case of infinite lattices, the singularities of
$K({\bf p})$ do not contribute to Fourier integrals of the type
(\ref{W_eval}) because the corresponding residues are identically zero
[these singularities fall in between the peaks of the delta-function
fence given by the right-hand side of (\ref{Poisson})].
We will compute the contributions of the different families of poles
to the integral (\ref{E_Ewald_rn_1}) separately. If the vector of
position ${\bf r}$ is inside one of the inclusions, the integration
contour must be closed in the upper half of the complex $p_z$-plane.
Correspondingly, only the poles with positive imaginary parts
contribute to the integral (\ref{E_Ewald_rn_1}) in this case.  The
Ewald-Oseen field can also be computed in the left half-space. If the
point of observation ${\bf r}$ is further away from the interface than
$h/2$, so that the inequality $\hat{\bf z}\cdot {\bf r} < -h/2$ holds,
the integration contour must be closed in the lower half of the
complex $p_z$-plane. In what follows, it will be shown that the poles
at $p_z = q_z + g_z$ yield the Bloch-wave field ${\bf E}_B({\bf r})$,
the pole at $p_z = {\mathcal P}_0$ yields the extinction field ${\bf
  E}_{\rm ext}({\bf r})$, the pole at $p_z = -{\mathcal P}_0$ yields
the reflected wave, and, finally, the poles $p_z \approx \pm {\mathcal
  P}_{{\bf g}_\perp}$ with ${\bf g}_\perp\neq 0$ yield the
fast-decaying surface wave.

\subsubsection{Bloch wave}

We start by computing the Bloch-wave contribution to the Ewald-Oseen
field, ${\bf E}_B({\bf r})$. We place the point of observation in
$\Omega_{\rm tot}$, use the expression (\ref{f_pz_b}) for $f(p_z)$ and
evaluate the contributions of the poles $p_z = q_z + g_z$ to the
integral (\ref{E_Ewald_rn_1}). This results in the following
expression:
\begin{align}
\label{E_EO_B}
{\bf E}_B({\bf r}) = \frac{4\pi}{3h^3}\sum_{\bf g} \exp\left[ i
  \left({\bf q} + {\bf g}\right) \cdot {\bf r}\right] K({\bf q} + {\bf
  g}) \nonumber \\
\times \int_{\Omega} {\bf F}({\bf R})\exp\left[-i\left({\bf q} + {\bf
      g}\right) \cdot {\bf
    R} \right] d^3R \ , \ \
{\bf r} \in \Omega_{\rm tot} \ .
\end{align}
\noindent
Here we have used the equalities ${\bf g}_\perp + \hat{\bf z}g_z =
{\bf g}$ and $\sum_{{\bf g}_\perp}\sum_{g_z} = \sum_{\bf g}$. Now, if
${\bf F}({\bf R})$ is expanded according to (\ref{F_g_def}), and if
the expansion coefficients ${\bf F}_{\bf g}$ satisfy (\ref{eq_F}),
then the field given by Eq.~(\ref{E_EO_B}) satisfies ${\bf E}_B({\bf
  r}) = (4\pi/3\chi) {\bf P}_B({\bf r})$ for ${\bf r} \in \Omega_{\rm
  tot}$, where ${\bf P}_B$ is of the form (\ref{BW_1}). Thus,
(\ref{EB_PB_a}) is satisfied if the Bloch wave of the polarization ${\bf
  P}_B$ is the same as one would find by solving the
eigenproblem (\ref{eq_F}) for an infinite medium. This justifies the
use of the Poisson summation formula in Sec.~\ref{subsec:inf_3D}.

Eq.~(\ref{eq_F}) applies to general photonic crystals that are not
necessarily describable by EMPs. As was discussed in
Sec.~\ref{subsec:inf_3D}, homogenization is obtained by taking the
limit $h\rightarrow 0$. This limit must be computed separately for the
equations with ${\bf g}=0$ and ${\bf g}\neq 0$, which results in
(\ref{sys_3D}). This system of equations defines an eigenproblem for
the Bloch wave vector ${\bf q}$, while the polarization vector ${\bf
  F}_0$ is obtained as an eigenvector of (\ref{F_0}). The higher-order
expansion coefficients ${\bf F}_{\bf g}$ are uniquely determined by
${\bf F}_0$ but ${\bf F}_0$ itself is defined by (\ref{sys_3D}) only
up to a multiplicative factor. Next, we will show that this factor is
fixed by the condition (\ref{EB_PB_b}).

\subsubsection{Extinction wave}

We now compute the contribution of the pole located at $p_z =
{\mathcal P}_0$. The function $f(p_z)$ is analytic in the vicinity
of ${\mathcal P}_0$; therefore, we can use the expression
(\ref{f_pz_a}) for $f(p_z)$. Since Eqs.~(\ref{EB_PB_b}) should hold
only for ${\bf r}\in \Omega_{\rm tot}$, we close the integration
contour in the upper half-plane. A straightforward calculation yields
\begin{align}
\label{E_EO_ext}
{\bf E}_{\rm ext}({\bf r}) & = \frac{4\pi i}{h^2} \frac{\exp\left[i
    \left({\bf q}_\perp + \hat{\bf z}{\mathcal P}_0\right) 
  \cdot {\bf r} \right]}{\exp\left[i\left({\mathcal P}_0 - q_z
  \right)h\right] - 1} \nonumber \\ 
& \times \frac{ k_b^2 - ({\bf q}_\perp + \hat{\bf z}{\mathcal P}_0) \otimes ({\bf
    q}_\perp + \hat{\bf z}{\mathcal P}_0)}{2{\mathcal P}_0} \nonumber \\
& \times  \int_\Omega d^3R {\bf F}({\bf R}) \exp \left[ -i \left( {\bf
      q}_\perp + \hat{\bf z}{\mathcal P}_0\right)\cdot {\bf R} \right]
\ , \\
& \hspace*{5cm} {\bf r} \in \Omega_{\rm tot} \ . \nonumber 
\end{align}
\noindent
We seek the condition under which ${\bf E}_{\rm ext}({\bf r}) = -
{\bf E}_i({\bf r})$ for ${\bf r} \in \Omega_{\rm tot}$, where ${\bf
  E}_i({\bf r})$ is given by (\ref{E_i_def}). It immediately
transpires that the above equality can hold only if ${\bf q}_\perp =
{\bf k}_\perp$.  The continuity of the tangential components of all wave
vectors, including the incident wave vector ${\bf k}_i$, the reflected
wave vector ${\bf k}_r$ and the Bloch wave vector of the transmitted
wave ${\bf q}$ follows from the discrete translational symmetry of the
problem. We now find from (\ref{k_i_r}) that ${\mathcal P}_0 = k_{iz}$
and ${\bf q}_\perp + \hat{\bf z}{\mathcal P}_0 = {\bf k}_i$. With the
use of these equalities and the notation
\begin{equation}
\label{F_k}
\tilde{\bf F}({\bf k}) = \int_\Omega {\bf F}({\bf R}) \exp\left( -i
  {\bf k} \cdot {\bf R}\right) d^3R \ ,
\end{equation}
\noindent
we can simplify Eq.~(\ref{E_EO_ext}) as
\begin{align}
\label{E_EO_ext_1}
{\bf E}_{\rm ext}({\bf r}) & = \frac{4\pi i}{h^2} \frac{\exp\left(i
    {\bf k}_i \cdot {\bf r} \right)}{\exp\left[i\left(k_{iz} - q_z
    \right)h\right] - 1} \nonumber \\
& \times \frac{k_b^2 - {\bf k}_i \otimes {\bf k}_i}{2 k_{iz}}
\tilde{\bf F}({\bf k}_i) \ , \ \ {\bf r} \in \Omega_{\rm tot} \ .
\end{align}
\noindent
The extinction condition then takes the form
\begin{align}
\label{extinction_exact}
{\bf A}_i = - \frac{2\pi i}{h^2} 
\frac{k_b^2 - {\bf k}_i \otimes {\bf k}_i}
{\exp\left[i\left(k_{iz} - q_z \right)h\right] - 1} 
\frac{\tilde{\bf F}({\bf k}_i)}{k_{iz}} \ .
\end{align}
\noindent
So far, no approximations have been made. The homogenization limit is
obtained by observing that
\begin{subequations}
\label{limits}
\begin{align}
\label{limit_1}
& \lim_{h\rightarrow 0} \exp\left[i\left(\pm k_{iz} - q_z
    \right)h\right]  = 1 + i(\pm k_{iz} - q_z)h \ , \\
\label{limit_2}
& \lim_{h\rightarrow 0} \tilde{\bf F}({\bf k}_i) = \lim_{h\rightarrow 0}
\tilde{\bf F}({\bf k}_r) = V(1 + \Sigma){\bf F}_0 \ .
\end{align}
\end{subequations}
\noindent
Once the above limiting expressions are used, the extinction condition
becomes of the form
\begin{align}
\label{extinction_hmg}
{\bf A}_i = - 2 \pi \rho
\frac{k_b^2 - {\bf k}_i \otimes {\bf k}_i}
{k_{iz} \left( k_{iz} - q_z \right) } (1 + \Sigma){\bf F}_0 \ .
\end{align}
\noindent
This equation couples the amplitude of the incident field, ${\bf
  A}_i$, and the amplitude of the Bloch polarization wave, ${\bf
  F}_0$. The vector ${\bf F}_0$ must simultaneously satisfy the
following two conditions: (i) be an eigenvector of the tensor in the
square brackets in Eq.~(\ref{F_0}) and (ii) satisfy
(\ref{extinction_hmg}). These two conditions determine both the
direction and the length of ${\bf F}_0$.

\subsubsection{Reflected wave}

Consider now the case when the point of observation ${\bf r}$ in the
left half-space. As discussed above, we will place ${\bf r}$ at least
$h/2$ away from the interface. This will allow us to close the
integration contour in (\ref{E_Ewald_rn_1}) in the lower half of the
complex $p_z$-plane. The reflected wave is obtained by computing the
input of the pole $p_z = -{\mathcal P}_0$. We find that the electric
field of the reflected wave is of the form (\ref{E_r_def}) where the
amplitude ${\bf A}_r$ is given by
\begin{align}
\label{reflextion_exact}
{\bf A}_r = \frac{2\pi i}{h^2} \frac{k_b^2 - {\bf k}_r \otimes {\bf
    k}_r}{\exp\left[-i\left(k_{iz} + q_z \right)h\right] - 1}
\frac{\tilde{\bf F}({\bf k}_r)}{k_{iz}} \ .
\end{align}
\noindent
This expression contains no approximations. In the homogenization
limit, we use the limiting expressions (\ref{limits}) and obtain
\begin{align}
\label{refl_hmg}
{\bf A}_r = -2\pi\rho
\frac{k_b^2 - {\bf k}_r \otimes {\bf k}_r}{k_{iz}\left(k_{iz} + q_z \right)}
  \left( 1 + \Sigma \right) {\bf F}_0 \ .
\end{align}
\subsubsection{Surface wave}

Finally, let us evaluate the contribution of the poles $p_z =
{\mathcal P}_{{\bf g}_\perp}$ with ${\bf g}_\perp \neq 0$.  For ${\bf
  r} \in \Omega_{\rm tot}$, we have
\begin{align}
{\bf E}_S({\bf r}) & = \frac{2\pi i}{h^2}\sum_{{\bf g}_\perp\neq 0}
f\left({\mathcal P}_{{\bf g}_\perp}\right) 
\exp\left(i {\bf k}_{{\bf g}_\perp} \cdot {\bf
  r}\right) \nonumber \\
& \times \frac{k_b^2 - {\bf k}_{{\bf g}_\perp} \otimes {\bf k}_{{\bf g}_\perp}}{{\mathcal
  P}_{{\bf g}_\perp}}
\tilde{\bf F}\left( {\bf k}_{{\bf g}_\perp} \right) \ , \
\ {\bf r} \in \Omega_{\rm tot} 
\label{surf_exact}
\end{align}
\noindent
where
\begin{equation}
{\bf k}_{{\bf g}_\perp} = {\bf q}_\perp + {\bf g}_\perp + \hat{\bf
  z}{\mathcal P}_{{\bf g}_\perp} \ .
\end{equation}
\noindent
If the condition (\ref{no_diffr}) holds, the quantities ${\mathcal
  P}_{{\bf g}_\perp}$ have nonzero imaginary parts even if the host is
transparent. Therefore, the surface wave decays exponentially away
from the interface. In the homogenization limit, the exponential decay
is fast. Indeed, in the limit $h\rightarrow 0$, we have (for ${\bf
  g}_\perp\neq 0$): ${\mathcal P}_{{\bf g}_\perp} \rightarrow
ig_\perp$, ${\bf k}_{{\bf g}_\perp} \rightarrow {\bf g}_\perp +
i\hat{\bf z} g_\perp$, $f\left( {\mathcal P}_{{\bf g}_\perp}\right)
\rightarrow -1/g_\perp h$. With these limits taken into account, the
surface wave takes the following form:
\begin{align}
{\bf E}_S({\bf r}) & = - \frac{2\pi i}{h^3}\sum_{{\bf g}_\perp\neq 0}
\frac{k_b^2 - \left({\bf g}_\perp + i\hat{\bf z}g_\perp\right) \otimes
  \left({\bf g}_\perp + i\hat{\bf z}g_\perp\right)}{g_\perp^2} \nonumber \\ 
&\times \exp \left[ \left( i {\bf g}_\perp - \hat{\bf z}g_\perp \right)
  \cdot {\bf r}\right] \tilde{\bf F}\left( {\bf g}_\perp + \hat{\bf
    z}g_\perp \right) \ , \\
& \hspace{5cm}  {\bf r} \in \Omega_{\rm tot} \ . \nonumber
\label{surf_hmg}
\end{align}
\noindent
It can be seen that ${\bf E}_S$ decays exponentially on the
scale of $h$.  So does the wave of polarization ${\bf P}_S$,
as both fields are related by the integral equation (\ref{surf_eq}).

Solving Eq.~(\ref{surf_eq}) numerically can be a very difficult task.
Fortunately, doing so is not necessary if one is only concerned with
far-field measurements.

\subsection{Reflection coefficient}
\label{subsec:refl}

We will now utilize the results of the previous subsection to compute
the reflection coefficients for the half-space. We will use the
assumption of Sec.~\ref{subsec:iso}, namely, that the crystallographic
and optical axes of the medium coincide so that the tensor
$\Sigma$ is diagonal in the laboratory frame. Apart from other
simplifications, media of this type are non-chiral and do not rotate
the polarization of the transmitted and reflected waves. This property
holds even beyond the homogenization limit, since it is a straightforward
consequence of the elementary cell symmetries, and it will enable us
to consider the s- and p-polarizations separately.

In this subsection, we will explicitly use the reference frame shown
in Fig.~\ref{fig:sketch_2D}. That is, we will assume that the plane of
incidence is the $xz$-plane and that the projection of the wave
vectors ${\bf k}_i$, ${\bf k}_r$ and ${\bf q}$ onto the interface is
${\bf k}_\perp = k_x \hat{\bf x}$.

\subsubsection{S-polarization}

In the case of s-polarization, the incident and reflected waves are
polarized perpendicularly to the plane of incidence. Consequently, we
have ${\bf A}_i,{\bf A}_r \propto \hat{\bf y}$, and the {\em exact}
reflection coefficient is given by
\begin{equation}
\label{rs_1}
r = \frac{{\bf A}_r \cdot \hat{\bf y}}{{\bf A}_i \cdot \hat{\bf y}} =
- \frac{\tilde{\bf F}({\bf k}_r) \cdot \hat{\bf y}}{\tilde{\bf F}({\bf
    k}_i) \cdot
  \hat{\bf y}} \frac{\exp\left[ i(k_{iz} - q_z)h\right] -
  1}{\exp\left[ -i(k_{iz} + q_z)h\right] - 1} \ .
\end{equation}
\noindent
To derive the second equality, we have used the expressions
(\ref{extinction_exact}) and (\ref{reflextion_exact}) for the
amplitudes ${\bf A}_i$ and ${\bf A}_r$. This is an exact expression
that retains its physical meaning as long as (\ref{no_diffr}) holds.
In the homogenization limit, we use the expressions (\ref{limits}) to
obtain
\begin{equation}
\label{rs_2}
r = \frac{k_{iz} - q_z}{k_{iz} + q_z} \ .
\end{equation}
\noindent
Here $q_z$ is given by
\begin{equation}
\label{qz_c_s}
q_z = \sqrt{k^2 \eta_y - k_x^2} \ ,
\end{equation}
\noindent
which follows from the dispersion relation (\ref{disp_2D_s}), in which
we must take $q_x=k_x$. The square root branch in (\ref{qz_c_s}) is
determined by the condition ${\rm Im}(q_z) >0$.

The expressions (\ref{rs_2}) and (\ref{qz_c_s}) should be compared to
the corresponding Fresnel coefficient $r_F$ and the dispersion
relation for a homogeneous medium characterized by the permittivity
and permeability tensors $\bar{\epsilon}$ and $\bar{\mu}$:
\begin{equation}
\label{rs_F}
r_F = \frac{k_{iz} - q_z/\bar{\mu}_{xx}}{k_{iz} + q_z/\bar{\mu}_{xx}}
\ ,
\end{equation}
\noindent
The wave number $q_z$ in an effective medium satisfies the dispersion
relation
\begin{equation}
\label{qz_h_s}
q_z = \sqrt{k^2 \bar{\epsilon}_{yy} \bar{\mu}_{xx}  -
  k_x^2\frac{\bar{\mu}_{xx}}{\bar{\mu}_{zz}}} \ .
\end{equation}
As was discussed in Sec.~\ref{subsubsec:q_general}, we must impose the
condition (\ref{eff_pars}) on the EMPs $\bar{\epsilon}$ and
$\bar{\mu}$ in order to obtain the same laws of dispersion in the
composite and in the continuous medium models. In particular, this
condition guarantees that the quantities $q_z$ given by
Eqs.~(\ref{qz_c_s}) and (\ref{qz_h_s}) are equal for all values of
$k_x$. But if this is so, the only way the two expression (\ref{rs_2})
and (\ref{rs_F}) can yield the same reflection coefficient is if we
set $\xi=1$ in (\ref{eff_pars}), which corresponds to $\bar{\mu}=1$.

We note that to reach the above conclusion, it is sufficient to
consider the reflection coefficient for s-polarization only. We will
show next that the same conclusion can be reached by considering
p-polarization only and that the homogenization results obtained in
these two cases are consistent.

\subsubsection{P-polarization}

In the case of p-polarization, the reflection coefficient can be
conveniently defined by using the ratio of tangential components of
the magnetic field for the reflected and incident waves. The magnetic
field amplitudes of these waves are given by
\begin{equation}
\label{B_def}
{\bf B}_{i,r} = \frac{1}{k} {\bf k}_{i,r} \times {\bf A}_{i,r} \ .
\end{equation}
\noindent
As could be anticipated, the amplitudes ${\bf B}_{i,r}$ are aligned
with the $y$-axis. We can now use the expressions
(\ref{extinction_exact}) and (\ref{reflextion_exact}) for the
amplitudes ${\bf A}_{i,r}$ to find the {\em exact} reflection
coefficient:
\begin{align}
  r =& \frac{{\bf B}_r \cdot \hat{\bf y}}{{\bf B}_i \cdot \hat{\bf y}}
  \nonumber \\
  =- &  \frac{\left[{\bf k}_r \times \tilde{\bf F}({\bf k}_r) \right]
    \cdot \hat{\bf y}}{\left[ {\bf k}_i \times \tilde{\bf F}({\bf
        k}_i) \right] \cdot \hat{\bf y}} \frac{\exp\left[ i(k_{iz} -
      q_z)h\right] - 1}{\exp\left[ -i(k_{iz} + q_z)h\right] - 1} \ .
\label{rp_1}
\end{align}
\noindent
In the homogenization limit, this expression is simplified by using
(\ref{limits}), which leads to
\begin{equation}
\label{rp_2}
r = 
- \frac{\left[{\bf k}_r \times (1+\Sigma){\bf F}_0 \right] \cdot
  \hat{\bf y}}{\left[ {\bf k}_i \times (1 + \Sigma){\bf F}_0 \right] \cdot
  \hat{\bf y}} \frac{k_{iz} - q_z}{k_{iz} + q_z} \ .
\end{equation}
\noindent
As shown in Appendix~\ref{app:p-pol}, Eq.~(\ref{rp_2}) can be further
simplified to read
\begin{equation}
\label{rp_3}
r = \frac{k_{iz}/\epsilon_b - q_z/\eta_x}{k_{iz}/\epsilon_b + q_z/\eta_x} \ .
\end{equation}
In (\ref{rp_2}),(\ref{rp_3}), $q_z$ satisfies the dispersion relation
for the p-polarized wave, (\ref{disp_2D_p}). With the substitution
$q_x=k_x$, the latter reads
\begin{equation}
\label{qz_c_p}
q_z = \sqrt{k^2 \eta_x - k_x^2 \frac{\eta_x}{\eta_z}} \ .
\end{equation}
\noindent
As in the case of s-polarization, the branch of the square root is
determined by the condition ${\rm Im}(q_z) >0$.

We wish to compare the expressions (\ref{rp_3}) and (\ref{qz_c_p}) to
the analogous expressions in a continuous medium with the EMPs
$\bar{\epsilon}$ and $\bar{\mu}$. The Fresnel reflection coefficient
for a p-polarized incident wave is given by
\begin{equation}
\label{rp_F}
r_F = \frac{k_{iz}/\epsilon_b -
  q_z/\bar{\epsilon}_{xx}}{k_{iz}/\epsilon_b + q_z/\bar{\epsilon}_{xx}}
\ , 
\end{equation}
and the dispersion relation in the effective medium is
\begin{equation}
\label{qz_h_p}
q_z = \sqrt{k^2 \bar{\epsilon}_x\bar{\mu}_y - k_x^2 \frac{\bar{\epsilon}_x}{\bar{\epsilon}_z}}
\end{equation}
\noindent
As in the case of s-polarization, the condition (\ref{eff_pars}) with
an arbitrary parameter $\xi$ guarantees that the two expressions
(\ref{qz_c_p}) and (\ref{qz_h_p}) yield the same wave number $q_z$ for
all values of $k_x$. However, the expressions (\ref{rp_3}) and
(\ref{rp_F}) yield the same reflection coefficient only if we set
$\xi=1$ in (\ref{eff_pars}).

This completes the proof that the correct choice of the parameter
$\xi$ in (\ref{eff_pars}) is $\xi=1$ and, correspondingly, the correct
homogenization result is $\bar{\mu}=1$. A similar proof has been given
by us for a one-dimensional layered medium in Ref.~\cite{markel_10_1}
for both s- and p-polarizations.

\section{Comparison of point-dipole and continuous-medium
  models}
\label{sec:point-bulk}

The model of point-like polarizable particles arranged on a
three-dimensional infinite lattice possesses an intuitive physical
appeal. Historically, many authors have used this model and, although
an exhaustive review is outside of the scope of this paper,
Refs.~\cite{mahan_69_1,sipe_74_1,lamb_80_1,draine_93_1,belov_05_1,abajo_07_1}
can be mentioned. Unfortunately, the model is haunted by divergences.
In this section, we will discuss the nature and origins of these
divergences and some of the commonly-used methods for their
regularization. We will also attempt, to the degree it is possible, to
establish a correspondence between the model of point dipoles and the
model of a continuous two-component medium, which is the subject of
this paper.

Most previous works on electromagnetic waves in point-dipole lattices
assume that the background medium is vacuum. For compatibility of
results and simplicity of notations, we will also make this assumption
(in this section only) and set $\epsilon_b=1 + i0$, $k_b=k=\omega/c +
i0$.

The model of point dipoles considers an array of point-like particles
which have well-defined locations, but no shape or size. Instead of
the latter two physical characteristics, the electric dipole
polarizability $\alpha$ is used. In some generalizations of the model,
the magnetic dipole polarizability is also included. The basic idea of
this approach is that the electromagnetic response of a particle is
completely characterized by its polarizability.

If only the electric polarizability is retained, one arrives, in lieu
of the integral equation (\ref{main_eq}), at the set of algebraic
equations
\begin{equation}
\label{CDE}
\frac{1}{\alpha}{\bf d}_n = {\bf E}_i({\bf r}_n) +
  \sum_{m\neq n} G({\bf r}_n,{\bf r}_m) {\bf d}_m \ .
\end{equation}
\noindent
Here ${\bf d}_n$ is the electric dipole moment of the $n$-th particle.
Now two important points should be made. First, the summation on the
right-hand side of (\ref{CDE}) is restricted only to the indices $m$
which are not equal to $n$. This reflects the idea that the electric
field at the site of the $n$th dipole is a superposition of the
incident wave ${\bf E}_i({\bf r}_n)$ and the waves scattered by all
other dipoles.  Second, energy conservation requires
that~\cite{draine_88_1,markel_92_1,markel_95_1} ${\rm Im}(1/\alpha)
\leq -2 k^3/3$. If the equality holds, the particles are
non-absorbing. It is convenient to decompose the inverse
polarizability as
\begin{equation}
\label{alpha_alpha_LL}
\frac{1}{\alpha} = \frac{1}{\alpha_{\rm LL}} - i \frac{2k^3}{3} \ ,
\end{equation}
\noindent
where $\alpha_{\rm LL}$ is the ``Lorenz-Lorentz'' quasistatic
polarizability and $-i2 k^3/3$ is the first
non-vanishing radiative correction to the imaginary part of
$1/\alpha$. Radiative corrections to the real part of $1/\alpha$ also
exist and are, in fact, of a lower order in $k$, but it is the
correction to the imaginary part which is physically important and
should be retained even in the limit $k h \rightarrow 0$. We will see
momentarily that the two seemingly unrelated facts mentioned above are
mathematically connected.

We now consider an infinite lattice, set the incident field to zero
and seek the solution to (\ref{CDE}) in the form ${\bf d}_n = {\bf d}
\exp(i {\bf q} \cdot {\bf r}_n)$. This results in the eigenproblem
\begin{equation}
\label{CDE_eigen}
\frac{1}{\alpha} {\bf d} = S({\bf q}) {\bf d} \ ,
\end{equation}
\noindent
where
\begin{equation}
\label{W_CDE_def}
S({\bf q}) = \sum_{m\neq n} G({\bf r}_n,{\bf r}_m) \exp\left[ -i {\bf
    q} \cdot ({\bf r}_n - {\bf r}_m) \right] 
\end{equation}
\noindent
is the dipole sum. Using the Fourier representation (\ref{GR_F}), we
rewrite (\ref{W_CDE_def}) as
\begin{equation}
\label{W_1}
S({\bf q}) = \frac{4\pi}{3} \int \frac{d^3p}{(2\pi)^3} K({\bf p})
\sum_{m\neq n}  \exp\left[ i ({\bf p} - {\bf q}) \cdot ({\bf r}_m - {\bf r}_n)
\right] \ .
\end{equation}
The first complication encountered in the above is that the summation on the
right-hand side of (\ref{W_1}) is incomplete. We can easily fix this
problem by adding and subtracting unity to the series, which leads to
\begin{equation}
\label{W_2}
S({\bf q}) = \frac{4\pi}{3}\left[\frac{1}{h^3} \sum_{\bf g} K({\bf q} +
  {\bf g}) - \int \frac{d^3p}{(2\pi)^3} K({\bf p}) \right] \ ,
\end{equation}
\noindent
where we have used the Poisson summation formula (\ref{Poisson}).
Still, both terms on the right-hand side of (\ref{W_2}) are divergent.
We will deal with the integral first. To this end, we utilize the
expression for $K({\bf p})$ given in (\ref{K_def}) and notice that the
angular integral of the term $p^2 - 3{\bf p}\otimes {\bf p}$ is zero
in three dimensions.  Therefore, we have
\begin{equation}
\label{int_1}
I \equiv \int \frac{d^3p}{(2\pi)^3} K({\bf p}) = 4\pi \int_0^\infty
\frac{p^2 dp}{(2\pi)^3} \frac{2 k^2}{p^2 - k^2} \ .
\end{equation}
\noindent
This is still a divergent integral. We can regularize (\ref{int_1}) by writing
\begin{equation}
\label{int_2}
I = \lim_{\lambda \rightarrow 0} \left\{ 4\pi \int_0^\infty
\frac{p^2 dp}{(2\pi)^3} \frac{2 k^2}{p^2 - k^2} \exp[-(\lambda p)^2]
\right\} \ .
\end{equation}
\noindent
The above limit indeed exists and is equal to $ik^3 / 2\pi$, assuming
that ${\rm Im} k > 0$ (which is true if we take $k = \omega/c + i0$).
Upon substitution of this result into (\ref{W_2}), we find that
\begin{equation}
\label{cancellation}
\frac{4\pi}{3}\left[ - \int \frac{d^3p}{(2\pi)^3} K({\bf p}) \right] =
- i \frac{2k^3}{3} \ .
\end{equation}
\noindent
We now use the decomposition (\ref{alpha_alpha_LL}) and notice that
the above term is canceled by a similar term on the left-hand side
of (\ref{CDE_eigen}). Taking into account this cancellation,
(\ref{CDE_eigen}) becomes
\begin{equation}
\label{CDE_eigen_1}
\frac{1}{\alpha_{\rm LL}} {\bf d} =  \frac{4\pi}{3h^3} \sum_{\bf g}
K({\bf q} + {\bf g})  {\bf d} \ .
\end{equation}
The mathematical tricks used so far are not very objectionable. The
result (\ref{cancellation}) is a reflection of the fact that
\begin{equation}
\label{G_int}
\lim_{\lambda \rightarrow 0} \left[ \frac{3}{4\pi\lambda^3} \int_{\vert {\bf r}^\prime
  - {\bf r} \vert \leq \lambda} G({\bf r},{\bf r}^\prime)d^3r^\prime \right] =
-i\frac{2k^3}{3} \ .
\end{equation}
\noindent
Here we have assumed that the particle is spherically symmetric. The
use of a different integration volume in (\ref{G_int}), or of a
different regularization function in (\ref{int_2}), would certainly
yield a different result. Fortunately, if $k h \ll 1$, only the real
part of $I$ is affected by the choice of the regularization function
in (\ref{int_2}) while the imaginary part is relatively stable. If
${\rm Re}I$ is unimportant, e.g., if it is small compared to the sum
of real parts all other contributions in (\ref{W_2}), then
(\ref{CDE_eigen_1}) is a good approximation, regardless of the true
shape of the particles.

However, the divergence of the series in the right-hand side of
(\ref{CDE_eigen_1}) is truly problematic. One can attempt to
regularize this divergence by the same mathematical trick that was
used above. However, the result of such a manipulation would indeed
depend on the regularization function in a nontrivial way. One can
conclude that knowledge of the particle polarizability is, in fact,
insufficient for solving the problem at hand. The shape of the
particles is also important and can not be disregarded.

Another way to look at this is the following. The polarizability
$\alpha$ defines the response of a particle to an external electric
field which is almost uniform over the particle volume.  However, in
an infinite three-dimensional lattice, the electric field is not
uniform over the particle volume, no matter how small the particle is.
This is because the lattice Green's function $W({\bf R},{\bf
  R}^\prime)$ given by (\ref{W_eval}) experiences an integrable
divergence when ${\bf R}={\bf R}^\prime$. However, in the point-dipole
model, we are attempting to evaluate this function exactly at ${\bf
  R}={\bf R}^\prime=0$, which is not mathematically reasonable.

It appears that the only feasible approach to regularize the summation
in (\ref{CDE_eigen_1}) is to endow the particles with a finite volume,
as was done, for example, in Ref.~\cite{sipe_74_1}. This would
naturally lead to a modification of (\ref{CDE_eigen_1}) in which the
right-hand side is multiplied by a decaying function $f({\bf g})$, 
ensuring convergence. Unfortunately, the exact form of
$f({\bf g})$ strongly depends on the particle shape and size. If the
regularization is carried out in a mathematically-consistent way, one
would end up with a set of equations that are identical to the
equations obtained here, for the model of a continuous two-component
medium.

Evidently, within the point-dipole model, one wishes to avoid introducing
the particle shape and size. Then the only conceivable approach to
regularization is simply to truncate the series in (\ref{CDE_eigen_1}),
by leaving only the ${\bf g}=0$ term in the summation, which leads to
the eigenproblem
\begin{equation}
\label{CDE_eigen_2}
\frac{1}{\alpha_{\rm LL}} {\bf d} =  \frac{4\pi}{3h^3}
K({\bf q})  {\bf d} \ .
\end{equation}
\noindent
Regularization of this type is, in fact, appropriate for small
spherical particles. If one also uses the quasistatic polarizability
of a sphere of radius $a$, namely,
\begin{equation}
\label{alpha_LL}
\alpha_{\rm LL} = a^3 \frac{\epsilon_a - 1}{\epsilon_a +  2} \ ,
\end{equation}
\noindent
then (\ref{CDE_eigen_2}) becomes equivalent to the Clausius-Mossotti
relation and the EMT that follows from it is the standard
Maxwell-Garnett approximation.

One may be tempted to forget about the limits of applicability
of Eq.~(\ref{CDE_eigen_2}). In other words, once (\ref{CDE_eigen_2})
is derived, it is technically possible to use it with {\em any}
polarizability $\alpha_{\rm LL}$. The latter can be obtained
independently, i.e., by solving the Laplace equation for a single
isolated particle of arbitrary shape. Unfortunately, this approach is
mathematically inconsistent. Eq.~(\ref{CDE_eigen_2}) was derived from
(\ref{CDE_eigen_1}) by applying a regularization method which is only
appropriate for small spheres. Applying (\ref{CDE_eigen_2}) to
particles of nonspherical shape is likely to result in errors.

In summary, the model of point dipoles is capable of reproducing the
standard Maxwell-Garnett mixing rule for small spheres. Radiative
corrections to this result can also be derived~\cite{draine_93_1}.
However, in three dimensions, the model breaks down and can not be
used when a substantial deviation from the Maxwell-Garnett
approximation is expected, i.e., for particles whose volume fraction
is not small or whose shape is different from a sphere. In other
words, the model does not provide a mathematically consistent way of
computing the self-energy $\Sigma$ which appears in equations
(\ref{det=0}) or (\ref{eps_eff}) and is, therefore, usable only in the
physical situations when $\Sigma$ can be neglected. 
Nevertheless, we note that in systems of lower dimensionality (e.g., in nanoparticle
chains), the point-dipole model is useful and can provide significant
physical insights.

\section{Continued-fraction expansion of the self-energy and the
  mean-field approximation}
\label{sec:CF} 

\subsection{Abstract notation}
\label{subsec:absnot}

In this section, we will find it convenient to rewrite
Eqs.~(\ref{g=/=0}) and (\ref{Sigma_def}) in Dirac notation.  First, we
note that, in order to recover all components of the tensor $\Sigma$,
one must solve (\ref{g=/=0}) for three different right-hand sides:
${\bf F}_0 = \hat{\bf x}$, ${\bf F}_0 = \hat{\bf y}$ and ${\bf F}_0 =
\hat{\bf z}$. To this end, we introduce a triplet of
infinite-dimensional vectors $\vert a_\beta \rangle$, operators $Q$,
$M$, $W$, and vectors $\vert b_\beta \rangle$ ($\beta=x,y,z$)
according to
\begin{subequations}
\label{abs_not}
\begin{align}
\langle \alpha {\bf g} \vert a_\beta \rangle &= M({\bf g})\delta_{\alpha\beta} \  , \\ 
\langle \alpha {\bf g} \vert Q \vert \alpha^\prime {\bf g}^\prime
\rangle &= \delta_{{\bf g}{\bf g}^\prime} \left( 1 - 3 \hat{g}_\alpha
  \hat{g}_{\alpha^\prime} \right) \ , \\ 
\langle \alpha {\bf g} \vert M \vert \alpha^\prime {\bf g}^\prime
\rangle &= \delta_{\alpha\alpha^\prime}M({\bf g} - {\bf g}^\prime) \ , \\ 
W &= QM \ , \\ 
\vert b_\beta \rangle &= Q \vert a_\beta \rangle \ .
\end{align}
\end{subequations}
\noindent
Note that $Q$ is diagonal in the index ${\bf g}$, $M$ is diagonal in
the index $\alpha$, but the product of the two, $W=QM$, is not diagonal. We
must also keep in mind that the index ${\bf g}$ in the above equations
is not allowed to take the zero value. We further define the vectors
$\vert F_\beta \rangle$ as the solutions to
\begin{equation}
\label{g=/=0_abs}
\left( \frac{1}{\rho\chi} - W \right) \vert F_\beta \rangle = \vert b_\beta \rangle \ .
\end{equation}
\noindent
The above is equivalent to the set (\ref{g=/=0}). The tensor elements
of $\Sigma$ are defined by
\begin{align}
\label{Sigma_abs}
\Sigma_{\alpha\beta} & = \langle a_\alpha \vert F_\beta \rangle =
\langle a_\alpha \vert \left( \frac{1}{\rho\chi} - W \right)^{-1}
\vert b_\beta \rangle \nonumber \\
& = \langle a_\alpha \vert \left( \frac{1}{\rho\chi} - QM
\right)^{-1}Q \vert a_\beta \rangle \ .
\end{align}
\noindent
It can be seen that $\Sigma$ is computed as the resolvent of the
operator $W=QM$ and plays the role of the self-energy, which accounts
for interactions between the inclusions.

\subsection{Mean-field approximation}
\label{subsec:MF}

The mean-field approximation is often misunderstood. In particular, it
is unrelated to Maxwell-Garnett theory. Rather, it allows one to
replace certain operators by appropriately chosen scalar multiples of
the identity. The approximation reproduces the exact zeroth and first
moments of the resolvent and serves as the first-order approximation
in its continued-fraction expansion. Here the approximation is
explained following Berry and Percival~\cite{berry_86_1}.

Let us seek the solution to Eq.~(\ref{g=/=0_abs}) in the form $\vert
F_\beta \rangle = \lambda \vert b_\beta \rangle$, where $\lambda$ is a
scalar to be determined. Upon substitution of this ansatz into
(\ref{g=/=0_abs}), we obtain the equation
\begin{equation}
\label{eq1}
\left( \frac{1}{\rho\chi} - \frac{1}{\lambda} \right)
  \vert b_\beta \rangle = W \vert b_\beta \rangle \ .
\end{equation}
\noindent
Because $\vert b_\beta \rangle$ is, generally, not an eigenvector of
$W$, there is no such value of $\lambda$ for which Eq.~(\ref{eq1})
would hold.  The best we can hope for is that a {\em projection} of
this equation onto a given vector would hold for some $\lambda$. Since
we are interested not in the whole vector $\vert F_\beta\rangle$ but
in its projection onto $\vert a_\alpha \rangle$, it seems reasonable to
project Eq.~(\ref{eq1}) onto the latter. This yields
\begin{equation}
\label{eq2}
\lambda = \frac{\rho\chi}{1 - \rho\chi {\displaystyle \frac{\langle a_\alpha \vert W \vert b_\beta
  \rangle}{\langle a_\alpha \vert b_\beta \rangle}} } \ ,
\end{equation}
\noindent
and the corresponding mean-field approximation for the self-energy is
\begin{align}
\label{Sigma_MF_1}
\Sigma_{\alpha\beta} & = \frac{ \rho\chi \langle a_\alpha \vert b_\beta
  \rangle}{1 - \rho\chi {\displaystyle \frac{\langle a_\alpha \vert W \vert b_\beta
  \rangle}{\langle a_\alpha \vert b_\beta \rangle} } } 
& = \frac{ \rho\chi \langle a_\alpha \vert Q \vert a_\beta
  \rangle}{1 - \rho\chi {\displaystyle \frac{\langle a_\alpha \vert QMQ \vert a_\beta
  \rangle}{\langle a_\alpha \vert Q \vert a_\beta \rangle} } } \ .
\end{align}
\noindent
As was mentioned in Sec.~\ref{subsec:low-dens}, the matrix element
\begin{equation}
\langle a_\alpha \vert Q \vert a_\beta \rangle  = \sum_{{\bf g}\neq 0}
\left[ M(-{\bf g}) Q({\bf g}) M({\bf g}) \right]_{\alpha\beta}
\end{equation}
\noindent
is identically zero for inclusions with cubic symmetry (in
three-dimensional composites) so that Eq.~(\ref{Sigma_MF_1}) yields in
this case zero and is not useful. If $\langle a_\alpha \vert Q \vert
a_\beta \rangle$ is zero, a non-vanishing mean-field approximation can
be obtained by ``shifting'' the solution according to $\vert F_\beta
\rangle = \rho\chi \vert b_\beta \rangle + \vert F_\beta^\prime
\rangle$. The self-energy is then given by $\Sigma_{\alpha\beta} =
\langle a_\alpha \vert F_\beta^\prime \rangle$ where $\vert
F_\beta^\prime \rangle$ satisfies
\begin{equation}
\label{F_shifted}
\left( \frac{1}{\rho\chi} - W \right) \vert F_\beta^\prime \rangle =
\rho\chi W \vert b_\beta \rangle \ . 
\end{equation}
\noindent
The mean-field approximation for the ``shifted'' equation
(\ref{F_shifted}) is
\begin{align}
\label{Sigma_MF_2}
\Sigma_{\alpha\beta}
= \frac{ (\rho\chi)^2 \langle a_\alpha \vert QMQ \vert a_\beta
  \rangle}{1 - \rho\chi {\displaystyle \frac{\langle a_\alpha \vert
      (QM)^2 Q \vert a_\beta \rangle}{\langle a_\alpha \vert QMQ \vert
      a_\beta \rangle} } } \ .
\end{align}

\subsection{Continued-fraction expansion of the self-energy}
\label{subsec:CF}

Continued-fraction expansions (CFEs) are very useful in
physics~\cite{haydock_80_1,markel_04_3}. The mathematical underpinning
of all CFEs is the theory of the correspondence between the formal
Laurent series of meromorphic functions and certain continued
fractions~\cite{jones_book_80}. There exists a deep mathematical
relation between CFEs and the problem of moments, that is, the problem
of finding a distribution from the knowledge of its moments.

CFEs can be derived in different ways. Haydock~\cite{haydock_80_1} has
employed the Lanczos recursion to transform a certain Hamiltonian to
tridiagonal form. A diagonal element of the inverse of a tridiagonal
matrix can be written as a J-fraction (a continued fraction of Jacobi
type). In Ref.~\cite{haydock_80_1}, this procedure was applied
to a Hermitian operator to compute a diagonal matrix element of the
resolvent. In this paper, the operator $W$ in (\ref{g=/=0_abs}) or
(\ref{Sigma_abs}) is not symmetric or Hermitian and we are interested
in off-diagonal elements of the resolvent.  Therefore, the numerical
procedure used by Haydock is not directly applicable. Perhaps, it can
be generalized to become applicable to the problem at hand; we have
not explored this possibility. Instead, we will derive a CFE for the
right-hand side of Eq.~(\ref{Sigma_abs}) from the following theorem
which does not require any symmetry properties of the operators
involved, yields a CFE for arbitrary off-diagonal elements, and, to
the best of our knowledge, has not been reported in the literature.
The resultant expansion will be an S-fraction (a continued fraction of
Stieltjes type). Note that an S-fraction can always be transformed
into a J-fraction by the so-called equivalence transformation.

\newtheorem{theo}{Theorem} 
\begin{theo}
  Let $W$ be a linear operator acting on the Hilbert space ${\mathcal
    H}$ and ${\mathcal Z}$ be a complex number. Suppose that $\vert
  \phi \rangle, \vert \psi \rangle \in {\mathcal H}$. If (i) $\langle
  \phi \vert \psi \rangle \neq 0$ and (ii) $({\mathcal Z} - W)^{-1}$
  exists, then
\begin{equation}
\label{theo}
\langle \phi \vert ({\mathcal Z} - W)^{-1} \vert \psi \rangle =
\frac{{\mathcal Z}^{-1} \langle \phi \vert \psi \rangle}{1 - 
{\displaystyle \frac{\langle \phi \vert ({\mathcal Z} - WT)^{-1}W \vert \psi
  \rangle}{\langle \phi\vert \psi \rangle }} }  \ ,
\end{equation}
\noindent
where
\begin{equation}
\label{T_def}
T = 1 - \frac{\vert \psi \rangle \langle \phi \vert}{\langle \phi \vert \psi \rangle} \ . 
\end{equation}
\noindent
The proof is given in Appendix~\ref{app:proof}. Note that (\ref{theo})
has a finite limit when ${\mathcal Z}\rightarrow 0$.
\end{theo}

The factor $\langle \phi \vert ({\mathcal Z} - WT)^{-1} W \vert \psi
\rangle$ in the denominator of (\ref{theo}) can be written as $\langle
\phi \vert ({\mathcal Z} - W_1)^{-1} \vert \psi_1 \rangle$, where
$W_1= WT$ and $\vert \psi_1 \rangle = W \vert \psi \rangle$. The formula
(\ref{theo}) can now be applied to $\langle \phi \vert ({\mathcal Z} -
W_1)^{-1} \vert \psi_1 \rangle$, and so on iteratively. After some
manipulation, this yields the following expansion:
\begin{equation}
\label{CFE}
\langle \phi \vert ({\mathcal Z} - W)^{-1} \vert \psi \rangle =
\frac{\kappa_1}{{\mathcal Z} - {\displaystyle \frac{\kappa_2}{1 -
    {\displaystyle \frac{\kappa_3}{{\mathcal Z} - \cdots}} }} } \ ,
\end{equation}
\noindent
Note the interlacing factors of ${\mathcal Z}$ and $1$. The
coefficients $\kappa_j$ ($j=1,2,\ldots$) are obtained from a three-point
recursion. Namely, starting from $\vert \psi_0 \rangle = 0$, $\vert
\psi_1 \rangle = \vert \psi \rangle$ and $\kappa_1 = \langle \phi \vert
\psi \rangle$, we compute for $j=1,2,\ldots$
\begin{equation}
\label{three-point}
\vert \psi_{j+1} \rangle = W \Big{(} \vert \psi_j \rangle - \kappa_j
  \vert \psi_{j-1} \rangle \Big{)} \ , \ \ \kappa_{j+1} = \frac{\langle
    \phi \vert \psi_{j+1} \rangle}{\langle \phi \vert \psi_j \rangle}
  \ . 
\end{equation}
\noindent
To obtain a CFE of the right-hand side of Eq.~(\ref{Sigma_abs}), we
identify ${\mathcal Z} = 1/\rho\chi$, $W = QM$, $\vert \phi \rangle =
\vert a_\alpha \rangle$ and $\vert \psi \rangle = \vert b_\beta
\rangle = Q \vert a_\beta \rangle$.

With the above substitutions taken into account, it transpires that
the coefficients $\kappa_j$ are determined only by the geometry of the
composite. Once a set of $\kappa_j$ have been found for a given
geometry, the EMPs can be easily computed for any material parameters
of the composite constituents. This is a characteristic feature of a
spectral theory and the CFE (\ref{CFE}) is, in fact, a spectral
representation of the self-energy $\Sigma$.

Finally, we note that, in the case of three-dimensional composites
with cubic symmetry, the first condition of the Theorem does not hold
when the theorem is applied directly to (\ref{Sigma_abs}). In this
case, one can build a CFE staring from the ``shifted'' equation
(\ref{F_shifted}).

\section{Numerical simulations}
\label{sec:num}

\subsection{General setup}
\label{subsec:sim_setup}

Numerical simulations have been performed for a two-dimensional
composite. The composite is periodic in the $xy$-plane while the
inclusions form infinitely-long fibers which are oriented parallel to
the $z$-axis and can have different cross sections. The case when the
electric field is parallel to the fibers is not considered here, since
this polarization results in a simple arithmetic average of the type
(\ref{eps_eff_Z}). However, when the electric field is polarized in
the $xy$-plane, the homogenization problem is nontrivial and can be
numerically challenging. We will consider inclusions with circular and
square cross sections, as is illustrated in Fig.~\ref{fig:cells}.  The
functions $M({\bf g})$ for these shapes are given in
Appendix~\ref{app:M}.

\begin{figure}
\centerline{\psfig{file=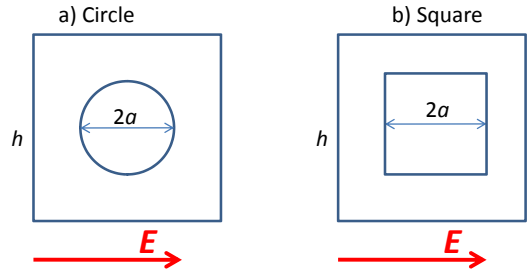,width=9cm,bbllx=136bp,bblly=320bp,bburx=656bp,bbury=576bp,clip=t}}
\caption{\label{fig:cells} (color online) Two types of elementary cells used in
  numerical simulations.}
\end{figure}

It is assumed that the host medium is vacuum and the inclusions are
metallic and characterized by a frequency-dependent Drude
permittivity of the form
\begin{equation}
\label{eps_a_b}
\epsilon_a = 1 - \frac{3\omega_F^2}{\omega(\omega + i \gamma)} \ , \ \ \epsilon_b = 1 \ .
\end{equation}
\noindent
In Eq.~(\ref{eps_a_b}), $\omega_F=\omega_p/\sqrt{3}$ is the Frohlich
frequency, $\omega_p$ is the plasma frequency, and $\gamma$ is the
Drude relaxation constant. We will compute the effective permittivity
of the composite $\bar{\epsilon}$ as a function of frequency for
$0.1 \leq \omega/\omega_F \leq 2$ and for the fixed ratio
$\gamma/\omega_F=0.1$. It is assumed that, for all frequencies used in
the simulations, the basic condition for the validity of a standard
EMT, $k_bh,qh \ll 1$, is satisfied.

Numerical simulations will be performed by truncating the infinite set
of equations (\ref{g=/=0}) so that the vectors ${\bf g}$ fill the box
\begin{equation}
\label{Lbox}
-2\pi L/h \leq g_x, g_y \leq 2\pi L /h \ ,
\end{equation}
\noindent
where $L$ is an integer. The total number of ${\bf g}$-vectors which
satisfy the above inequality is $(2L+1)^2$ and the total number of
algebraic equations to be solved is $N=2[(2L+1)^2 -1]$, where we have
accounted for the fact that the vector ${\bf g}=0$ is excluded in the
set of equations (\ref{g=/=0}). It can be seen that $N \rightarrow
8L^2$ when $L\rightarrow \infty$. In the simulations, we will use
integer powers of $2$ for $L$, up to $L=2^8=256$. The latter case
corresponds to $N=526,366$ equations.

The truncated set of equations (\ref{g=/=0}) can be solved by any
direct numerical method. The computational complexity of direct
methods is $O(N^3)$ and the solution must be obtained anew for every
frequency used (we sample the frequency at $200$ equidistant points in
the interval $0.1\leq \omega/\omega_F\leq 2$). This is time-consuming
but possible for $L\leq 64$. For larger values of $L$, direct methods
become impractical. We will use, therefore, the CFE of
Sec.~\ref{subsec:CF}.  The computational complexity of this expansion
is $O(j_{\rm max} N^2)$, where $j_{\rm max}$ is the order of
truncation of the continued fraction. More specifically, the continued
fraction is truncated by assuming that $\kappa_j=0$ for $j>j_{\rm max}$,
so that only the first $j_{\rm max}$ coefficients are used in
Eq.~(\ref{CFE}).  For the problem at hand, $j_{\rm max} \approx 50$
will prove sufficient.  Other iterative methods, such as the conjugate
gradient method, also have computational complexity $O(j_{\rm
  max}N^2)$, $j_{\rm max}$ being the number of iterations. However,
the computationally-intensive part of the conjugate-gradient solver
(when applied to Eq.~(\ref{g=/=0})) must be repeated for every value
of $\omega$, while the coefficients $\kappa_j$ in (\ref{CFE}) need to be
computed only once for a given geometry.

The inclusions shown in Fig.~\ref{fig:cells} have cubic symmetry.  As
was discussed in Sec.~\ref{subsec:iso}, the self-energy $\Sigma$ is
reduced in this case to a scalar. As a result, the effective medium is
isotropic in the $xy$-plane. Of course, anisotropy can still be
revealed if the polarization vector has a component along the
$z$-axis. In the simulations reported below, we have computed $\Sigma$
by solving Eqs.~(\ref{g=/=0}) and using the definition
(\ref{Sigma_def}). The effective permittivity for
transversely-polarized waves was then computed by using
Eq.~(\ref{eps_eff}).

\subsection{Convergence and stability}
\label{subsec:conv}

\begin{figure}
\psfig{file=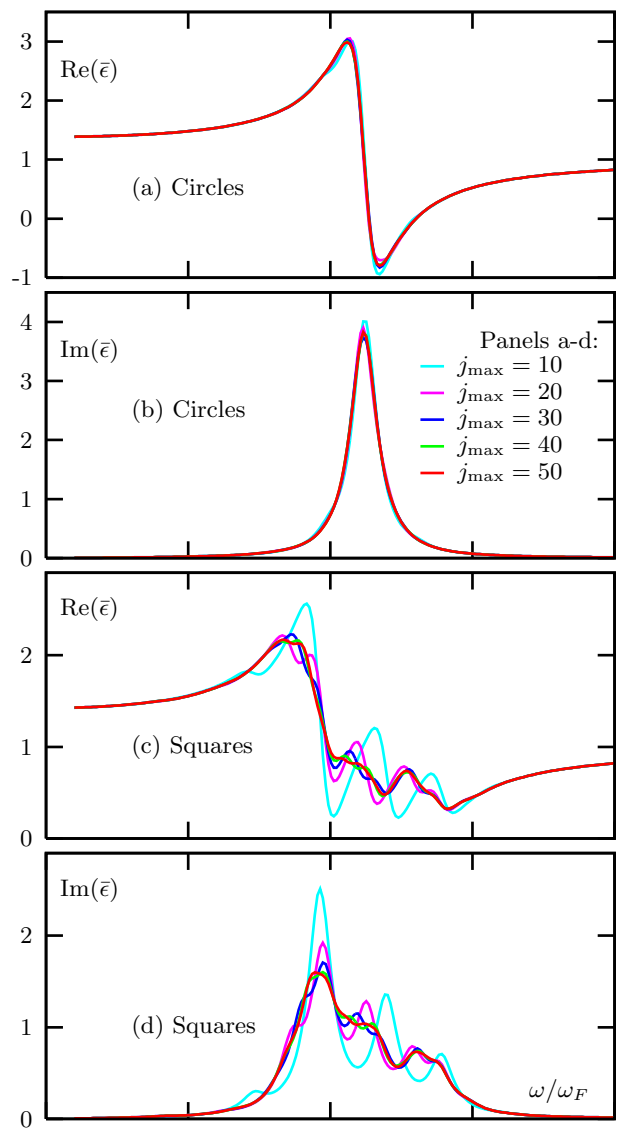,width=8.2cm,bbllx=60bp,bblly=300bp,bburx=290bp,bbury=725bp,clip=}
\caption{\label{fig:5} (color online) Convergence
  of the CFE (\ref{CFE}) with the truncation order $j_{\rm max}$ for
  circular (a,b) and quadratic (c,d) inclusions with the same volume
  density $\rho=0.16$. The set of equations (\ref{g=/=0}) has been
  truncated using $L=64$. In panels (a,b), the curves with $j_{\rm
    max}=30,40,50$ are indistinguishable.}
\end{figure}

The convergence of the CFE (\ref{CFE}) with the truncation order of the
continued fraction, $j_{\rm max}$, is illustrated in Fig.~\ref{fig:5}.
Here the real and imaginary parts of the effective permittivity are
plotted as functions of frequency. It can be seen that the convergence
is very fast for circular inclusions and somewhat slower for square
inclusions.  In all cases, $j_{\rm max}=50$ is sufficient for
convergence.

\begin{figure}
\psfig{file=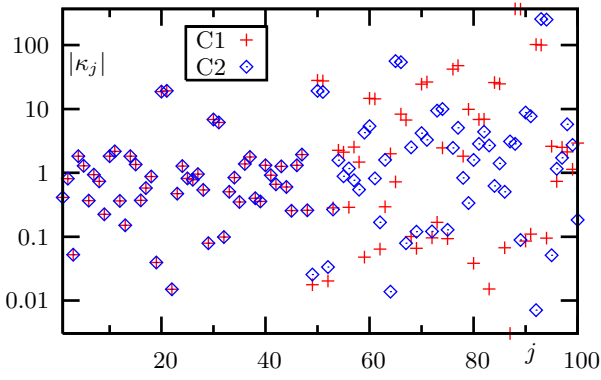,width=8.2cm,bbllx=70bp,bblly=580bp,bburx=300bp,bbury=725bp,clip=}
\caption{\label{fig:6} (color online) Absolute values of the
  coefficients $\kappa_j$ computed for circular inclusions with
  $\rho=0.16$ and $L=64$ on two different computers (C1 and C2).
  The same FORTRAN code and input data have been used in
  both cases.}
\end{figure}

The three-point recurrence relation (\ref{three-point}) is numerically
unstable for large values of $j$.  This is illustrated in
Fig.~\ref{fig:6}. Shown in this figure are the coefficients $\kappa_j$
obtained on two different computers for the geometry described in the
figure caption. The same code and input data were
used in both cases. The coefficients from the two sets coincide for $j
\lesssim 50$ with high precision. However, differences start to appear
at $j \sim 50$ and, at $j \sim 100$, the coefficients are
unreliable. The instability occurs when an iteration step in
(\ref{three-point}) asks for a relatively small difference of two
large numbers and the numerical precision of the floating-point
arithmetic is exceeded.

\begin{figure}
\psfig{file=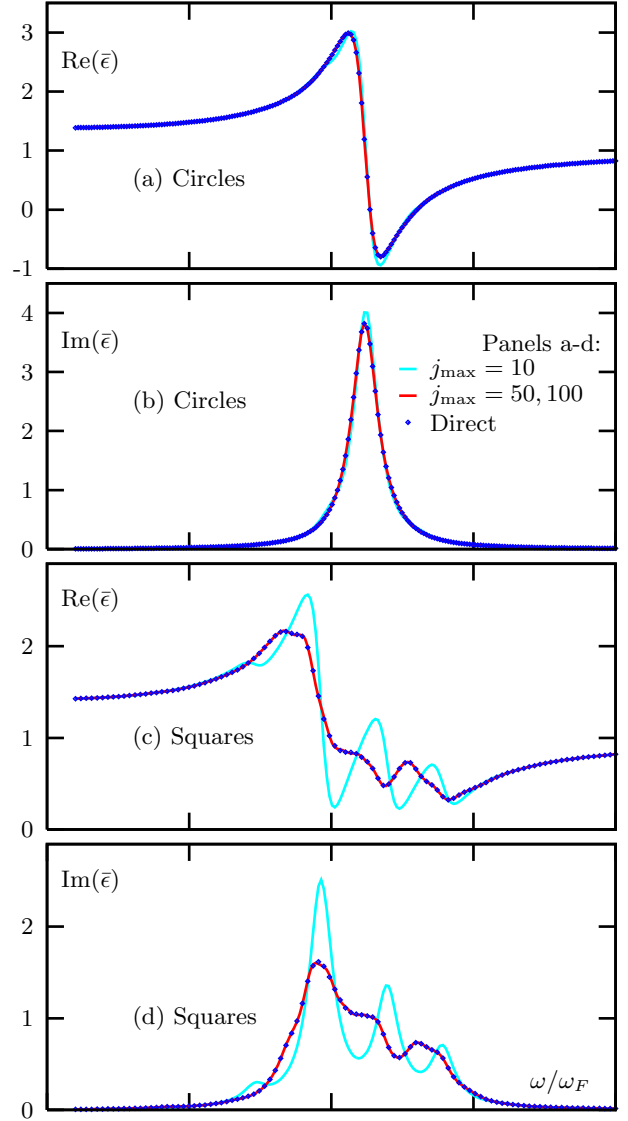,width=8.2cm,bbllx=60bp,bblly=300bp,bburx=290bp,bbury=725bp,clip=}
\caption{\label{fig:7} (color online) Effective permittivity of
  circular (a,b) and square (b,c) inclusions computed using the CFE
  (\ref{CFE}) with the truncation orders $j_{\rm max} = 10, 50, 100$,
  and by direct inversion of Eqs.~(\ref{g=/=0}). In all cases,
  $\rho=0.16$ and $L=64$. The data points for $j_{\rm max} = 50$ and
  $j_{\rm max} = 100$ are visually indistinguishable and, therefore,
  represented with the same curve.}
\end{figure}

The instability illustrated in Fig.~\ref{fig:6} appears to be
troublesome but is, in fact, of little concern. This is illustrated in
Fig.~\ref{fig:7}, which displays the effective permittivity computed
by the CFE (\ref{CFE}) for various truncation orders $j_{\rm max}$,
and the same quantity computed by solving Eqs.~(\ref{g=/=0}) directly.
One of the sets of $\kappa_j$'s displayed in Fig.~\ref{fig:6} has been
used for computing the data points for panels (a,b) of
Fig.~\ref{fig:7}.  Despite the instability, the curves with $j_{\rm
  max} = 50$ and $j_{\rm max} = 100$ are indistinguishable and very
close to the data points obtain by direct inversion of (\ref{g=/=0}).
Thus, the unreliable coefficients $\kappa_j$ do not influence the final
result.  This is one of the nice properties of all CFEs: a numerical
instability does not result in numerical imprecision. It is true
that increasing the truncation order beyond $j_{\rm max} = 50$ is not
useful, but it is not harmful either.  This point and some related
issues are discussed in more detail in Sec.~\ref{sec:disc} below.

\begin{figure}
\psfig{file=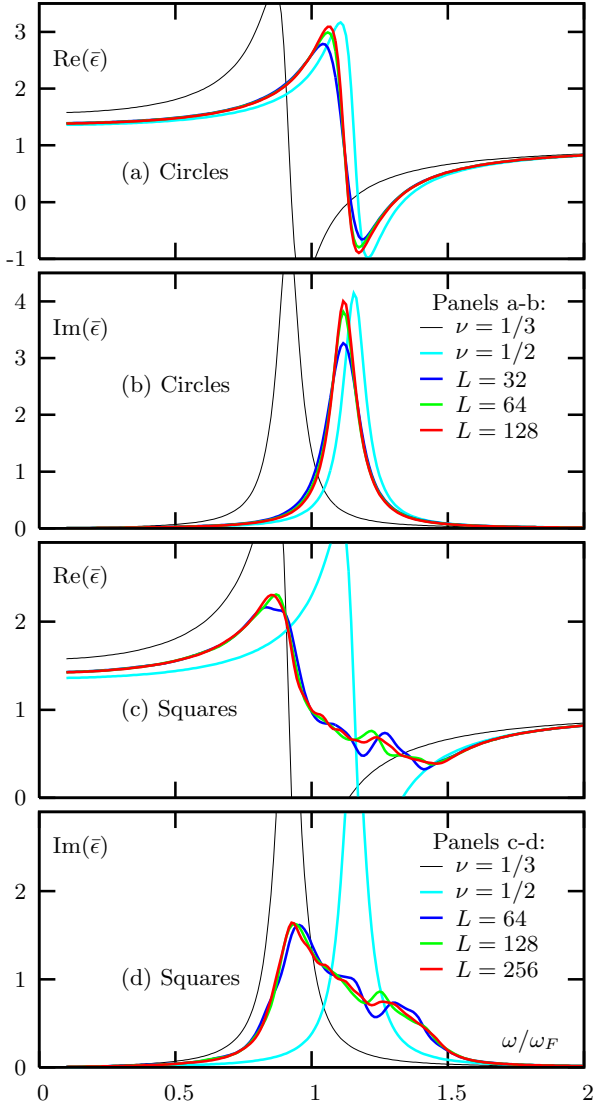,width=8.2cm,bbllx=60bp,bblly=290bp,bburx=300bp,bbury=725bp,clip=}
\caption{\label{fig:8} (color online) Convergence of the effective
  permittivity $\bar{\epsilon}$ with the size of the box, $L$, for
  circular (a,b) and square (c,d) inclusions with $\rho=0.16$. The
  curves labeled as $\nu=1/2$ and $\nu=1/3$ have been obtained from
  the generalized Maxwell-Garnett mixing formula (\ref{MG_nu}) for the
  values of $\nu$ indicated.}
\end{figure}

\begin{figure}
\psfig{file=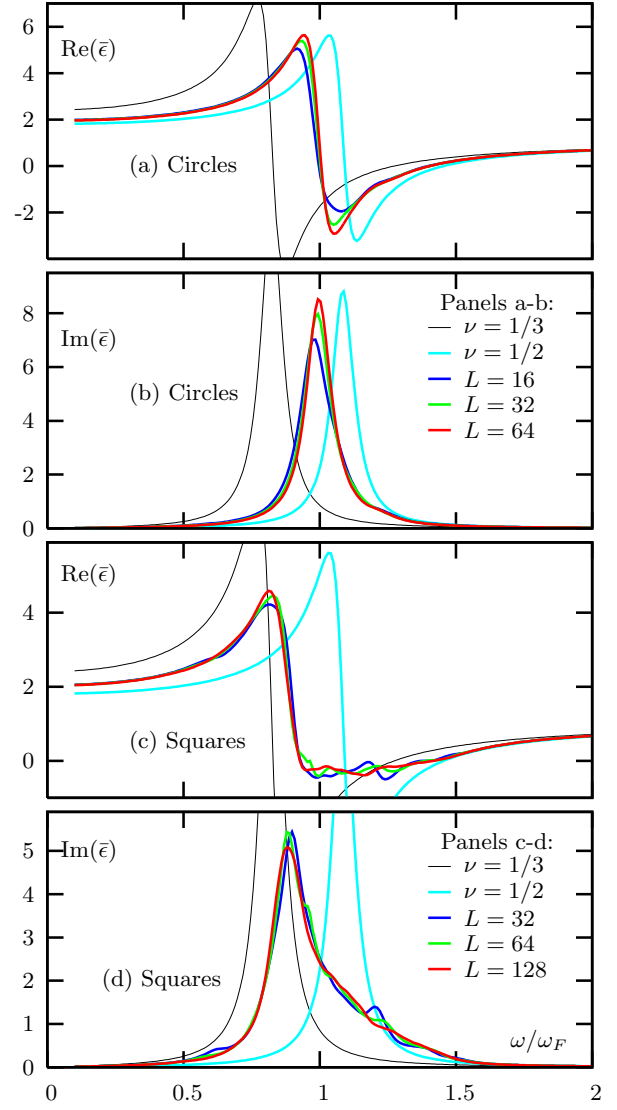,width=8.2cm,bbllx=60bp,bblly=290bp,bburx=300bp,bbury=725bp,clip=}
\caption{\label{fig:9} (color online) Same as in Fig.~\ref{fig:8} but for
  $\rho=0.32$ and different values of $L$, as indicated.}
\end{figure}

Having established the convergence properties of the CFE, we next
consider convergence with the size of the box, $L$ (up to now, all
plots have been computed for $L=64$). In
Figs.~\ref{fig:8},\ref{fig:9}, $\bar{\epsilon}$ is plotted as
functions of frequency for various values of the density, $\rho$, and
the box size, $L$. Also shown in these figures are the results
obtained from the generalized Maxwell-Garnett formula

\begin{equation}
\label{MG_nu}
\epsilon_\nu = \epsilon_b 
\frac{1 + {\displaystyle \frac{2\rho}{3}\frac{\epsilon_a - \epsilon_b}{\epsilon_b
    + \nu (\epsilon_a - \epsilon_b)}}}
{1 - {\displaystyle \frac{\rho}{3}\frac{(\epsilon_a - \epsilon_b)}{\epsilon_b
    + \nu(\epsilon_a - \epsilon_b)}}} \ ,
\end{equation}

\noindent 
which applies to ellipsoids, $\nu$ being the appropriate
depolarization factor. In the case of three-dimensional spheres,
$\nu=1/3$ and Eq.~(\ref{MG_nu}) coincides with Eq.~(\ref{eps_eff}) in
which the self-energy $\Sigma$ is set to zero. In the case of infinite
circular cylinders, the depolarization factor, which corresponds to
the orthogonal electric polarization, is $\nu=1/2$.

Several conclusions can be drawn from Figs.~\ref{fig:8} and
\ref{fig:9}.  First, convergence is obtained for boxes of reasonable
size. In all cases shown, $L=256$ yields very accurate results, and in
some cases $L=64$ is sufficient. However, it is important to note that
we have verified the convergence by doubling the size of the box.
Determination of convergence by using linearly sampled values of $L$,
(say, $L=10,11,12\ldots$) can be misleading.  This is a typical
situation when boundary-value problems are solved numerically.
Convergence must be established by at least doubling the size of the
mesh used.

Second, it can be seen that convergence is faster for $\rho=0.32$ than
for $\rho=0.16$. Although the electromagnetic interaction is stronger in
the second case, the faster convergence is to be expected. Indeed, the
size of the box should be selected so that the sum rules (\ref{B1})
are satisfied with some reasonable precision, and that is achieved at
smaller values of $L$ for larger values of $\rho$. Even faster
convergence is obtain for $\rho=64$ (data not shown). However, at the
percolation threshold ($\rho=\pi/4\approx 0.79$ for circular
inclusions), the convergence is relatively slow.

Third, the generalized Maxwell-Garnett formula (\ref{MG_nu}) with
$\nu=1/2$ yields a reasonable result for circular inclusions with
$\rho=0.16$. Even better agreement has been obtained for $\rho=0.08$
and $\rho=0.04$ (data not shown). However, as the size of circular
inclusions increases, the Maxwell-Garnett approximation becomes less
accurate.  For square inclusion, the approximation is inaccurate even
for very small values of $\rho$. In all cases, the electromagnetic
interaction tends to shift the absorption peaks from the
Maxwell-Garnett's prediction towards the lower frequencies. At
$\rho=0.32$, the effect is already quite pronounced.

\subsection{Comparison of inclusions of various size}
\label{subsec:eps_a}

\begin{figure}
\psfig{file=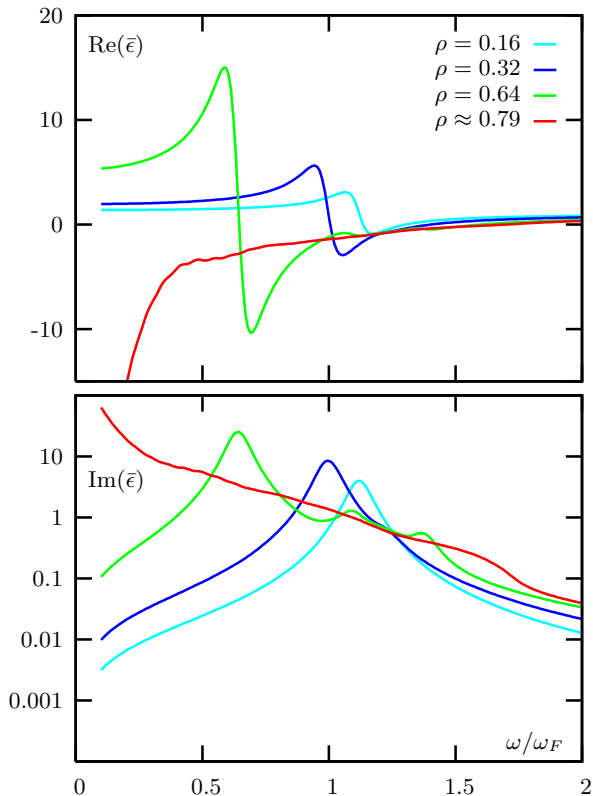,width=8.2cm,bbllx=60bp,bblly=410bp,bburx=300bp,bbury=725bp,clip=}
\caption{\label{fig:10} (color online) Effective permittivity for circular
  inclusions of different volume densities. The $\rho\approx 0.79$
  case corresponds to the percolation threshold (touching circles).}
\end{figure}

\begin{figure}
\psfig{file=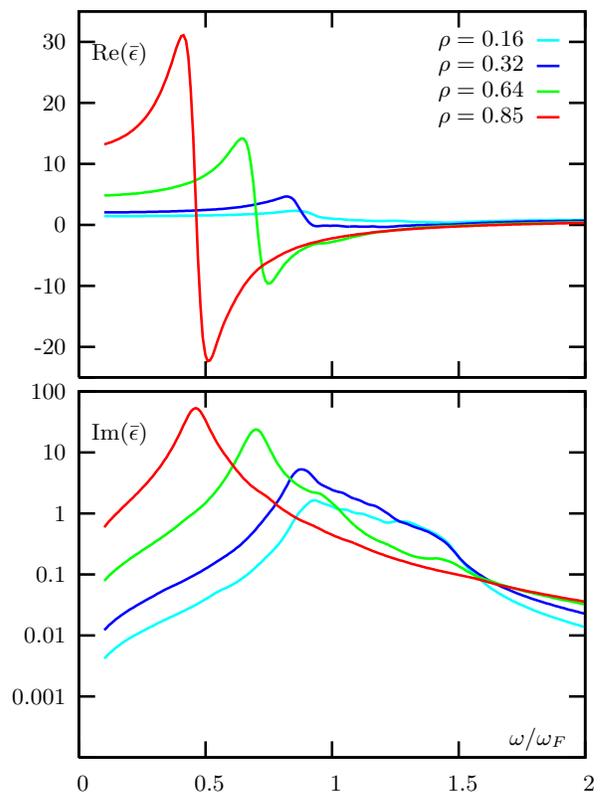,width=8.2cm,bbllx=60bp,bblly=410bp,bburx=300bp,bbury=725bp,clip=}
\caption{\label{fig:11} (color online) Same as in Fig.~\ref{fig:10}
  but for square inclusions.}
\end{figure}

We finally compare the effective permittivity for circular and square
inclusions of different sizes. The results are displayed in
Figs.~\ref{fig:10},\ref{fig:11}. In the case of circular inclusions,
there exists a pronounced spectral peak which shifts towards lower
frequencies when $\rho$ is increased. However, once the inclusions
touch (this happens at $\rho=\pi/4\approx 0.79$, the single resonance
is destroyed and a broad absorption band develops. The lower-frequency
behavior of $\bar{\epsilon}$ is in this case metallic, since the
percolating sample is characterized by a nonzero static conductivity.
This result can not be obtained within the Maxwell-Garnet
approximation, or the Bruggemann approximation, even at a qualitative
level.

The square inclusions do not touch for $\rho<1$. Correspondingly, the
low-frequency behavior of $\bar{\epsilon}$ is not metallic even
for large filling fractions, e.g., for $\rho=0.85$. Interestingly, at
relatively small values of $\rho$, the absorption spectrum forms a
band with one main resonance and many minor resonances which are
shifted towards the shorter waves. However, as $\rho$ increases, the
minor resonances become less pronounced. At $\rho=0.85$, the spectrum
is dominated by a single Lorentzian-type resonance. In the case of
circular inclusions, the picture is somewhat different. A single
Lorentzian resonance exists at small values of $\rho$ and additional
minor resonances develop as $\rho$ increases. These additional
resonances are clearly visible in the $\rho=0.64$ curve shown in the
left column of Fig.~\ref{fig:10}.

\section{Discussion}
\label{sec:disc}

A few points that deserve additional discussion are addressed in this
section, in no particular order.

\subsection{Conditions of applicability}

The EMT derived in this paper describes a composite medium accurately
if $qh,kh\ll 1$. There are no additional conditions. In particular,
there is no requirement that the permittivity (or conductivity) of any
constituent of the composite be bounded. However, if a metallic
inclusion has very small losses (very high conductivity), then the
effective permittivity computed according to the formulas of this
paper can have one or more sharp spectral peaks. These peaks are
caused by electromagnetic resonances in the inclusions (which we have
not disregarded by any means) and can be seen in
Figs.~\ref{fig:5}-\ref{fig:11}. In the spectral regions where these
resonances take place, it is possible that $q \gg k$. This effect is
known as the resonance wavelength shortening. Conceivably, the Bloch
wave number $q$ can become so large due to this effect that the
condition $qh \ll 1$ would break. In this case, our theory is
inapplicable.

The above consideration can be construed as a justification for
development of extended EMTs, e.g., by taking a limit in which the
conductivity of metal inclusions goes to infinity
first~\cite{nicorovici_95_1,nicorovici_95_2,poulton_04_1}, or by using
other trajectories in the parameter space~\cite{felbacq_05_1}.
However, two important caveats exist.  First, in many known
applications, EMPs of the order of unity are required, e.g.,
$\bar{\epsilon}\approx \bar{\mu} \approx -1$ is required for operation
of a superlens. In this case, of course, $q\approx k$, there is no
resonant wavelength shortening, and our theory applies. The second
caveat is that, even if metal inclusions have very high conductivity,
the imaginary part of the obtained effective permittivity is not small
close to a resonance.  This can be clearly seen in
Figs.~\ref{fig:5}-\ref{fig:11}. Therefore, there is not much hope to
obtain a resonant effect without having, simultaneously, strong
absorption in the medium. This observation is in agreement with
Stockman~\cite{stockman_07_1}, although we do not pursue here a
rigorous mathematical consideration of this point.

Finally, in the case when $qh$ is not actually small compared to unity
and our theory does not apply, it appears from considering the {\em
  exact} reflection coefficients (\ref{rs_1}),(\ref{rp_1}) that any
EMPs that can be introduced in any theory would depend on the angle of
incidence. More generally, the EMPs would depend on the type of
illumination. We conclude that the medium is simply not
electromagnetically homogeneous in this case.

\subsection{The case of small losses}

Another problem associated with high conductivity of metallic
inclusions is numerical stability and convergence.

The simulations of Sec.~\ref{sec:num} have been performed for a
relatively large loss parameter, $\gamma/\omega_F = 0.1$. If this
number is substantially reduced, the convergence with the truncation
order of the continued fraction, $j_{\rm max}$, is expected to become
slower. A general rule of thumb is that the truncation order should
not be less than the number of clearly discernible peaks in the
function ${\rm Im}\bar{\epsilon}(\omega)$ (the absorption spectrum).
This is because the CFE truncated at the order $j_{\rm max}$ captures
correctly the first $j_{\rm max}$ moments of the above function. At
sufficiently large values of $j$, the three-point recursion
(\ref{three-point}) becomes numerically unstable, as is illustrated in
Fig.~\ref{fig:6}. If the required value of $j_{\rm max}$ is larger
than the value of $j$ at which the onset of numerical instability
occurs, then the CFE will not yield an accurate numerical result.

The situation outlined above is common for all iterative methods. For
example, the convergence of the conjugate-gradient method becomes
extremely slow for small ratios of $\gamma/\omega_F$; at some point,
the recurrence relations used in the conjugate-gradient iterations
also become numerically unstable. One can hope to improve stability by
noting that the $n$-th order tail of the CFE (\ref{CFE}), that is, the
expression 
\begin{eqnarray*}
\frac{\kappa_{n+1}}{{\mathcal Z} - {\displaystyle \frac{\kappa_{n+2}}{{\mathcal Z} -
\ldots }}}
\end{eqnarray*}
\noindent
is also an expansion of a certain resolvent, and the instability
occurs because the parameter $\varepsilon$ (defined in the proof of
Theorem 1, Appendix~\ref{app:proof}) becomes numerically small. This
can be fixed by ``shifting'' the operator $A$ as described in
Sec.~\ref{subsec:MF}. In this way, a nested set of CFEs can be
obtained, where each CFE is numerically stable, as well as the whole
expression.

\subsection{Consideration of chirality and polarization conversion}

Although the general formalism of this paper allows one to take chiral
media into consideration, all derivations which were brought to a
logical conclusion have been carried out for the non-chiral case.
This has provided a mathematical simplification, yet left untouched a
wealth of interesting physical phenomena which are associated with
chirality. This shortcoming will be addressed by us in the future.

Even if the medium is non-chiral, it can exhibit the effect of
polarization conversion~\cite{elston_91_1}, which has been recently
predicted and experimentally observed in deeply-subwavelength
nanostructures in Ref.~\cite{feng_11_2}. In Sec.~\ref{subsec:refl}, we
have made an assumption that the plane of incidence coincides with one
of the crystallographic planes of the medium. In this case, the s- and
p-polarized waves are independent and polarization conversion does not
occur. However, the homogenization result obtained in this paper is
more general and, in particular, it is applicable to any direction of
incidence. If the plane of incidence does not coincide with any
crystallographic plane, the geometry of the problem becomes similar to
that considered in Ref.~\cite{feng_11_2} and polarization conversion
can occur. In other words, the reflected and transmitted (in the case
of a finite slab) waves due to a purely s- or p-polarized incident
wave can have both s- and p-polarized components and, at least
theoretically, it is possible to design a medium with the conversion
coefficient close to unity.

\subsection{3D vs 2D simulations}

So far, we have performed simulations only for 2D media. One can argue
that in the 3D case the size of the algebraic problem would become so
large as to render the method unusable. Of course, three-dimensional
electromagnetic problems are always challenging. However, there is
reason for optimism. Namely, the formula for the effective
permittivity (\ref{eps_eff}) uses the three-dimensional
Maxwell-Garnett approximation as the point of departure. In other
words, a nonzero value of $\Sigma$ provides a correction to the
three-dimensional Maxwell-Garnett formula. This happens to be true
even for two-dimensional media. However, the three-dimensional
Maxwell-Garnet formula is inaccurate in the 2D case even for very thin
cylinders, as is clearly illustrated in Figs.~\ref{fig:8},\ref{fig:9}.
In the numerical simulations of Sec.~\ref{sec:num} (for circular
inclusions), a lot of effort was spent to compute accurately the
self-energy $\Sigma$ whose effect was, essentially, to transform the
Maxwell-Garnett from a 3D to a 2D form.

In the case of small three-dimensional inclusions, one can expect a
much faster convergence with $L$. For example, if the inclusions are
small spheres, an accurate result is obtained by starting with
$\Sigma=0$. As the spheres increase in size, the Maxwell-Garnett
approximation becomes less accurate and a nonzero value of $\Sigma$
must be used. However, as we have seen in the numerical simulations,
the required values of $L$ are, in fact, smaller for larger sizes of
the inclusions.

Mathematically, the above considerations are related to an interesting
fact which was mentioned in Sec.~\ref{sec:CF}. Namely, the matrix
element $\langle a_\alpha \vert Q \vert a_\beta \rangle$ is
identically zero for three-dimensional cells with cubic symmetry.
Consequently, the mean-field approximation and the continued-fraction
expansion must be derived for the ``shifted'' equation
(\ref{F_shifted}). As a result, the mean-field formula
(\ref{Sigma_MF_2}) contains an overall factor of $(\rho\chi)^2$ while
in the 2D simulations of Sec.~\ref{sec:num}, this factor was equal to
$\rho\chi$.

\section{Summary}
\label{sec:summ}

We can draw the following conclusions:

\begin{enumerate}
    
\item A medium constructed from nonmagnetic components is also
  nonmagnetic in the limit $h\rightarrow 0$. This result is in line
  with arguments put forth in~\cite{bohren_86_1}, the simulations
  in~\cite{menzel_10_1} and the more formal mathematical theory
  of~\cite{wellander_03_1}.
  
\item The model of point-like polarizable particles is ill-suited for
  homogenization of three-dimensional periodic composites due to
  inherent divergences. The point-dipole approximation can be still a
  useful theoretical tool for studying systems in lower dimensions.
    
\item In agreement with the previous conclusion, we have found
  numerically that the EMPs are sensitive to the shape of inclusions
  even if the volume fraction is small. Thus, circular and square
  inclusions in Figs.~\ref{fig:7},\ref{fig:8} have very different
  spectra of EMPs, even though the volume fraction of the inclusions
  is $\rho=0.16$. When the volume fraction becomes larger, the
  differences between the circular and the square shapes are dramatic.
  Thus, it is shown in Figs.~\ref{fig:10},\ref{fig:11} that the
  percolation phenomenon occurs for the circular inclusions at the
  volume fraction $\rho=\pi/4\approx 0.79$, when the inclusions touch.
  The composite in this case is conducting. The composite consisting
  of square inclusions of the volume fill fraction (which do not
  touch) is still a dielectric.
  
\item We believe that the goal of homogenization theory is to describe
  a given physical composite. Therefore, rather than studying
  different limits, which correspond to different trajectories in the
  parameter space, it is important to delineate regions of the
  parameter space and to determine, to which one of these regions the
  particular composite belongs. Along similar lines, we note that a
  satisfactory theory of homogenization requires error estimates. That
  is, it is critical to understand how the error in the homogenization
  limit depends upon contrast. We plan to investigate this question in
  future work.

\end{enumerate}

\section*{Acknowledgments}

The authors are grateful to Profs.~Shari Moskow and Igor Tsukerman for
valuable discussions. This work was supported in part by the NSF grant
DMR-1120923.

\bibliographystyle{apsrev} 
\bibliography{abbrev,master,book}

\begin{thebibliography}{55}
\expandafter\ifx\csname natexlab\endcsname\relax\def\natexlab#1{#1}\fi
\expandafter\ifx\csname bibnamefont\endcsname\relax
  \def\bibnamefont#1{#1}\fi
\expandafter\ifx\csname bibfnamefont\endcsname\relax
  \def\bibfnamefont#1{#1}\fi
\expandafter\ifx\csname citenamefont\endcsname\relax
  \def\citenamefont#1{#1}\fi
\expandafter\ifx\csname url\endcsname\relax
  \def\url#1{\texttt{#1}}\fi
\expandafter\ifx\csname urlprefix\endcsname\relax\def\urlprefix{URL }\fi
\providecommand{\bibinfo}[2]{#2}
\providecommand{\eprint}[2][]{\url{#2}}

\bibitem[{\citenamefont{Simovski}(2009)}]{simovski_09_1}
\bibinfo{author}{\bibfnamefont{C.~R.} \bibnamefont{Simovski}},
  \bibinfo{journal}{Opt. Spectrosc.} \textbf{\bibinfo{volume}{107}},
  \bibinfo{pages}{766} (\bibinfo{year}{2009}).

\bibitem[{\citenamefont{Simovski}(2011)}]{simovski_11_1}
\bibinfo{author}{\bibfnamefont{C.~R.} \bibnamefont{Simovski}},
  \bibinfo{journal}{J. Opt.} \textbf{\bibinfo{volume}{13}},
  \bibinfo{pages}{103001} (\bibinfo{year}{2011}).

\bibitem[{\citenamefont{Bensoussan et~al.}(1978)\citenamefont{Bensoussan,
  Lions, and Papanicolaou}}]{bensoussan_78_1}
\bibinfo{author}{\bibfnamefont{A.}~\bibnamefont{Bensoussan}},
  \bibinfo{author}{\bibfnamefont{J.~L.} \bibnamefont{Lions}}, \bibnamefont{and}
  \bibinfo{author}{\bibfnamefont{G.}~\bibnamefont{Papanicolaou}},
  \emph{\bibinfo{title}{{Asymptotic Analysis for Periodic Structures}}}
  (\bibinfo{publisher}{N. Holland}, \bibinfo{year}{1978}).

\bibitem[{\citenamefont{Oleinik et~al.}(1992)\citenamefont{Oleinik, Shamaev,
  and Yosifian}}]{oleinik_book_92}
\bibinfo{author}{\bibfnamefont{O.~A.} \bibnamefont{Oleinik}},
  \bibinfo{author}{\bibfnamefont{A.~S.} \bibnamefont{Shamaev}},
  \bibnamefont{and} \bibinfo{author}{\bibfnamefont{G.~A.}
  \bibnamefont{Yosifian}}, \emph{\bibinfo{title}{{Mathematical Problems in
  Elasiticity and Homogenization}}} (\bibinfo{publisher}{Elsevier},
  \bibinfo{year}{1992}).

\bibitem[{\citenamefont{Milton}(2002)}]{milton_book_02}
\bibinfo{author}{\bibfnamefont{G.~W.} \bibnamefont{Milton}},
  \emph{\bibinfo{title}{{The Theory of Composites}}}
  (\bibinfo{publisher}{Cambridge University Press}, \bibinfo{year}{2002}).

\bibitem[{\citenamefont{Tartar}(2009)}]{tartar_book_09}
\bibinfo{author}{\bibfnamefont{L.}~\bibnamefont{Tartar}},
  \emph{\bibinfo{title}{{The General Theory of Homogenization}}}
  (\bibinfo{publisher}{Springer}, \bibinfo{year}{2009}).

\bibitem[{\citenamefont{Silveirinha}(2007)}]{silveirinha_07_1}
\bibinfo{author}{\bibfnamefont{M.~G.} \bibnamefont{Silveirinha}},
  \bibinfo{journal}{Phys. Rev. B} \textbf{\bibinfo{volume}{75}},
  \bibinfo{pages}{115104} (\bibinfo{year}{2007}).

\bibitem[{\citenamefont{Tsukerman}(2011)}]{tsukerman_11_1}
\bibinfo{author}{\bibfnamefont{I.}~\bibnamefont{Tsukerman}},
  \bibinfo{journal}{J. Opt. Soc. Am. B} \textbf{\bibinfo{volume}{28}},
  \bibinfo{pages}{577} (\bibinfo{year}{2011}).

\bibitem[{\citenamefont{Pors et~al.}(2011)\citenamefont{Pors, Tsukerman, and
  Bozhevolnyi}}]{pors_11_1}
\bibinfo{author}{\bibfnamefont{A.}~\bibnamefont{Pors}},
  \bibinfo{author}{\bibfnamefont{I.}~\bibnamefont{Tsukerman}},
  \bibnamefont{and} \bibinfo{author}{\bibfnamefont{S.~I.}
  \bibnamefont{Bozhevolnyi}}, \bibinfo{journal}{Phys. Rev. E}
  \textbf{\bibinfo{volume}{84}}, \bibinfo{pages}{016609}
  (\bibinfo{year}{2011}).

\bibitem[{\citenamefont{Silveirinha}(2011)}]{silveirinha_11_1}
\bibinfo{author}{\bibfnamefont{M.~G.} \bibnamefont{Silveirinha}},
  \bibinfo{journal}{Phys. Rev. B} \textbf{\bibinfo{volume}{83}},
  \bibinfo{pages}{165104} (\bibinfo{year}{2011}).

\bibitem[{\citenamefont{Feng et~al.}(2010)\citenamefont{Feng, Liu, Lomakin, and
  Fainman}}]{feng_10_2}
\bibinfo{author}{\bibfnamefont{L.}~\bibnamefont{Feng}},
  \bibinfo{author}{\bibfnamefont{Z.}~\bibnamefont{Liu}},
  \bibinfo{author}{\bibfnamefont{V.}~\bibnamefont{Lomakin}}, \bibnamefont{and}
  \bibinfo{author}{\bibfnamefont{Y.}~\bibnamefont{Fainman}},
  \bibinfo{journal}{Appl. Phys. Lett.} \textbf{\bibinfo{volume}{96}},
  \bibinfo{pages}{041112} (\bibinfo{year}{2010}).

\bibitem[{\citenamefont{Feng et~al.}(2011{\natexlab{a}})\citenamefont{Feng,
  Liu, and Fainman}}]{feng_11_1}
\bibinfo{author}{\bibfnamefont{L.}~\bibnamefont{Feng}},
  \bibinfo{author}{\bibfnamefont{Z.}~\bibnamefont{Liu}}, \bibnamefont{and}
  \bibinfo{author}{\bibfnamefont{Y.}~\bibnamefont{Fainman}},
  \bibinfo{journal}{Appl. Opt.} \textbf{\bibinfo{volume}{50}},
  \bibinfo{pages}{G1} (\bibinfo{year}{2011}{\natexlab{a}}).

\bibitem[{\citenamefont{Feng et~al.}(2011{\natexlab{b}})\citenamefont{Feng,
  Mizrahi, Zamek, Liu, Lomakin, and Fainman}}]{feng_11_2}
\bibinfo{author}{\bibfnamefont{L.}~\bibnamefont{Feng}},
  \bibinfo{author}{\bibfnamefont{A.}~\bibnamefont{Mizrahi}},
  \bibinfo{author}{\bibfnamefont{S.}~\bibnamefont{Zamek}},
  \bibinfo{author}{\bibfnamefont{Z.}~\bibnamefont{Liu}},
  \bibinfo{author}{\bibfnamefont{V.}~\bibnamefont{Lomakin}}, \bibnamefont{and}
  \bibinfo{author}{\bibfnamefont{Y.}~\bibnamefont{Fainman}},
  \bibinfo{journal}{ACS NANO} \textbf{\bibinfo{volume}{5}},
  \bibinfo{pages}{5100} (\bibinfo{year}{2011}{\natexlab{b}}).

\bibitem[{\citenamefont{Krokhin et~al.}(2002)\citenamefont{Krokhin, Halevi, and
  Arriaga}}]{krokhin_02_1}
\bibinfo{author}{\bibfnamefont{A.~A.} \bibnamefont{Krokhin}},
  \bibinfo{author}{\bibfnamefont{P.}~\bibnamefont{Halevi}}, \bibnamefont{and}
  \bibinfo{author}{\bibfnamefont{J.}~\bibnamefont{Arriaga}},
  \bibinfo{journal}{Phys. Rev. B} \textbf{\bibinfo{volume}{65}},
  \bibinfo{pages}{115208} (\bibinfo{year}{2002}).

\bibitem[{\citenamefont{Krokhin and Reyes}(2004)}]{krokhin_04_1}
\bibinfo{author}{\bibfnamefont{A.~A.} \bibnamefont{Krokhin}} \bibnamefont{and}
  \bibinfo{author}{\bibfnamefont{E.}~\bibnamefont{Reyes}},
  \bibinfo{journal}{Phys. Rev. Lett.} \textbf{\bibinfo{volume}{93}},
  \bibinfo{pages}{023904} (\bibinfo{year}{2004}).

\bibitem[{\citenamefont{Cherednichenko and
  Guenneau}(2007)}]{cherednichenko_07_1}
\bibinfo{author}{\bibfnamefont{K.~D.} \bibnamefont{Cherednichenko}}
  \bibnamefont{and} \bibinfo{author}{\bibfnamefont{S.}~\bibnamefont{Guenneau}},
  \bibinfo{journal}{Waves in Random Media} \textbf{\bibinfo{volume}{17}},
  \bibinfo{pages}{627} (\bibinfo{year}{2007}).

\bibitem[{\citenamefont{Guenneau et~al.}(2007)\citenamefont{Guenneau, Zolla,
  and Nicolet}}]{guenneau_07_1}
\bibinfo{author}{\bibfnamefont{S.}~\bibnamefont{Guenneau}},
  \bibinfo{author}{\bibfnamefont{F.}~\bibnamefont{Zolla}}, \bibnamefont{and}
  \bibinfo{author}{\bibfnamefont{A.}~\bibnamefont{Nicolet}},
  \bibinfo{journal}{Waves in Random Media} \textbf{\bibinfo{volume}{17}},
  \bibinfo{pages}{653} (\bibinfo{year}{2007}).

\bibitem[{\citenamefont{Guenneau and Zolla}(2011)}]{guenneau_11_1}
\bibinfo{author}{\bibfnamefont{S.}~\bibnamefont{Guenneau}} \bibnamefont{and}
  \bibinfo{author}{\bibfnamefont{F.}~\bibnamefont{Zolla}},
  \bibinfo{journal}{Prog. Electromagnetic Res.} \textbf{\bibinfo{volume}{27}},
  \bibinfo{pages}{91} (\bibinfo{year}{2011}).

\bibitem[{\citenamefont{Craster et~al.}(2011)\citenamefont{Craster, Kaplunov,
  E., and Guenneau}}]{craster_11_1}
\bibinfo{author}{\bibfnamefont{R.~V.} \bibnamefont{Craster}},
  \bibinfo{author}{\bibfnamefont{J.}~\bibnamefont{Kaplunov}},
  \bibinfo{author}{\bibfnamefont{N.}~\bibnamefont{E.}}, \bibnamefont{and}
  \bibinfo{author}{\bibfnamefont{S.}~\bibnamefont{Guenneau}},
  \bibinfo{journal}{J. Opt. Soc. Am. A} \textbf{\bibinfo{volume}{28}},
  \bibinfo{pages}{1032} (\bibinfo{year}{2011}).

\bibitem[{\citenamefont{Bohren}(2009)}]{bohren_09_1}
\bibinfo{author}{\bibfnamefont{C.~F.} \bibnamefont{Bohren}},
  \bibinfo{journal}{J. Nanophotonics} \textbf{\bibinfo{volume}{3}},
  \bibinfo{pages}{039501} (\bibinfo{year}{2009}).

\bibitem[{\citenamefont{Niklasson et~al.}(1981)\citenamefont{Niklasson,
  Granqvist, and Hunderi}}]{niklasson_81_1}
\bibinfo{author}{\bibfnamefont{G.~A.} \bibnamefont{Niklasson}},
  \bibinfo{author}{\bibfnamefont{C.~G.} \bibnamefont{Granqvist}},
  \bibnamefont{and} \bibinfo{author}{\bibfnamefont{O.}~\bibnamefont{Hunderi}},
  \bibinfo{journal}{Appl. Opt.} \textbf{\bibinfo{volume}{20}},
  \bibinfo{pages}{26} (\bibinfo{year}{1981}).

\bibitem[{\citenamefont{Doyle}(1989)}]{doyle_89_1}
\bibinfo{author}{\bibfnamefont{W.~T.} \bibnamefont{Doyle}},
  \bibinfo{journal}{Phys. Rev. B} \textbf{\bibinfo{volume}{39}},
  \bibinfo{pages}{9852} (\bibinfo{year}{1989}).

\bibitem[{\citenamefont{Nicorovici
  et~al.}(1995{\natexlab{a}})\citenamefont{Nicorovici, McPhedran, and
  Botten}}]{nicorovici_95_1}
\bibinfo{author}{\bibfnamefont{N.~A.} \bibnamefont{Nicorovici}},
  \bibinfo{author}{\bibfnamefont{R.~C.} \bibnamefont{McPhedran}},
  \bibnamefont{and} \bibinfo{author}{\bibfnamefont{L.~C.}
  \bibnamefont{Botten}}, \bibinfo{journal}{Phys. Rev. Lett.}
  \textbf{\bibinfo{volume}{75}}, \bibinfo{pages}{1507}
  (\bibinfo{year}{1995}{\natexlab{a}}).

\bibitem[{\citenamefont{Nicorovici
  et~al.}(1995{\natexlab{b}})\citenamefont{Nicorovici, McPhedran, and
  Botten}}]{nicorovici_95_2}
\bibinfo{author}{\bibfnamefont{N.~A.} \bibnamefont{Nicorovici}},
  \bibinfo{author}{\bibfnamefont{R.~C.} \bibnamefont{McPhedran}},
  \bibnamefont{and} \bibinfo{author}{\bibfnamefont{L.~C.}
  \bibnamefont{Botten}}, \bibinfo{journal}{Phys. Rev. E}
  \textbf{\bibinfo{volume}{52}}, \bibinfo{pages}{1135}
  (\bibinfo{year}{1995}{\natexlab{b}}).

\bibitem[{\citenamefont{Wellander and Kristensson}(2003)}]{wellander_03_1}
\bibinfo{author}{\bibfnamefont{N.}~\bibnamefont{Wellander}} \bibnamefont{and}
  \bibinfo{author}{\bibfnamefont{G.}~\bibnamefont{Kristensson}},
  \bibinfo{journal}{SIAM J. Appl. Math.} \textbf{\bibinfo{volume}{64}},
  \bibinfo{pages}{170} (\bibinfo{year}{2003}).

\bibitem[{\citenamefont{Felbacq and Bouchitte}(2005)}]{felbacq_05_1}
\bibinfo{author}{\bibfnamefont{D.}~\bibnamefont{Felbacq}} \bibnamefont{and}
  \bibinfo{author}{\bibfnamefont{G.}~\bibnamefont{Bouchitte}},
  \bibinfo{journal}{New J. Phys.} \textbf{\bibinfo{volume}{7}},
  \bibinfo{pages}{159} (\bibinfo{year}{2005}).

\bibitem[{\citenamefont{Poulton et~al.}(2004)\citenamefont{Poulton, Guenneau,
  and Movchan}}]{poulton_04_1}
\bibinfo{author}{\bibfnamefont{C.}~\bibnamefont{Poulton}},
  \bibinfo{author}{\bibfnamefont{S.}~\bibnamefont{Guenneau}}, \bibnamefont{and}
  \bibinfo{author}{\bibfnamefont{A.~B.} \bibnamefont{Movchan}},
  \bibinfo{journal}{Phys. Rev. B} \textbf{\bibinfo{volume}{69}},
  \bibinfo{pages}{195112} (\bibinfo{year}{2004}).

\bibitem[{\citenamefont{Bohren}(1986)}]{bohren_86_1}
\bibinfo{author}{\bibfnamefont{C.~F.} \bibnamefont{Bohren}},
  \bibinfo{journal}{J. Atmospheric Sci.} \textbf{\bibinfo{volume}{43}},
  \bibinfo{pages}{468} (\bibinfo{year}{1986}).

\bibitem[{\citenamefont{Menzel et~al.}(2010{\natexlab{a}})\citenamefont{Menzel,
  Paul, Rockstuhl, Pertsch, Tretyakov, and Lederer}}]{menzel_10_1}
\bibinfo{author}{\bibfnamefont{C.}~\bibnamefont{Menzel}},
  \bibinfo{author}{\bibfnamefont{T.}~\bibnamefont{Paul}},
  \bibinfo{author}{\bibfnamefont{C.}~\bibnamefont{Rockstuhl}},
  \bibinfo{author}{\bibfnamefont{T.}~\bibnamefont{Pertsch}},
  \bibinfo{author}{\bibfnamefont{S.}~\bibnamefont{Tretyakov}},
  \bibnamefont{and} \bibinfo{author}{\bibfnamefont{F.}~\bibnamefont{Lederer}},
  \bibinfo{journal}{Phys. Rev. B} \textbf{\bibinfo{volume}{81}},
  \bibinfo{pages}{035320} (\bibinfo{year}{2010}{\natexlab{a}}).

\bibitem[{\citenamefont{Menzel et~al.}(2010{\natexlab{b}})\citenamefont{Menzel,
  Rockstuhl, Iliew, Lederer, Andryieuski, Malureanu, and
  Lavrinenko}}]{menzel_10_2}
\bibinfo{author}{\bibfnamefont{C.}~\bibnamefont{Menzel}},
  \bibinfo{author}{\bibfnamefont{C.}~\bibnamefont{Rockstuhl}},
  \bibinfo{author}{\bibfnamefont{R.}~\bibnamefont{Iliew}},
  \bibinfo{author}{\bibfnamefont{F.}~\bibnamefont{Lederer}},
  \bibinfo{author}{\bibfnamefont{A.}~\bibnamefont{Andryieuski}},
  \bibinfo{author}{\bibfnamefont{R.}~\bibnamefont{Malureanu}},
  \bibnamefont{and} \bibinfo{author}{\bibfnamefont{A.~V.}
  \bibnamefont{Lavrinenko}}, \bibinfo{journal}{Phys. Rev. B}
  \textbf{\bibinfo{volume}{81}}, \bibinfo{pages}{195123}
  (\bibinfo{year}{2010}{\natexlab{b}}).

\bibitem[{\citenamefont{Simovski and Tretyakov}(2010)}]{simovski_10_1}
\bibinfo{author}{\bibfnamefont{C.~R.} \bibnamefont{Simovski}} \bibnamefont{and}
  \bibinfo{author}{\bibfnamefont{S.~A.} \bibnamefont{Tretyakov}},
  \bibinfo{journal}{Photonics and Nanostructures} \textbf{\bibinfo{volume}{8}},
  \bibinfo{pages}{254} (\bibinfo{year}{2010}).

\bibitem[{\citenamefont{Andryieuski et~al.}(2010)\citenamefont{Andryieuski,
  Menzel, Rockstuhl, Malureanu, Lederer, and Lavrinenko}}]{andryieuski_10_1}
\bibinfo{author}{\bibfnamefont{A.}~\bibnamefont{Andryieuski}},
  \bibinfo{author}{\bibfnamefont{C.}~\bibnamefont{Menzel}},
  \bibinfo{author}{\bibfnamefont{C.}~\bibnamefont{Rockstuhl}},
  \bibinfo{author}{\bibfnamefont{R.}~\bibnamefont{Malureanu}},
  \bibinfo{author}{\bibfnamefont{F.}~\bibnamefont{Lederer}}, \bibnamefont{and}
  \bibinfo{author}{\bibfnamefont{A.}~\bibnamefont{Lavrinenko}},
  \bibinfo{journal}{Phys. Rev. B} \textbf{\bibinfo{volume}{82}},
  \bibinfo{pages}{235107} (\bibinfo{year}{2010}).

\bibitem[{\citenamefont{Paul et~al.}(2011)\citenamefont{Paul, Menzel, Smigaj,
  Rockstuhl, Lalanne, and Lederer}}]{paul_11_1}
\bibinfo{author}{\bibfnamefont{T.}~\bibnamefont{Paul}},
  \bibinfo{author}{\bibfnamefont{C.}~\bibnamefont{Menzel}},
  \bibinfo{author}{\bibfnamefont{W.}~\bibnamefont{Smigaj}},
  \bibinfo{author}{\bibfnamefont{C.}~\bibnamefont{Rockstuhl}},
  \bibinfo{author}{\bibfnamefont{P.}~\bibnamefont{Lalanne}}, \bibnamefont{and}
  \bibinfo{author}{\bibfnamefont{F.}~\bibnamefont{Lederer}},
  \bibinfo{journal}{Phys. Rev. B} \textbf{\bibinfo{volume}{84}},
  \bibinfo{pages}{115142} (\bibinfo{year}{2011}).

\bibitem[{\citenamefont{Sipe and Van~Kranendonk}(1974)}]{sipe_74_1}
\bibinfo{author}{\bibfnamefont{J.~E.} \bibnamefont{Sipe}} \bibnamefont{and}
  \bibinfo{author}{\bibfnamefont{J.}~\bibnamefont{Van~Kranendonk}},
  \bibinfo{journal}{Phys. Rev. A} \textbf{\bibinfo{volume}{9}},
  \bibinfo{pages}{1806} (\bibinfo{year}{1974}).

\bibitem[{\citenamefont{Draine and Goodman}(1993)}]{draine_93_1}
\bibinfo{author}{\bibfnamefont{B.~T.} \bibnamefont{Draine}} \bibnamefont{and}
  \bibinfo{author}{\bibfnamefont{J.}~\bibnamefont{Goodman}},
  \bibinfo{journal}{Astrophys. J.} \textbf{\bibinfo{volume}{405}},
  \bibinfo{pages}{685} (\bibinfo{year}{1993}).

\bibitem[{\citenamefont{Belov and Simovski}(2005)}]{belov_05_1}
\bibinfo{author}{\bibfnamefont{P.~A.} \bibnamefont{Belov}} \bibnamefont{and}
  \bibinfo{author}{\bibfnamefont{C.~R.} \bibnamefont{Simovski}},
  \bibinfo{journal}{Phys. Rev. E} \textbf{\bibinfo{volume}{72}},
  \bibinfo{pages}{026615} (\bibinfo{year}{2005}).

\bibitem[{\citenamefont{Bergman}(1978)}]{bergman_78_1}
\bibinfo{author}{\bibfnamefont{D.~J.} \bibnamefont{Bergman}},
  \bibinfo{journal}{Phys. Rep.} \textbf{\bibinfo{volume}{43}},
  \bibinfo{pages}{377} (\bibinfo{year}{1978}).

\bibitem[{\citenamefont{Bergman}(1979{\natexlab{a}})}]{bergman_79_1}
\bibinfo{author}{\bibfnamefont{D.~J.} \bibnamefont{Bergman}},
  \bibinfo{journal}{J. Phys.: Condens. Matter} \textbf{\bibinfo{volume}{12}},
  \bibinfo{pages}{4947} (\bibinfo{year}{1979}{\natexlab{a}}).

\bibitem[{\citenamefont{Bergman}(1979{\natexlab{b}})}]{bergman_79_2}
\bibinfo{author}{\bibfnamefont{D.~J.} \bibnamefont{Bergman}},
  \bibinfo{journal}{Phys. Rev. B} \textbf{\bibinfo{volume}{19}},
  \bibinfo{pages}{2359} (\bibinfo{year}{1979}{\natexlab{b}}).

\bibitem[{\citenamefont{Abajo}(2007)}]{abajo_07_1}
\bibinfo{author}{\bibfnamefont{F.~J.~G.} \bibnamefont{Abajo}},
  \bibinfo{journal}{Rev. Mod. Phys.} \textbf{\bibinfo{volume}{79}},
  \bibinfo{pages}{1267} (\bibinfo{year}{2007}).

\bibitem[{\citenamefont{Landau and Lifshitz}(1984)}]{landau_ess_84}
\bibinfo{author}{\bibfnamefont{L.~D.} \bibnamefont{Landau}} \bibnamefont{and}
  \bibinfo{author}{\bibfnamefont{L.~P.} \bibnamefont{Lifshitz}},
  \emph{\bibinfo{title}{{Electrodynamics of Continuous Media}}}
  (\bibinfo{publisher}{Pergamon Press}, \bibinfo{address}{Oxford},
  \bibinfo{year}{1984}).

\bibitem[{\citenamefont{Itin}(2010)}]{itin_10_1}
\bibinfo{author}{\bibfnamefont{Y.}~\bibnamefont{Itin}}, \bibinfo{journal}{Phys.
  Lett. A} \textbf{\bibinfo{volume}{374}}, \bibinfo{pages}{1113}
  (\bibinfo{year}{2010}).

\bibitem[{\citenamefont{Markel and Schotland}(2010)}]{markel_10_1}
\bibinfo{author}{\bibfnamefont{V.~A.} \bibnamefont{Markel}} \bibnamefont{and}
  \bibinfo{author}{\bibfnamefont{J.~C.} \bibnamefont{Schotland}},
  \bibinfo{journal}{J. Opt.} \textbf{\bibinfo{volume}{12}},
  \bibinfo{pages}{015104} (\bibinfo{year}{2010}).

\bibitem[{\citenamefont{Maradudin and Mills}(1975)}]{maradudin_75_1}
\bibinfo{author}{\bibfnamefont{A.~A.} \bibnamefont{Maradudin}}
  \bibnamefont{and} \bibinfo{author}{\bibfnamefont{D.~L.} \bibnamefont{Mills}},
  \bibinfo{journal}{Phys. Rev. B} \textbf{\bibinfo{volume}{11}},
  \bibinfo{pages}{1392} (\bibinfo{year}{1975}).

\bibitem[{\citenamefont{Mahan and Obermair}(1969)}]{mahan_69_1}
\bibinfo{author}{\bibfnamefont{G.~D.} \bibnamefont{Mahan}} \bibnamefont{and}
  \bibinfo{author}{\bibfnamefont{G.}~\bibnamefont{Obermair}},
  \bibinfo{journal}{Phys. Rev.} \textbf{\bibinfo{volume}{183}},
  \bibinfo{pages}{834} (\bibinfo{year}{1969}).

\bibitem[{\citenamefont{Lamb et~al.}(1980)\citenamefont{Lamb, Wood, and
  Ashcroft}}]{lamb_80_1}
\bibinfo{author}{\bibfnamefont{W.}~\bibnamefont{Lamb}},
  \bibinfo{author}{\bibfnamefont{D.~M.} \bibnamefont{Wood}}, \bibnamefont{and}
  \bibinfo{author}{\bibfnamefont{N.~W.} \bibnamefont{Ashcroft}},
  \bibinfo{journal}{Phys. Rev. B} \textbf{\bibinfo{volume}{21}},
  \bibinfo{pages}{2248} (\bibinfo{year}{1980}).

\bibitem[{\citenamefont{Draine}(1988)}]{draine_88_1}
\bibinfo{author}{\bibfnamefont{B.~T.} \bibnamefont{Draine}},
  \bibinfo{journal}{Astrophys. J.} \textbf{\bibinfo{volume}{333}},
  \bibinfo{pages}{848} (\bibinfo{year}{1988}).

\bibitem[{\citenamefont{Markel}(1992)}]{markel_92_1}
\bibinfo{author}{\bibfnamefont{V.~A.} \bibnamefont{Markel}},
  \bibinfo{journal}{J. Mod. Opt.} \textbf{\bibinfo{volume}{39}},
  \bibinfo{pages}{853} (\bibinfo{year}{1992}).

\bibitem[{\citenamefont{Markel}(1995)}]{markel_95_1}
\bibinfo{author}{\bibfnamefont{V.~A.} \bibnamefont{Markel}},
  \bibinfo{journal}{J. Opt. Soc. Am. B} \textbf{\bibinfo{volume}{12}},
  \bibinfo{pages}{1783} (\bibinfo{year}{1995}).

\bibitem[{\citenamefont{Berry and Percival}(1986)}]{berry_86_1}
\bibinfo{author}{\bibfnamefont{M.~V.} \bibnamefont{Berry}} \bibnamefont{and}
  \bibinfo{author}{\bibfnamefont{I.~C.} \bibnamefont{Percival}},
  \bibinfo{journal}{Optica Acta} \textbf{\bibinfo{volume}{33}},
  \bibinfo{pages}{577} (\bibinfo{year}{1986}).

\bibitem[{\citenamefont{Haydock}(1980)}]{haydock_80_1}
\bibinfo{author}{\bibfnamefont{R.}~\bibnamefont{Haydock}},
  \emph{\bibinfo{title}{{Solid State Physics}}} (\bibinfo{publisher}{Academic
  Press}, \bibinfo{year}{1980}), vol.~\bibinfo{volume}{35}, chap.
  \bibinfo{chapter}{The recursive solution of the Schrodinger equation}, pp.
  \bibinfo{pages}{215--294}.

\bibitem[{\citenamefont{Markel et~al.}(2004)\citenamefont{Markel, Pustovit,
  Karpov, Obuschenko, Gerasimov, and Isaev}}]{markel_04_3}
\bibinfo{author}{\bibfnamefont{V.~A.} \bibnamefont{Markel}},
  \bibinfo{author}{\bibfnamefont{V.~N.} \bibnamefont{Pustovit}},
  \bibinfo{author}{\bibfnamefont{S.~V.} \bibnamefont{Karpov}},
  \bibinfo{author}{\bibfnamefont{A.~V.} \bibnamefont{Obuschenko}},
  \bibinfo{author}{\bibfnamefont{V.~S.} \bibnamefont{Gerasimov}},
  \bibnamefont{and} \bibinfo{author}{\bibfnamefont{I.~L.} \bibnamefont{Isaev}},
  \bibinfo{journal}{Phys. Rev. B} \textbf{\bibinfo{volume}{70}},
  \bibinfo{pages}{054202} (\bibinfo{year}{2004}).

\bibitem[{\citenamefont{Jones and Thron}(1980)}]{jones_book_80}
\bibinfo{author}{\bibfnamefont{W.~B.} \bibnamefont{Jones}} \bibnamefont{and}
  \bibinfo{author}{\bibfnamefont{W.~J.} \bibnamefont{Thron}},
  \emph{\bibinfo{title}{{Continued Fractions. Analytic Theory and
  Applications}}} (\bibinfo{publisher}{Addison-Wesley Pub.},
  \bibinfo{year}{1980}).

\bibitem[{\citenamefont{Stockman}(2007)}]{stockman_07_1}
\bibinfo{author}{\bibfnamefont{M.~I.} \bibnamefont{Stockman}},
  \bibinfo{journal}{Phys. Rev. Lett.} \textbf{\bibinfo{volume}{98}},
  \bibinfo{pages}{177404} (\bibinfo{year}{2007}).

\bibitem[{\citenamefont{Elston et~al.}(1991)\citenamefont{Elston, Bryan-Brown,
  and Sambles}}]{elston_91_1}
\bibinfo{author}{\bibfnamefont{S.~J.} \bibnamefont{Elston}},
  \bibinfo{author}{\bibfnamefont{G.~P.} \bibnamefont{Bryan-Brown}},
  \bibnamefont{and} \bibinfo{author}{\bibfnamefont{J.~R.}
  \bibnamefont{Sambles}}, \bibinfo{journal}{Phys. Rev. B}
  \textbf{\bibinfo{volume}{44}}, \bibinfo{pages}{6393} (\bibinfo{year}{1991}).

\end{thebibliography}

\appendix

\section{Mathematical properties of $M({\bf g})$ and some special
  cases}
\label{app:M}

From the definition (\ref{M_def}), it follows that 

\begin{equation}
\label{B0}
M(0) = 1 \ , \ \ M(-{\bf g}) = M^*({\bf g}) \ . 
\end{equation}

\noindent
For the case of inclusions whose center of symmetry coincides with the
center of the unit cell, we have $M(-{\bf g}) = M({\bf g})$ and,
therefore, $M({\bf g})$ is real. If the center of symmetry is
displaced by a vector ${\bf a}$, the function ${\bf M}({\bf g})$ is
transformed according to ${\bf M}({\bf g}) \rightarrow \exp(-i{\bf
  a}\cdot {\bf g}) {\bf M}({\bf g})$.

By applying the Poisson summation formula, we can derive the following
sum rules:

\begin{subequations}
\label{B1}
\begin{align}
\sum_{\bf g}M({\bf g}) & = \left\{ \begin{array}{ll}
1/\rho \ , & 0\in \Omega \ , \\
0      \ , & 0 \notin \Omega \ , 
\end{array} \right.  \\
\sum_{\bf g} M(-{\bf g})M({\bf g}) &= \frac{1}{\rho} \ , \\
\sum_{{\bf g}^\prime}M({\bf g}-{\bf g}^\prime) M({\bf g}^\prime) &=
\frac{1}{\rho} M({\bf g}) \ .
\end{align}
\end{subequations}

\noindent
These equations hold for inclusions of arbitrary shape.

Now define a complimentary function $N({\bf g})$ by 

\begin{equation}
\label{B2}
N({\bf g}) = \frac{1}{h^3 - V} \int_{C\backslash\Omega} \exp(-i{\bf
  g}\cdot {\bf R}) d^3R \ .
\end{equation}

\noindent
Here $C$ denotes the unit cell and $C\backslash\Omega$ is the region
complimentary to the inclusion. It can be seen that $N({\bf g})$ has
all the properties of $M({\bf g})$ with the substitution $\rho
\rightarrow 1-\rho$. Additionally, the functions $M({\bf g})$ and
$N({\bf g})$ are related by

\begin{equation}
\label{B3}
\rho M({\bf g}) + (1 - \rho) N({\bf g}) = \delta_{{\bf g}0} \ .
\end{equation}

\noindent
From this, we obtain the low and high-density limits:

\begin{equation}
\label{B4}
\lim_{\rho \rightarrow 0}N({\bf g}) = \lim_{\rho \rightarrow 1} M({\bf
  g}) = \delta_{{\bf g}0} \ . 
\end{equation}

\noindent
Of course, the high-density limit is unreachable for most regular
shapes (with the exception of cubes). For example, in the case of
spheres, the maximum allowed value of $\rho$ is $\pi/6$.

Some special cases of $M({\bf g})$ are given below. For an inclusion
in the shape of either a 3D sphere or a 2D circle of radius $a\leq
h/2$,

\begin{subequations}
\label{B5}
\begin{align}
& M_{\rm 3D}({\bf g}) = \frac{3[\sin(ga) - ga\cos(ga)]}{(ga)^3} \ , \\
& M_{\rm 2D}({\bf g}) = \frac{2J_1(ga)}{ga} \ ,
\end{align}
\end{subequations}

\noindent
where $J_1(x)$ is the cylindrical Bessel function of the first kind.
For a parallelepiped or rectangle centered at the origin with all
faces parallel to the crystallographic planes and sides of length $2a_x$,
$2a_y$ and $2a_z$,

\begin{subequations}
\label{B6}
\begin{align}
& M_{\rm 3D}({\bf g}) = \frac{\sin(g_x a_x)}{g_x a_x}  \frac{\sin(g_y a_y)}{g_y
  a_y}  \frac{\sin(g_z a_z)}{g_z a_z} \ , \\
& M_{\rm 2D}({\bf g}) = \frac{\sin(g_x a_x)}{g_x a_x}  \frac{\sin(g_y a_y)}{g_y
  a_y} \ .
\end{align}
\end{subequations}

\section{Details of some calculations pertaining to the case of p-polarization}
\label{app:p-pol}

To simplify notations, we will denote (in this Appendix only)
\begin{equation}
\label{C00}
1+\Sigma \equiv S \ , \ \  \rho\chi \equiv \kappa \ , 
\end{equation}
\noindent
so that
\begin{equation}
\label{C0}
\eta_\alpha = \epsilon_b\frac{1 + 2\kappa S_{\alpha\alpha}}{1 - \kappa
  S_{\alpha\alpha}} \ .
\end{equation}
We start by deriving Eq.~(\ref{Fx_over_Fz}). To this end, we write the
wave vector of the p-polarized wave as ${\bf q} = q_x\hat{\bf x} +
q_z\hat{\bf z}$ (note that $q_y$=0) and seek a nontrivial solution to
Eq.~(\ref{F_0}). Multiplying (\ref{F_0}) by the non-zero factor $q^2 -
k_b^2$ and using (\ref{K_def}), we obtain the following equation:
\begin{equation}
\label{C1}
(q^2 - k_b^2){\bf F}_0 - \kappa (2 k_b^2 + q^2)S{\bf F}_0
+ 3\kappa {\bf q} \left( {\bf q} \cdot S{\bf F}_0\right) = 0 \ .
\end{equation}
We now account for the fact that the tensors $\Sigma$ and $S=1+\Sigma$
are diagonal in the laboratory frame and write
\begin{equation}
\label{C2a}
\left( S{\bf F}_0 \right)_\alpha = S_{\alpha\alpha} F_{0\alpha} \ , \
\ \alpha=x,y,z 
\end{equation}
\noindent
and
\begin{equation}
\label{C2b}
{\bf q} \cdot S{\bf F}_0 = q_x S_{xx}F_{0x} + q_z S_{zz}F_{0z} \ .
\end{equation}
\noindent
Using this result, and projecting Eq.~(\ref{C1}) onto the $y$-axis, we
immediately obtain $F_{0y}=0$. The two remaining Cartesian components
of ${\bf F}_0$ satisfy a system of two linear equations, which are
obtainable by projecting (\ref{C1}) onto the $x$- and $z$-axes. These
two equation are not linearly independent, provided that the
dispersion relation (\ref{disp_2D_p}) holds [otherwise, the only
solution to (\ref{C1}) is trivial]. It is, therefore, sufficient to
consider one of these equations, say, by projecting (\ref{C1}) onto
the $x$-axis. The resultant equation is
\begin{equation}
\label{C3}
A F_{0x} +B F_{0z} = 0 \ ,
\end{equation}
\noindent
where
\begin{subequations}
\label{C4}
\begin{align}
A & = \left(1 - \kappa S_{xx} \right) q^2 + 3\kappa S_{xx} q_x^2 -
\left( 1 + 2\kappa S_{xx} \right) k_b^2  \ , \label{C4a} \\ 
B & = 3\kappa S_{zz} q_x q_z \ .
\label{C4b}
\end{align}
\end{subequations}
We now simplify the expression (\ref{C4a}) for the coefficient $A$.
Specifically, we substitute into this expression $q^2 = q_z^2 + q_x^2$
and $k_b^2 = \epsilon_b k^2 = \epsilon_b(q_z^2/\eta_x +
q_x^2/\eta_z)$, where we have used the dispersion relation
(\ref{disp_2D_p}). This yields
\begin{align}
A & = \left(1 - \kappa S_{xx} \right) (q_z^2 + q_x^2)  +
 3\kappa S_{xx} q_x^2  \nonumber \\
   & - \epsilon_b \left(1 + 2 \kappa S_{xx} \right)
   \left(\frac{q_z^2}{\eta_x} + \frac{q_x^2}{\eta_z} \right) \ . 
\label{C4c}
\end{align}
We now use (\ref{C0}) to write out the quantities $\eta_x$ and
$\eta_z$ in (\ref{C4c}) in terms of $S_{xx}$ and $S_{zz}$. It can be
seen that the terms proportional to $q_z^2$ cancel, and we obtain
\begin{equation}
\label{C5}
A = 3\kappa S_{zz} \frac{1 + 2\kappa S_{xx}}{1 + 2\kappa S_{zz} } q_x^2 \ .
\end{equation}
We use this result and the expression (\ref{C4b}) for $B$ to compute
\begin{equation}
\label{C6}
\frac{F_{0x}}{F_{0z}} = -\frac{B}{A} = -\frac{1 + 2\kappa S_{zz}}{1 +
  2\kappa S_{xx}} \frac{q_z}{q_x} \ .
\end{equation}
\noindent
Returning to the original notations (\ref{C00}), we obtain
(\ref{Fx_over_Fz}).

Next, we show how to derive Eq.~(\ref{rp_3}) from
(\ref{rp_2}). Eq.~(\ref{rp_2}) contains the factor
\begin{equation}
\label{C7}
R \equiv \frac{\left[{\bf k}_r \times (1+\Sigma){\bf F}_0 \right] \cdot
  \hat{\bf y}}{\left[ {\bf k}_i \times (1 + \Sigma){\bf F}_0 \right] \cdot
  \hat{\bf y}} = \frac{\left[{\bf k}_r \times S{\bf F}_0 \right] \cdot
  \hat{\bf y}}{\left[ {\bf k}_i \times S{\bf F}_0 \right] \cdot
  \hat{\bf y}} \ ,
\end{equation}
which we will now evaluate. To compute the vector products, we note
that ${\bf k}_i = \hat{\bf x} k_x + \hat{\bf z} k_{iz}$, ${\bf k}_r =
\hat{\bf x} k_x - \hat{\bf z} k_{iz}$ and $S{\bf F}_0 = \hat{\bf x}
S_{xx}F_{0x} + \hat{\bf z} S_{zz}F_{0z}$. From this, we find
\begin{equation}
\label{C8}
R = \frac{k_x S_{zz}F_{0z} + k_{iz} S_{xx}F_{0x}}
         {k_x S_{zz}F_{0z} - k_{iz} S_{xx}F_{0x}} \ .
\end{equation}
Next, we use the ratio $F_{0x}/F_{0z}$ given by (\ref{C6}), account
for the conservation of the wave vector projection onto the interface,
that is, $q_x=k_x$, and re-write (\ref{C8}) as
\begin{equation}
\label{C9}
R =
\frac{k_x^2 S_{zz} \left(1 + 2\kappa S_{xx} \right) - k_{iz}q_z
  S_{xx} \left(1 + 2\kappa S_{zz} \right)} 
     {k_x^2 S_{zz} \left(1 + 2\kappa S_{xx} \right) + k_{iz}q_z
  S_{xx} \left(1 + 2\kappa S_{zz} \right)} \ .
\end{equation}
To proceed, we need to exclude the variable $k_x^2$ from (\ref{C9}).
Using the dispersion relations (\ref{disp_2D_p}) and (\ref{k_ir_2})
for the refracted and the incident waves (in the geometry considered,
$q_x^2 = k_\perp^2 = k_x^2$), we write
\begin{equation}
\label{C10}
\frac{q_z^2}{\eta_x} + \frac{k_x^2}{\eta_x} = k^2 =
\frac{1}{\epsilon_b}k_b^2 = \frac{1}{\epsilon_b}\left(k_x^2 +
  k_{iz}^2\right) .
\end{equation}
\noindent
Solving (\ref{C10}) for $k_x^2$, we obtain
\begin{align}
k_x^2 &= \frac{k_{iz}^2/\epsilon_b - q_z^2/\eta_x}{1/\eta_z -
  1/\epsilon_b} \nonumber \\
      &= \frac{1 + 2\kappa S_{zz}}{3\kappa S_{zz}} \left( \frac{1 -
          \kappa S_{xx}}{1 + 2\kappa S_{xx}} q_z^2 - k_{iz}^2 \right) \ ,
\label{C11}
\end{align}
\noindent
where we have used (\ref{C0}) to obtain the second expression from the
first. We now substitute the result given in (\ref{C11}) into
(\ref{C9}). The factors of $1 + 2\kappa S_{zz}$ in the numerator and
the denominator cancel, and we obtain
\begin{equation}
\label{C12}
R = \frac{(1-\kappa S_{xx})q_z^2 - (1+2\kappa S_{xx})k_{iz}^2 - 3\kappa
S_{xx} k_{iz}q_z}{(1-\kappa S_{xx})q_z^2 - (1+2\kappa S_{xx})k_{iz}^2 + 3\kappa
S_{xx} k_{iz}q_z} \ .
\end{equation}
At the next step, we divide the numerator and the denominator in
(\ref{C12}) by the factor $1 + 2\kappa S_{xx}$ and, accounting for the
identity
\begin{equation}
\label{C13}
\frac{3\kappa S_{xx}}{1 + 2\kappa S_{xx}} = \epsilon_b \left(
  \frac{1}{\epsilon_b} - \frac{1}{\eta_x} \right) \ ,
\end{equation}
\noindent
obtain
\begin{equation}
\label{C14}
R = 
\frac{\displaystyle \frac{q_z^2}{\eta_x} - \frac{k_{iz}^2}{\epsilon_b} -
  \left(\frac{1}{\epsilon_b} - \frac{1}{\eta_x} \right)k_{iz}q_z
}{\displaystyle \frac{q_z^2}{\eta_x} - \frac{k_{iz}^2}{\epsilon_b} +
  \left(\frac{1}{\epsilon_b} - \frac{1}{\eta_x} \right)k_{iz}q_z} \ . 
\end{equation}
The expressions in the numerator and denominator can now be
factorized, and we arrive at the final result
\begin{equation}
\label{C15}
R = 
- \frac{\displaystyle \left(q_z + k_{iz} \right) \left( \frac{k_{iz}}{\epsilon_b} -
      \frac{q_z}{\eta_x} \right) }{\displaystyle \left(q_z - k_{iz} \right) \left( \frac{k_{iz}}{\epsilon_b} +
      \frac{q_z}{\eta_x} \right)} \ .  
\end{equation}
Substitution of this expression into (\ref{rp_2}) immediately results
in (\ref{rp_3}). 

\section{Proof of Theorem 1}
\label{app:proof}

\subsection*{C.1. An equivalence transformation}

To derive the equality (\ref{theo}), we first introduce some notation.
Let

\begin{subequations}
\label{A1}
\begin{align}
\label{vareps_def}
\varepsilon &\equiv \langle \phi \vert \psi \rangle \ , \\
\label{Proj_def}
P &\equiv \vert \psi \rangle \langle \phi \vert \ , \\
\label{R_def}
R({\mathcal Z};A) & \equiv ({\mathcal Z} - A)^{-1} \ , \\
\label{B_def-app}
B & \equiv R({\mathcal Z};W)W \ , \\
\label{sigma_def}
\sigma & \equiv \langle \phi \vert R({\mathcal Z}; W) \vert \psi \rangle \ .
\end{align}
\end{subequations}

\noindent
Here $R({\mathcal Z};A)$ is the resolvent of the linear
operator $A$ and ${\mathcal Z}$ is a complex number. In the new
notation, the operator $T$ defined in (\ref{T_def}) takes the form

\begin{equation}
\label{A1a}
T = 1 - \frac{1}{\varepsilon} P
\end{equation}

\noindent
and Eq.~(\ref{theo}) is rewritten as

\begin{equation}
\label{A3}
\sigma = \frac{1}{\mathcal Z} \frac{\varepsilon}{1 - {\displaystyle \frac{1}{\varepsilon}
  \langle \phi \vert R({\mathcal Z}; WT)W \vert \psi \rangle} } \ . 
\end{equation}

\noindent
Note that, by the first hypothesis of the Theorem, $\varepsilon\neq
0$.

We now write the following chain of equalities in which the second
hypothesis of the Theorem, namely, that $R({\mathcal Z};
W)$ exists, has been used:

\begin{align}
\label{A4}
R({\mathcal Z};WT) & = \left( {\mathcal Z} - WT \right)^{-1} \nonumber
\\
& = \left( {\mathcal Z} - W +
\frac{1}{\varepsilon}WP \right)^{-1} \nonumber \\
& = \left(R^{-1}({\mathcal Z};W) +
\frac{1}{\varepsilon}WP \right)^{-1} \nonumber \\ 
& = \left[R^{-1}({\mathcal Z}; W)\left( 1 + \frac{1}{\varepsilon} R({\mathcal Z}; W)
 WP\right) \right]^{-1} \nonumber \\
& = \varepsilon \left[\varepsilon + R({\mathcal Z}; W)
 WP \right]^{-1} R({\mathcal Z};W) \ . 
\end{align}

\noindent
Using the last equality in (\ref{A4}) and the notation (\ref{B_def-app}),
we rewrite (\ref{A3}) identically as

\begin{equation}
\label{A2}
\sigma  = \frac{1}{\mathcal Z}
\frac{\varepsilon}{1 - \langle \phi \vert (\varepsilon + BP)^{-1} B \vert \psi \rangle} \ .
\end{equation}

\subsection*{C.2. A useful identity} 

Below, we will frequently use the following identity:

\begin{equation}
\label{a_B_b}
\langle \phi \vert B \vert \psi \rangle = {\mathcal Z} \sigma -
\varepsilon \ .
\end{equation}

\noindent
The above equation is easily derived by noting that

\begin{subequations}
\label{A6a}
\begin{align}
& \langle \phi \vert B \vert \psi \rangle = \langle \phi \vert ({\mathcal Z} -
W)^{-1}W \vert \psi \rangle \nonumber \\
 &= \langle \phi \vert ({\mathcal Z} -
W)^{-1}(W - {\mathcal Z}) \vert \psi \rangle + {\mathcal Z} \langle
\phi \vert ({\mathcal Z} - W)^{-1} \vert \psi \rangle \nonumber \\
 &= -\varepsilon + {\mathcal Z}\sigma \ .
\end{align}
\end{subequations}

\subsection*{C.3. The main derivation}

To proceed, we need to express the operator $(\varepsilon + BP)^{-1}$,
which appears in the right-hand side of (\ref{A2}), in a more
tractable form. To this end, consider the equation

\begin{equation}
\label{A5}
(\varepsilon + BP) \vert x \rangle = \vert b \rangle \ ,
\end{equation}

\noindent
where $\vert x \rangle$ is viewed as the unknown and $\vert b \rangle
\neq 0$ is an otherwise arbitrary element of the same Hilbert space.
Using the definition of $P$ (\ref{Proj_def}), we transform (\ref{A5})
to

\begin{equation}
\label{A6}
\varepsilon \vert x \rangle + B \vert \psi \rangle \langle \phi \vert x
\rangle = \vert b \rangle \ ,
\end{equation}

\noindent
project the result onto $\vert \phi \rangle$, and find that

\begin{equation}
\label{A7}
\langle \phi \vert x \rangle = \frac{\langle \phi \vert b
  \rangle}{\varepsilon + \langle \phi \vert B \vert \psi \rangle} \ .
\end{equation}

\noindent
We now use the previously-derived identity (\ref{a_B_b}) in the
right-hand side of (\ref{A7}) to obtain

\begin{equation}
\label{A8}
\langle \phi \vert x \rangle = \frac{\langle \phi \vert b
  \rangle}{{\mathcal Z}\sigma} \ .
\end{equation}

\noindent
Upon substitution of (\ref{A8}) into (\ref{A6}), we find the solution
to (\ref{A5}) or (\ref{A6}), namely,

\begin{equation}
\label{A9}
\vert x \rangle = \frac{1}{\varepsilon} \left( 1 - \frac{B\vert \psi
    \rangle \langle \phi \vert}{{\mathcal Z}\sigma} \right) \vert b
\rangle \ .
\end{equation}

\noindent
Since the vector $\vert b \rangle$ in (\ref{A5}) is arbitrary, we
conclude that

\begin{equation}
\label{A10}
(\varepsilon + BP)^{-1} = \frac{1}{\varepsilon} \left( 1 - \frac{B
    \vert \psi \rangle \langle \phi \vert}{{\mathcal Z} \sigma} \right) \ .
\end{equation}

\noindent
This equality can be verified directly by substitution.

\subsection*{C.4. Putting everything together}

We can now put everything together and obtain (\ref{A2}). From
(\ref{A10}), we have

\begin{equation}
\langle \phi \vert (\varepsilon + BP)^{-1} =  \frac{1}{\varepsilon}
\left( 1 - \frac{\langle \phi \vert B \vert \psi \rangle}{{\mathcal Z}
    \sigma}\right) \langle \phi \vert = \frac{\langle \phi \vert }{{\mathcal
      Z}\sigma} \ ,
\end{equation}

\noindent
where we have, again, used (\ref{a_B_b}). Now, we can write

\begin{equation}
\langle \phi \vert (\varepsilon + BP)^{-1} B \vert \psi \rangle =
\frac{\langle \phi \vert B \vert \psi \rangle}{{\mathcal Z}\sigma} = 1 -
\frac{\varepsilon}{{\mathcal Z}\sigma} \ .
\end{equation}

\noindent
Upon substitution of this result into the right-hand side of
(\ref{A2}), we find that the latter is, indeed, an identity, and so
are (\ref{A3}) and (\ref{theo}).

\end{document}